\documentclass[iop,revtex4]{emulateapj}
\usepackage{amsmath}

\tighten
\usepackage[hyperfootnotes=false,naturalnames=true,pdfstartview=FitH,pdfpagemode=UseNone]{hyperref}

\newcommand{\asec}{\hbox to 1pt{}\rlap{$^{\prime\prime}$}.\hbox to 2pt{}}
\newcommand{\amin}{\hbox to 1pt{}\rlap{$^{\prime}$}.\hbox to 2pt{}}

\slugcomment{Accepted for Publication in {\it The Astrophysical Journal}}
\shortauthors{Lauer et al.}
\shorttitle{Present Epoch Brightest Cluster Galaxies}

\begin{document}

\title{Brightest Cluster Galaxies at the Present Epoch}

\author{Tod R. Lauer}
\affil{National Optical Astronomy Observatory,\footnote{The
National Optical Astronomy Observatory is operated by AURA, Inc.,
under cooperative agreement with the National Science Foundation.}
P.O. Box 26732, Tucson, AZ 85726}

\author{Marc Postman}
\affil{Space Telescope Science Institute,
\footnote{Operated by the Association of Universities for Research in Astronomy, Inc., for the National Aeronautics and Space Administration.}
3700 San Martin Drive, Baltimore, MD 21218}

\author{Michael A. Strauss, Genevieve J. Graves, and Nora E. Chisari}
\affil{Department of Astrophysical Sciences, Princeton University,
Princeton, NJ}

\begin{abstract} 
We have obtained photometry and spectroscopy of 433 $z\leq0.08$
brightest cluster galaxies (BCGs) in a full-sky survey of Abell clusters
to construct a BCG sample suitable for probing
deviations from the local Hubble flow.
The BCG Hubble diagram over $0<z<0.08$ is consistent to within 2\%\ of
the Hubble relation specified by a $\Omega_m=0.3,$ $\Lambda=0.7$ cosmology.
This sample allows us to explore the structural and photometric properties
of BCGs at the present epoch, their location in their hosting
galaxy clusters, and the effects of the cluster environment
on their structure and evolution.
We revisit the $L_m-\alpha$ relation for BCGs,
which uses $\alpha,$ the log-slope of the BCG photometric curve of growth,
to predict the metric luminosity
in an aperture with 14.3 kpc radius, $L_m,$ for use as a distance indicator.
Residuals in the relation are 0.27 mag rms.
We measure central stellar velocity dispersions, $\sigma,$
of the BCGs, finding the Faber-Jackson relation to flatten
as the metric aperture grows to include an increasing fraction
of the total BCG luminosity.
A 3-parameter ``metric plane'' relation using $\alpha$ and $\sigma$ together
gives the best prediction of $L_m,$ with 0.21 mag residuals.
The distribution of projected spatial offsets, $r_x$ of BCGs
from the X-ray-defined cluster center is a
steep $\gamma=-2.33$ power-law over $1<r_x<10^3$ kpc.
The median offset is $\sim10$ kpc,
but $\sim15$\%\ of the BCGs have $r_x>100$ kpc.
The absolute cluster-dispersion normalized BCG peculiar velocity
$|\Delta V_1|/\sigma_c$ follows an exponential distribution
with scale length $0.39\pm0.03.$
Both $L_m$ and $\alpha$ increase with $\sigma_c.$
The $\alpha$ parameter is further moderated by both the
spatial and velocity offset from the cluster center, with
larger $\alpha$ correlated with the proximity
of the BCG to the cluster mean velocity or potential center.
At the same time, position in the cluster has little effect on $L_m.$
Likewise, residuals from the metric plane show no correlation
with either the spatial or velocity offset from the cluster center.
The luminosity difference between
the BCG and second-ranked galaxy, M2, increases as the peculiar
velocity of the BCG within the cluster decreases.
Further, when M2 is a close luminosity ``rival'' of the BCG,
the galaxy that is closest to either the velocity
or X-ray center of the cluster is most likely to have the larger $\alpha.$
We conclude that the
inner portions of the BCGs are formed outside the cluster, but interactions
in the heart of the galaxy cluster grow and extend the envelopes of the BCGs.
\end{abstract}

\keywords{galaxies: clusters: general --- galaxies: distances and redshifts ---
galaxies: elliptical and lenticular, cD --- galaxies: fundamental parameters ---
galaxies: photometry}

\section{The Most Massive Galaxies in the Universe}

The brightest and most massive galaxies
in the present-day Universe are the
first-ranked or brightest cluster galaxies ({\rm BCG}) in rich galaxy clusters.
The first studies of BCGs
focussed on their high and almost ``standard-candle'' luminosities,
which allowed the Hubble-flow to be characterized out to
large distances \citep{s72a,s72b,go75}.
The dispersion about the mean luminosity was shown to be
significantly smaller than would be the case had the BCGs
simply been the brightest galaxies drawn from
a standard luminosity function \citep{tr, ls}.
The narrowness of the BCG luminosity distribution does not extend to
less massive galaxy groups \citep{gp83}, however;
and more recent work argues that only
the more luminous BCGs may be special \citep{lom}.
These results highlight the need to understand at what mass scale
the unique formation and evolution mechanisms that shape BCGs come into play.

The acronym ``BCG" underscores that these galaxies are
tied to the galaxy clusters that host them.
If BCGs are indeed special it is likely to be because their
formation and evolution is tied to physical mechanisms unique
to rich galaxy clusters.
Cannibalism, whereby a BCG sitting in the middle of the cluster potential
tends to engulf and merge with its neighbors, has been invoked to explain
the high luminosity of these systems (e.g., \citealt{ot75};
\citealt{ho}, but see also \citealt{rich75}), but it remains
unclear why this leads to such uniform properties, especially when we
know that many clusters undergo interactions and merging.
BCG growth by cannibalism does appear to take place in clusters at some level
\citep{l88}, however, dynamical arguments suggest that most of
the BCG assembly takes place outside the cluster \citep{m85}.

The properties of BCGs are distinct
from those of the other galaxies in clusters, and any model for their
formation has to acknowledge this.
BCGs generally sit close to the X-ray centers of their hosting clusters
and usually have small ``peculiar'' velocities relative to the cluster mean.
\citet{sastry}, \citet{bing}, \citet{lambas}, and others
showed that BCGs tend to be aligned with their parent cluster.
This has been explored in detail with data from the
SDSS by \citet{no10} and \citet{hao}, who
found that this alignment is marked only in clusters in which the
BCG is {\it dominant}, i.e., more than 0.65 mag brighter than the
average of the second and third-ranked galaxies.
Those clusters in which the BCG is not
strongly dominant may be systems that recently
underwent a merger, and are therefore not completely relaxed.
In short, in many ways the BCG reflects the environment of the
cluster that hosts it.

Our approach to understanding the origin
of BCGs is to conduct an extensive examination
of their present-day structure, luminosity, and cluster environments.
We organize our thinking around three broad questions:

\subsection{What Are the Present-Day Properties of BCGs?}

The dispersion in the luminosities of the BCGs
about the mean Hubble relation, measured by the first studies
to use BCGs as distance indicators,
was typically 0.3 to 0.4 magnitudes \citep{s72a,s72b,go75}.
An important refinement of the use of BCG as
distance indicators was developed by \citet{h80},
who showed that BCG metric luminosity, $L_m,$ was correlated 
with the logarithmic slope, $\alpha,$ of the photometric curve-of-growth.
The $L_m-\alpha$ relation is a form of a luminosity-radius relation
that side-steps the difficulties of characterizing the
extended envelopes of BCG at large radii and faint isophotal levels.
\citet[hereafter PL95]{pl95} reinvestigated the
use of BCG as distance indicators, using the $L_m-\alpha$ relation
for a full-sky characterization of the linearity of the local Hubble-flow
\citep{lp92} and providing a distant reference-frame to measure the relative
peculiar velocity of the Local Group \citep{lp94}.
Residuals about the PL95 $L_m-\alpha$ relation were only 0.25 mag (rms).

This paper presents a large full-sky sample of BCGs
in Abell clusters over the redshift range $0<z\leq0.08.$
The original goal for obtaining this sample was to
extend the bulk-flow analysis of \citet{lp94} to greater distances.
That work implied that the Abell clusters
within $z\leq0.05$ participated in a coherent motion in excess of
$689\pm178~{\rm km~s^{-1}}$ superimposed on the background cosmological
expansion or ``Hubble flow'' within the volume containing the sample.
This analysis will be presented in a separate work.
Requirements for measuring accurate bulk flows, however,
specify much of the sample definition, observational methodologies,
and analysis of the BCG properties undertaken in this work.
A full sky sample allows for the optimal determination
of any large-scale bulk mass flow.
The relatively low redshift-limit of the sample and its
overall size is dictated by the scale out to which the BCGs
can be used as accurate distance indicators.
The observational methodology is driven by the need to obtain
highly uniform photometry over the angular and spatial extent of the sample.
Much of the analysis is a reinvestigation of the use
of BCGs as distance indicators,
with a substantially larger sample and new observations that go well beyond
the material available to PL95.

Regardless of the bulk-flow analysis,
the present sample offers an excellent opportunity to 
assess the structural properties of BCG
to understand their origin, evolution over time, and
their particular uniqueness as the luminous endpoint of galaxy formation,
problems that were not addressed by the smaller sample and
less-complete cluster information available to \citet{pl95}.
\citet{oh} and \citet{l07}, for example, found that the central
stellar velocity dispersions, $\sigma,$ of BCGs increase very slowly
if at all with the {\it total} BCG luminosity
(also see \citealt{b07,von,liu}).
Typical BCG $\sigma$ values are modest
for their large luminosities, which may reflect the origin
of BCGs in ``dry'' mergers \citep{boylan}.
In contrast, BCGs are unusually extended as compared to giant ellipticals,
as is seen in the relation between effective radius, $R_e,$
and total luminosity of the BCGs \citep{l07,b07}.
We will use the structure of BCGs as a probe
of the effects of cluster environment on their evolution.
In a companion paper \citep{p2} we will compare
the structure of BCGs to those of other highly luminous elliptical galaxies.
The mutual relations between
$L,$ $\sigma,$ and $R_e$ for elliptical galaxies overall are
understood as various projections
of the ``fundamental plane'' \citep{7sfp, dnd}.
Understanding the relationship of BCGs to the fundamental plane
will be explored in \citet{p2}.

\subsection{Where Are the BCGs Located in Their Galaxy Clusters?}

The ``textbook'' picture of a galaxy cluster is that it is a swarm of
galaxies anchored by a massive cD residing at rest in the exact center
of the potential as marked by hot, X-ray emitting, gas.
Early work on the X-ray morphology of galaxy clusters \citep{j84}
and their velocity structure \citep{ql} indeed show that
the BCG is likely to be centrally located.
There are certainly examples of such clusters in our sample.
At the same time, there are also
massive galaxy clusters, like Coma (Abell 1656), in which neither the BCG,
nor second-ranked galaxy, M2, are at the center of the potential.
Coma may be the recent merger of two clusters, and this is the point ---
the position of the BCG with respect to the center of the potential,
X-ray emission, may testify to the evolutionary state of both
the BCG and the cluster.  More recent work \citep{patel,h14} shows
that the BCG is often displaced from the center of the
cluster potential as defined by the X-ray emission.
For the present sample of clusters we have
quantified the distribution of projected spatial-offsets of the BCGs,
finding that it is a steep power-law over three decades in radius.

For the BCG to reside at the spatial center of the cluster, it must also be
at rest there.  It has long been known that there are
BCGs with ``significant'' peculiar velocities within the cluster
\citep{zab90, mal, zab93, oh01}.
PL95 described the overall distribution of BCG peculiar
velocities within their sample as a Gaussian with dispersion, $\sigma_1=
264~{\rm km~s^{-1}},$ comparing this to the substantially larger
mean cluster velocity-dispersion $\sigma_c=666~{\rm km~s^{-1}}.$
We attempted to verify this result with our much larger present sample,
finding now that the distribution of peculiar velocities is exponential,
extending out to galaxies with $\Delta V_1>\sigma_c.$

Both the distributions of the spatial and velocity offsets
of the BCGs are particularly interesting when compared to the
\citet{martel} simulations of galaxy cluster formation and evolution.
These simulations emphasize that the location of the BCG within a cluster
bears witness to its history of formation from smaller accreted
groups and clusters.  The dark matter, galaxy, and X-ray-emitting gas
distributions in any cluster all have different timescales and
physical mechanisms for responding to the accretion or interaction
with another cluster.  The locations of the BCGs reflect this.
\citet{skibba} studied the peculiar velocities and spatial
positions of a large sample of clusters and groups,
showing that the location of the brightest galaxy in the systems
provides a sharp test of the mechanisms that formed them.

\subsection{How Does the Cluster Environment Relate to the Properties
of the BCGs?}

The relationship of the structure and luminosity of the BCG
to the properties of the cluster has proven to be a multi-faceted problem.
Initial work showed the BCG luminosity to be only weakly related
to the richness of the clusters \citep{s72b,sh73,s75,s76}.
We re-investigated this relation in PL95,
and saw no relation between the metric luminosity and cluster richness.

BCG luminosity and structure, however, do appear
to be related to the X-ray properties of the clusters.
\citet{schom} found that the envelope luminosity of cD galaxies, a subset
of the BCGs, increases with total cluster X-ray luminosity.
\citet{edge1} and \citet{edge2} found a strong relationship
between BCG luminosity and cluster X-ray temperature,
which itself is closely related to the cluster velocity dispersion \citep{st72}.
\citet{he97} and \citet{cm98} also found that BCG luminosity
increases with cluster X-ray luminosity.
Lastly, \citet{brough} found the structure of the BCG to also
correlate with cluster X-ray luminosity, with the BCG envelope becoming
more extended in more luminous clusters.

In this paper we show that both the BCG luminosity,
and the radial extent (as characterized
by $\alpha$) of their envelopes (where by ``envelope" we mean
simply the outer portions of the galaxies)
correlate with cluster velocity dispersion.
We take this a step further, however, finding that the extent of the
envelope is related to both the spatial and velocity positions of
the BCG {\it within} its hosting cluster.  The luminosity and structural
difference between the BCG and the second ranked galaxy, M2, also appears
to depend on which galaxy has the smaller peculiar
velocity within the cluster or the smaller offset from
the center of the cluster potential as marked by X-ray emission.
\citet{bg83} found that early type galaxies with extended halos
(e.g., D or cD galaxies) lie on significant peaks in the cluster
galaxy distribution regardless of whether they are the BCG.
We now see how the structure of the BCG itself changes smoothly
as a function of how close to the center of the cluster it resides.

\subsection{This Paper}

We begin in $\S\ref{sec:sample}$
with the geometric and redshift selection of the Abell clusters
defining the present sample, detailing the imaging observations
used both to select the BCG for any given cluster and to provide
accurate surface photometry.  Spectroscopic observations are presented,
which provide BCG redshifts and central stellar velocity dispersions.
A crucial part of the sample definition is the derivation of
accurate mean redshifts and velocity dispersions for the galaxy clusters.
The projected spatial and velocity locations of the BCGs within
their clusters is presented in $\S\ref{sec:loc}.$
The photometric and kinematic properties of the BCGs
are presented in $\S\ref{sec:prop}$, with particular attention
to parametric relations between the metric luminosity and BCG structure.
This section also explores the relationship between BCG properties
and cluster environment.
Additional information about the BCGs is provided by the properties
of the second-ranked galaxies, M2, which are presented in $\S\ref{sec:m2}.$
We summarize what we have learned about
the origin and evolution of BCGs in the final section of the paper.

\begin{figure*}[t!]
\plotone{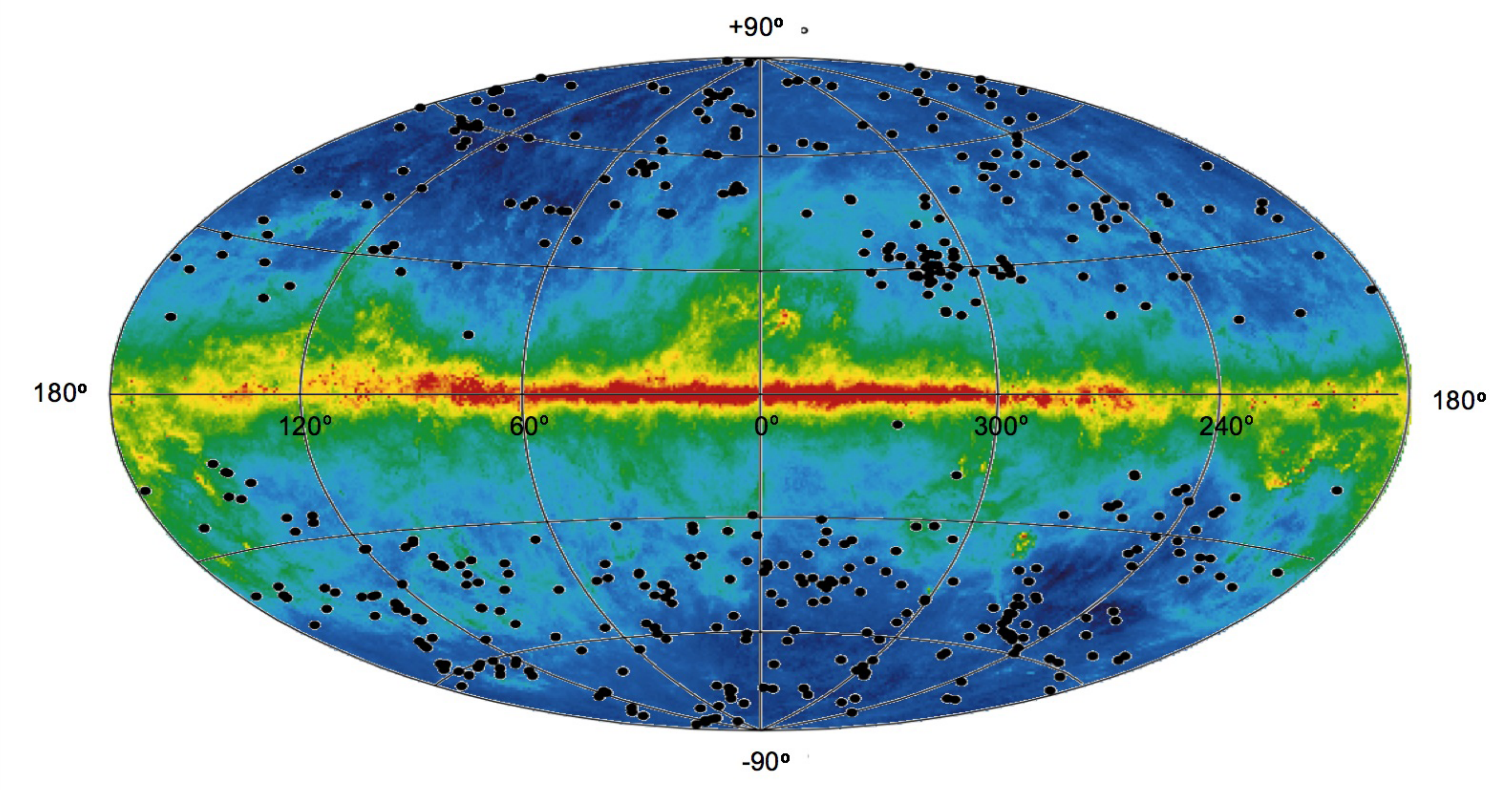}
\caption{The distribution of the present sample of BCGs is shown in galactic
coordinates superimposed over a predicted 94 GHz dust map derived from
IRAS and COBE. The dense concentration of clusters at
$l\sim315^\circ,\ b\sim+30^\circ$ is due to the combination of the 
Hydra-Centaurus and Shapley superclusters.
The dust map is a publicly available data 
product derived using the {\tt predict\_thermal} algorithm
by \citet{fds} and shows the predicted dust emission,
in mK antenna temperature units at 94 GHz, using their 2-component model 8.}
\label{fig:samp_sky}
\end{figure*}

\section{A Full-Sky Sample of Local BCGs}\label{sec:sample}

\subsection{Definition of the Sample}

The present sample of BCGs comprises 433 \citet{abell} and ACO
\citep{ACO} galaxy clusters with mean heliocentric velocities,
$V<24,000~{\rm km~s^{-1}}$ and galactic latitude, $|b|\ge15^\circ.$
There is no limit on the minimum richness class of the clusters.
Table \ref{tab:sample} lists the BCG coordinates,
and heliocentric velocities, $V_1,$
as well as the cluster velocities, $V_c,$ cluster velocity dispersions,
$\sigma_c,$ the number of galaxy velocities used to compute these, $N_g,$
and the \citet{sfd} $A_B$ values.
The distribution on the sky in galactic coordinates is shown in
Figure \ref{fig:samp_sky}.  We use a cosmological model
with $H_0=70~{\rm km~s^{-1}~Mpc^{-1}},$ $\Omega_m=0.3,$ and
$\Lambda=0.7$ throughout this paper.

\begin{figure}[!t]
\includegraphics[width=\columnwidth]{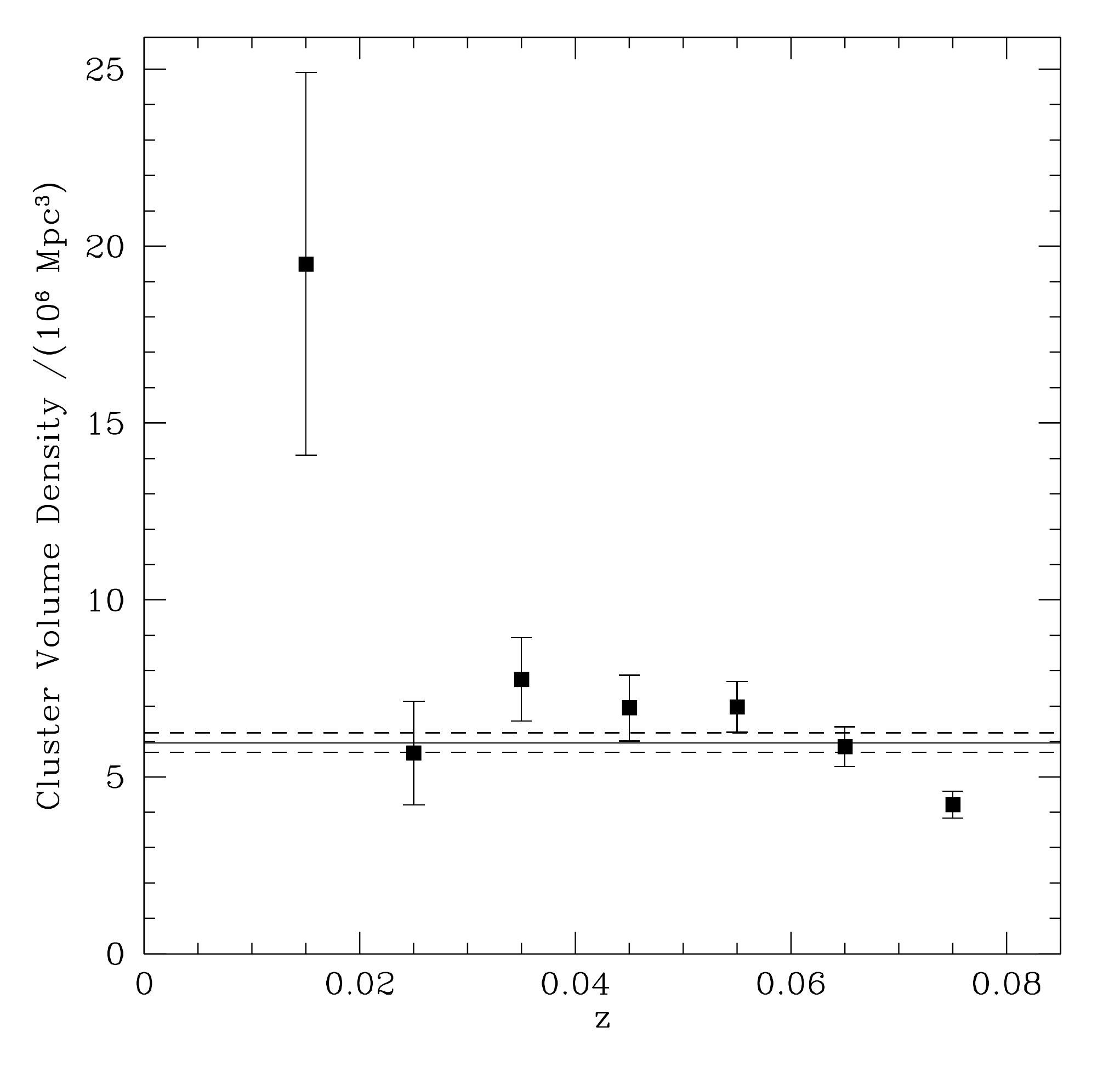}
\caption{The comoving volume density of the clusters in this study,
binned in $\Delta z=0.01$ shells
as a function of redshift is shown. 
The error bars reflect the Poisson errors in the number of clusters
in each shell. The horizontal lines give the density (and associated error)
of BCGs averaged over the entire volume,
but including 53 additional clusters lacking observations
or that were excluded for having non-elliptical BCGs.
The volume calculation accounts for
the $\pm15^\circ$ galactic-plane ``zone of avoidance'' and the counts in each bin are 
corrected for the Abell cluster galactic latitude selection
function $P(|b|) = {\rm dex}(0.3 [1 - csc|b|])$. }
\label{fig:samp_vol}
\end{figure}

The sample was originally designed to serve as a reference frame to measure
the peculiar velocity of the Milky Way.
The inferred luminosities of the BCGs serve as distance indicators,
following the methodology presented in \citet[hereafter LP94]{lp94}.
In that paper a volume-limited frame was constructed
from 119 clusters with $V<15,000~{\rm km~s^{-1}},$
again using a $\pm15^\circ$ galactic ``zone-of-avoidance.''
The present sample largely includes the LP94 set (as will be qualified
further below).
For convenience, we will refer to the LP94 set of clusters
as the 15K sample, while its present augmentation is the 24K sample.
The 24K outer limit of the survey was selected to provide
a significantly deeper reference frame than that constructed in LP94, but
one that would not be too strongly affected by the limited
depth of the Abell and ACO catalogues, which are heavily incomplete
beyond $z\sim0.1$ \citep{p92}.

The present sample is drawn from a considerably larger provisional
sample defined by us in the early 1990's based on a literature survey
of Abell clusters with measured or estimated redshifts.
Because we wanted to construct the best sampling of the local volume possible
within the limitations of the Abell and ACO catalogues, we were
liberal with accepting plausible candidates for the 24K sample.
As we describe below, the final cluster selection
is made with a combination of the latest published redshift surveys and
radial velocities that we measured ourselves.

The final set of 433 clusters are those of the larger candidate set
for which we observed the BCG, and had sufficient redshift information
to determine that the cluster was within the 24K redshift limit.
There are 38 additional clusters that had well-determined
redshifts that placed them within the 24K volume, but for which
we were unable to image the BCG or sufficient candidates
to be confident of the BCG selection.
There are 15 more clusters with non-elliptical BCGs,
which were also excluded from the sample.
These clusters are listed in Table \ref{tab:no_obs}.
These two sets include 5 clusters
observed as part of the 15K sample, but that are now deleted
from the present sample for a variety of reasons.
With the present richer data set we now
find that for two of the 15K clusters we in fact observed a
small foreground group in front of a rich cluster, selected
the M2 rather than the correct BCG in one cluster,
and that the BCG is non-elliptical in the remaining two clusters.

\begin{figure*}[!t]
\plotone{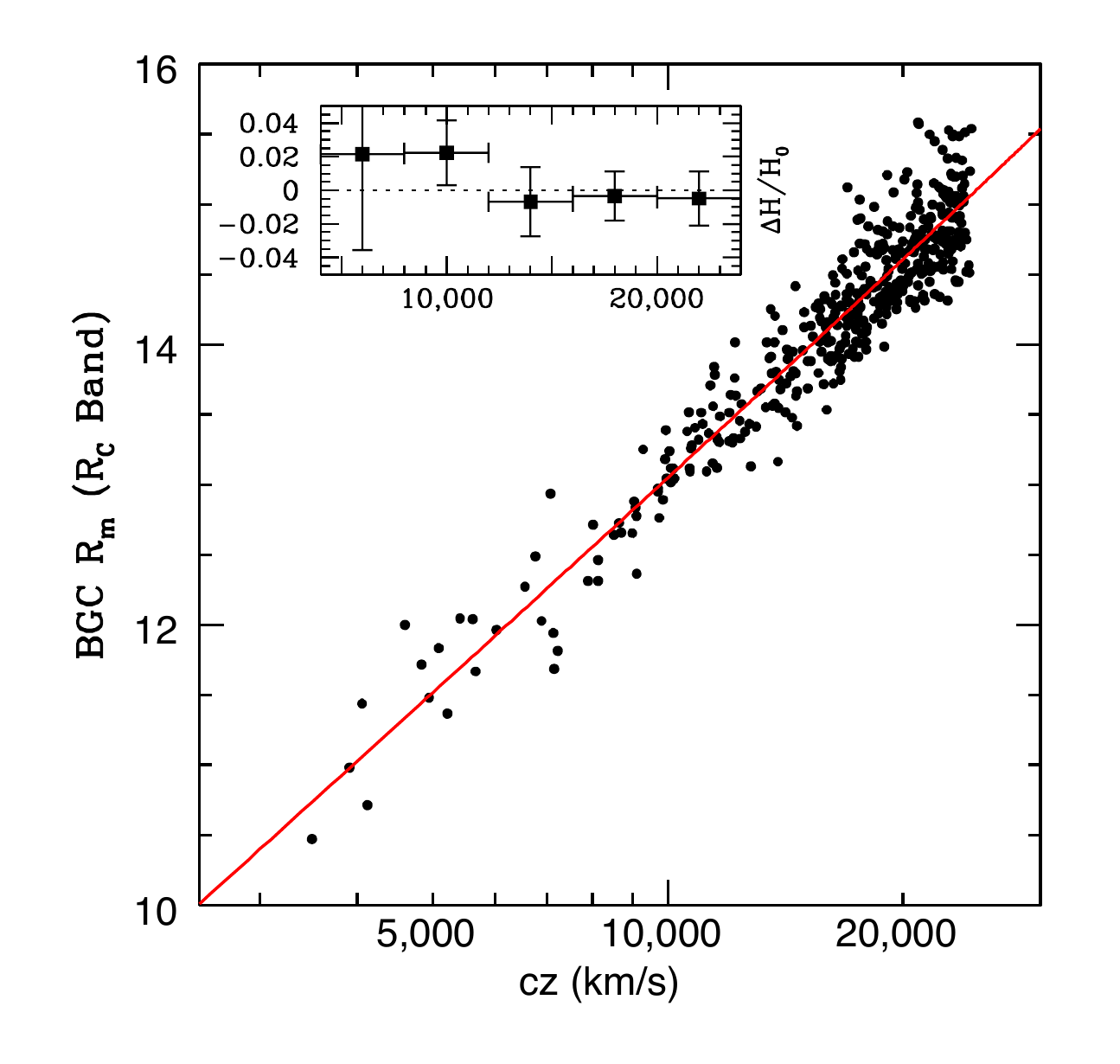}
\caption{The Hubble diagram derived from the
present 24K sample, showing BCG apparent metric luminosity as a function
of cluster redshift (CMB frame).
The red line is the mean theoretical Hubble relationship assuming
$\Omega_m=0.3,$ and $\Lambda=0.7.$
The inset shows the binned residuals about relation in shells of
4000 ${\rm km~s^{-1}}$ starting at 4000 ${\rm km~s^{-1}},$
expressed as $\Delta H/H_0.$
The largest deviation in a shell is $\Delta H/H_0=-0.022,$
but none is significantly different from zero.}
\label{fig:hubble}
\end{figure*}

The cluster space density as a function of redshift is shown in
Figure \ref{fig:samp_vol}. The cluster counts in each bin
have been weighted by the established Abell cluster galactic
latitude selection function:
$P(|b|) = {\rm dex}(0.3 [1 - csc|b|])$ \citep{bs83}. The volume computation
accounts for the $\pm15^\circ$ galactic-plane zone-of-avoidance.  
The average cluster comoving space density over the range
$z \leq 0.08$ is $(5.97 \pm0.28)\times10^{-6}~
{\rm Mpc^{-3}},$ including the 53 missed and non-elliptical BCG clusters.
The space density of Abell clusters used in this study is relatively constant,
with the exception of the last bin where a significant decline is seen.
The positive deviation in the first bin, 
centered at $z = 0.015$, is not statistically significant ($\sim 2.5\sigma$).

The completeness of the Abell catalog as a function of richness,
redshift, and galactic latitude has been extensively studied.
The trends are such that the completeness is lower at lower richness,
higher redshift, and lower galactic latitude.
More specifically, \cite{dr2002} show that the detection efficiency
as a function of richness class in the Abell catalog is $\sim$55\%\ for
RC $=\ 0$, $\sim$75\%\ for RC $=\ 1$ and is essentially 100\%\ for RC $\ge\ 2$.
The richness class distribution in the current 24K sample is
55\%\ RC=0, 35\%\ RC=1, and 10\%\ RC $\ge$ 2.
Our study here is immune to the known completeness trends so long
as the properties of the BCG in the detected Abell clusters
are representative of those in the clusters that were missed during
the construction of the northern and southern Abell catalogs.

\subsubsection {A BCG Hubble Diagram}

As a further illustration of the sample geometry and its
utility as a probe of the Hubble flow within the local volume,
we show a Hubble diagram derived from our BCG sample in
Figure \ref{fig:hubble}. The details of the sample selection,
reduction, and analysis of the photometry needed to
generate this figure are the subject of
much of the rest of this paper.
For now, the relevant details are that the velocities
are mean cluster velocities in the cosmic microwave background
(CMB) frame, and the photometry is the
{\it apparent} metric luminosity, $R_m,$ of the BCGs,
but with extinction and k-corrections applied.  The photometry
has also been corrected to $\alpha=0.5$ using the relationship
(equation \ref{eqn:lalplog})
between metric luminosity and $\log\alpha,$ a parameter
measuring the slope of the photometric curve-of-growth at
the metric radius, $r_m.$

The Hubble diagram shows that the number of galaxies
per velocity interval, rises with distance as $\sim D^2,$
as expected for a survey with a roughly constant cluster
density with redshift.
The sharp cut-off at the 24K velocity limit is also evident.
The rms scatter about the nominal theoretical relation specified by
$\Omega_m=0.3,$ and $\Lambda=0.7$ is 0.271 mag.
We can constrain any departures from the expected Hubble flow
as a function of redshift by binning the residuals about the Hubble relation.
Since our sample is full sky, this effectively tests
for monopole variations in the Hubble flow with distance.
Figure \ref{fig:hubble} shows the mean residuals
in shells of 4000 ${\rm km~s^{-1}}$ starting at 4000 ${\rm km~s^{-1}},$
expressed as $\Delta H/H_0.$
All but the innermost shell are consistent with
$\Delta H/H_0=0$ at the $<2.0\%$ level; the innermost shell
is also consistent with $\Delta H/H_0=0,$ but with a poorer
5.7\%\ error due to the small number of BCGs interior to 8000
${\rm km~s^{-1}}.$
The present result is completely
consistent with the BCG Hubble diagram derived from
the earlier 15K sample \citep{lp92},
but with errors nearly a factor of two smaller.

\subsection{Selection of the BCGs}

We define the BCG to be the brightest member (in the $R_C$-band) of the cluster
within a 14.3 kpc radius\footnote{This is the
same as the $10~h^{-1}$ kpc radius used in LP94 and PL95.}
``metric aperture'' centered on the
galaxy (and with close or embedded companions photometrically subtracted),
with the proviso that the galaxy must also be an elliptical.
The use of metric BCG luminosity as a distance indicator
was initially advanced by \citet{go75}, and developed
further by \citet{h80} and PL95.  As demonstrated in
PL95, our particular choice of the metric aperture minimizes
the scatter in the average BCG luminosity.
The aperture is large enough to
include a large fraction of the total luminosity of the BCG,
but avoids the difficulty of measuring a total magnitude
for the BCG, which requires surface photometry at very faint
levels and large angular radii in a rich-cluster environment.
This problem has limited the accuracy of a number of
recent studies of BCGs.
The photometry provided by the SDSS and 2MASS surveys, for example,
strongly underestimates the total luminosity of low-$z$ BCGs \citep{l07}.
The SDSS photometry suffers from
over-subtraction of the sky background, while the 2MASS total
magnitudes are based on a profile model
that fails to include the extensive envelopes of the galaxies.

As with the definition of the cluster sample,
we were liberal with observing all
plausible BCG candidates for any given cluster,
making the final choice only when all the observations were in hand.
The initial selection of BCG candidates
was done visually from digitized sky-survey plates,
augmented with velocity information when available.
Unless one galaxy was strongly dominant and known
to be in the cluster from its redshift,
we would typically select several bright elliptical galaxies
for imaging and spectroscopic observations, with the final selection based on
CCD aperture photometry and knowledge of the cluster redshift.

As noted in PL95, the BCGs in the 15K sample were often displaced
in angle and/or velocity from the nominal cluster center, thus we attempted
to select all bright elliptical galaxies within the nominal Abell radius
of the cluster, rather than the brightest ``central'' galaxy.
The Abell radius is $1.5h^{-1}~{\rm Mpc}$ or 2.1~Mpc for the present
cosmological parameters.\footnote{This is somewhat larger than the
over-density radius, $r_{200},$ \citep{r200} which has a median value
of 1.7~Mpc for the present sample, and is often used
as a proxy for the physical extent of a galaxy cluster.}
One of the questions that will be
considered in the later sections is the extent to which the BCG is
in fact displaced from the cluster spatial and velocity centroids ---
allowing for the possibility that the BCG may be significantly offset
from either is critical to the BCG selection.

The selection of the BCG can be complex,
and different surveys may disagree on which galaxy is the
BCG in any given cluster.
As one example, we compared our selection to those from
\citet{von}, who extracted their sample from the SDSS-based C4 cluster
catalogue \citep{c4}, using isophotal magnitudes for BCG luminosity.
Of the 429 C4 clusters selected by \citet{von}
that should be within our redshift limit, only 44 clusters are in our sample.
\footnote{There were 8 additional clusters that might have been
in the common set.  We were not able to obtain BCG photometry in 4 of them,
and the BCGs in the remaining 4 were not elliptical galaxies.}
Of the 44 cluster matches, we agreed on the BCG in 33 or 75\%\  of the clusters.
In 8 of the 11 clusters remaining \citet{von} selected a galaxy that we
classified as M2, the second-ranked galaxy, based on our photometry.
As noted in the next section, this choice may depend on the size of
the metric aperture, but we concluded that our M2 would be
the BCG based on total flux (see below) in only 3 of the 8 cases.
Lastly, for one cluster, A1142, the C4 catalogue identified two clusters,
with the BCG for one corresponding to our M2 for A1142.

\subsubsection{A Subset of Bright M2 Members}\label{sec:m2_sub}

As a natural consequence of imaging all plausible BCG candidates,
we also imaged a large sample of second-ranked cluster members, M2,
as based on their metric luminosities.
This set is presented in Table \ref{tab:m2_sample}.
We observed 179 M2 galaxies, corresponding
to $\sim41\%$ coverage over the total sample of 433 clusters.
Of course, we were most likely to observe
M2 when it was a close rival to the BCG.  We thus have constructed
a sample of M2s that are likely to have properties similar to the BCGs.
Indeed, many of the M2 galaxies in clusters with more luminous BCGs in fact are
more luminous than a significant fraction of the
BCGs in other clusters.
The M2 sample appears to be nearly complete for
galaxies within 0.3 mag of the BCG luminosity in any cluster.
Because we were not complete in observing M2s that were not close
rivals of the BCG, however, this sample must be used carefully.

We emphasize that because the BCG/M2 selection is based on the metric,
rather than total luminosity, there are 14 M2s (identified
in Table \ref{tab:m2_sample}) that would have been selected as the BCG had
a larger aperture been used.
These galaxies have $\alpha$ substantially larger than that of their
corresponding BCG, such that the integrated flux out to a given
radius ultimately exceeds that of the BCG when the radius is large enough.
Because this ambiguity affects only a small portion of the sample,
and the radial limits of the surface photometry are heterogeneous
at radii well outside the metric aperture, we prefer
to preserve the purely metric-aperture based BCG selection.\footnote{All
of the M2s flagged as ultimately exceeding the BCG in luminosity do so
at only very large radii.  In one case, however, for the BCG/M2 pair in
A3531, the transition occurred just outside the metric radius,
so we designated the initial M2 galaxy as the true BCG.}

\subsection{Imaging Observations and Photometry}

\subsubsection{Observations}

Images of the BCGs were obtained in 13 runs between 1989 and 1995 using
CCD cameras on the KPNO-4m, KPNO-2.1m, and CTIO-1.5m telescopes.  The runs
are listed in Table \ref{tab:imruns}.  The first set of runs from 1989 to 1991
were used for the observations of the 15K sample presented in PL95, but
are repeated here for convenience.  As compared to the first set of
runs, the cameras used in the later runs generally had larger fields,
allowing for more straight forward estimation of the sky level,
as well as improved efficiency for observing multiple BCG
candidates in a single observation.

For the PL95 observations of the 15K sample, we obtained images
in both the Kron-Cousins $R_C$ and Johnson $B$ filters. The $R_C$-band
imagery served as the primary material used for the photometry, with
the $B$-band providing auxiliary information to test the validity
of the extinction and k-corrections, as well as to test the BCG
$B-R_C$ color as diagnostic of the properties of the galaxies.
In PL95, however, we found that the BCGs had a very narrow range in color
($\left< B-R_C \right> = 1.51;\ \sigma_{(B-R_C)} = 0.06$ mag)
that showed no correlation with other properties of the BCGs or
with residuals in their photometric distance estimates. We thus
elected to only obtain $R_C$-band images for the present
sample, given the demands of observing a large number of galaxies in the
limits of the observing time available.\footnote{2MASS K-band photometry is
too shallow to provide reliable measurements over the metric aperture.
See Appendix B in \citet{l07} for full details.}

To allow for the use of BCGs as photometric distance indicators,
we could only obtain useful images under photometric conditions.
About 20\%\ of each night was dedicated to observing
\citet{landolt} standard stars.  Frequent observation of standards
not only allowed the photometric quality of the night to be monitored,
but also allowed for frequent characterization of the airmass-extinction term,
which often varied from night to night, or even over
the duration of a single night.
Given the very narrow range of color seen in BCGs,
we were less concerned with determining the color terms of the
cameras, and selected standard stars that closely matched the typical
$B-R_C$ colors of the BCGs.  The median scatter in the standard star
photometry over all nights was only 0.008 mag, with the two poorest
nights having residuals of 0.022 and 0.035 mags.

In addition to obtaining accurate photometric calibration,
we also were concerned with accurate flat-field calibration of the
images, such that accurate sky levels could be measured.  This
was done by observing a number of ``blank sky'' fields during the
night to correct for large-scale illumination patterns that were
not removed by the standard use of dome flat-field images.
We could not use the alternative of generating
a sky-flat from the stack of images obtained on any given night,
since the BCGs are extended and were typically centered in the CCD fields.
This procedure reduced the error in the sky levels from several percent
to a few tenths of a percent.  As we discuss below, the final total
error in the metric magnitudes as measured by cross validation
is only 0.01 mag, demonstrating that any errors
associated with the sky subtraction must be less than
those contributed by the photometric solution.

\subsubsection{Image Reduction and Surface Photometry}\label{sec:imred}

Reduction of the CCD images obtained in the
newer set of runs followed the same procedures as
were described in PL95.  Sky levels were determined from the intensity
modes measured in the corners of the images.  Surface photometry
of the BCG candidates was obtained using the least-squares
isophote-fitting algorithm of \citet{snuc}.
In brief, the algorithm describes the galaxies
as a nested set of concentric elliptical isophotes, which are allowed to
have arbitrary surface brightness, ellipticity, and position angle
as a function of radius.
The key feature of the algorithm is that it allows galaxies in the
images to overlap; indeed it was developed explicitly to decompose
``multiple-nucleus'' BCG into individual galaxies.
In multiple systems, overlapping, merging, or even luminous
galaxies completely embedded in the BCG are modeled and subtracted
from the envelope of the BCG prior to measurement of the metric luminosity.
Again, no assumed form of the surface brightness profile was imposed.
The algorithm also allows bad pixels, bright stars, compact galaxies,
dust patches, and so on, to be excluded from the surface photometry
solution.

\begin{figure*}[!t]
\plotone{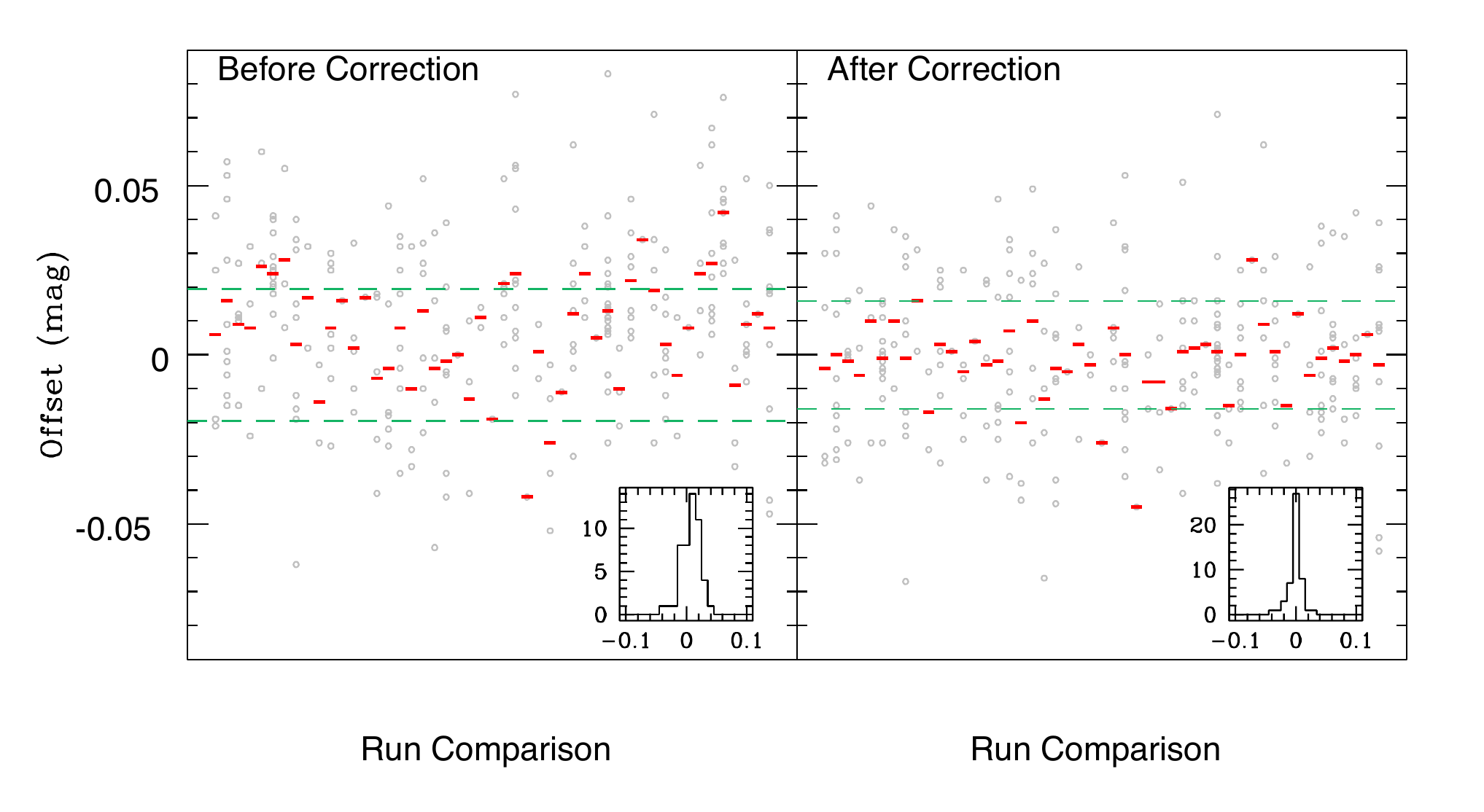}
\caption{The differences between the
single observations of the metric magnitude for galaxies
observed in common between different observing runs.
A total of 48 inter-run comparisons, derived from
260 overlap galaxy observations, are shown.
The light grey data points show the individual differences. Red lines show the mean offset for each inter-run comparison.
The dashed green lines show the total rms value. The plot on the left shows the results before the global offset corrections 
are applied to the photometry. The plot on the right shows the residual differences after the offset corrections are applied to each run. The distribution of the average metric magnitude offsets for all 48 inter-run comparisons is shown in the histograms in the lower right of each plot.
As the analysis is based on {\it differences} between pairs of observations,
the photometric error in any single observation will be
$\sqrt{2}\times$ smaller on average. After correction, the final {\it total}
photometric error for any metric magnitude is 0.011 mag.}
\label{fig:runcompare}
\end{figure*} 

Once the surface photometry for all the galaxies in an image
was completed, model images were reconstructed from the surface
brightness profiles and their total flux integrated in a geometric series
of circular apertures centered on the galaxy.  This is the final form of
the photometry used for the subsequent analysis.  The actual
value of the luminosity with the metric aperture is obtained by
using cubic splines to interpolate among the series of apertures,
based on the final velocity adopted for any given cluster.
This representation is highly accurate; the $1-\sigma$ difference
between the surface photometry integrated over the metric
aperture versus integrating over the galaxy image directly
(carefully cleaned of contaminating sources) is only 0.003 mag.

Lastly, the photometry is corrected for galactic extinction and
the filter K-correction, as was done in PL95;
for the $R_C$-band, $K_R=\log_{10}(1+0.96z),$ and $A_R=0.59A_B.$
Extinction values for the present work are provided by \citet{sfd},
in contrast with PL95, where extinctions from \citet{bh} were used.
Table \ref{tab:sample} gives the $A_B$ values used.

\subsubsection{Cross Validation of the  Photometry}

To provide additional validation of the galaxy photometry,
observations of several galaxies were repeated in multiple runs,
as was also done in PL95.  We also re-imaged most of the 15K BCGs
in the course of obtaining the 24K sample.
This not only allows the accuracy of any
given aperture measurement to be confirmed, but also provides a test for
any systematic differences in the photometric zeropoints
between the various runs.
Because we were concerned with obtaining consistent photometry
over the full sky, care was taken to ensure that extensive cross-validation
observations between the north/south hemispheres
and spring/fall observing seasons were obtained.
Figure~\ref{fig:runcompare} shows the measured differences
in the metric magnitude for galaxies observed across different runs.
The data allowed 48 separate inter-run comparisons to be performed,
assembled from 260 overlap galaxy observations 
obtained over the course of the imaging part of the survey.
This provided 62\%\ of the 78 potentially unique comparisons among
the 13 runs, thus densely populating the run/run cross-correlation matrix.

The total rms for all {\it differences}
between pairs of duplicate galaxy metric magnitudes is 0.0195 mag,
which implies that the average error in any single
measure is 0.0138 mag, a factor of $\sqrt{2}$ smaller.
By fitting for an average photometric offset correction for each run
(done as a simultaneous least-squares fit to the entire ensemble
of overlap observations),
we can reduce the total difference rms to 0.0159 mag,
or 0.0112 mag for any single observation.
Because the average photometric correction for any single run
is small, 0.0104 mag, we have chosen not to apply the corrections
in the present work; however, systematic differences between runs
are more important for the measurement of large scale deviations
from the Hubble flow, and will be considered in our use of the
present sample as a velocity reference frame.

\subsection{Spectroscopic Observations}

We obtained long-slit spectra of all BCG candidates in the sample
over the course of 14 observing runs, spanning a five year timeframe,
at NOAO's Cerro Tololo Inter-American Observatory (CTIO)
and Kitt Peak National Observatory (KPNO).
The CTIO observations were done primarily using the Blanco 4-m telescope,
except for the first two runs, which used the 1.5-m telescope.
All KPNO runs were done using the Goldcam spectrograph on the 2.1m telescope.
Table \ref{tab:specruns} summarizes the instrumental parameters.
The slit width was set to 2 arc-seconds.
For most observations, two or three independent exposures were obtained
(and coadded for further analysis),
although in some cases only a single exposure was acquired.
The exposure times for each individual exposure varied depending on telescope
aperture and the estimated target redshift.
As the overall objective was to use the spectra to obtain
both a measurement of the redshift
and the internal stellar velocity dispersion,
we set integrations to achieve a minimum signal-to-noise
ratio of 20 per pixel in the final co-added 1D spectrum.
A total of 842 co-added spectra were obtained for 689 unique galaxies.

Over the course of the survey, we repeatedly observed 13 bright nearby galaxies
as radial velocity reference standards.
These observations were designed to provide a cross-check
on our redshift measurement accuracy over the duration of the program.
The mean absolute value of the velocity difference between
the reference galaxy redshifts
from different runs was 32 km s$^{-1}$
($\langle \Delta {\rm v} / {\rm v} \rangle = 0.005$) with an rms
scatter of 38 km s$^{-1}$.
We also observed a subset of BCGs multiple times,
both from this survey and from the earlier LP94
survey to serve as cross-checks between different
observing runs and telescopes.
For the velocity dispersion estimates,
a series of spectra were repeatedly obtained of 27 K-giant stars.

The 2D spectra were corrected for basic instrumental signatures.
Bad columns were identified and interpolated over.
Bias subtraction was done by first using the overscan region to determine
the mean DC level, which was subtracted from the full frame.
Bias structure removal was then performed using a series
of zero-duration exposures acquired before the start of each night.
Quartz lamp exposures were co-added and normalized to provide
a flat-field correction frame.
Any cosmic ray hits that extended for more than
2 pixels were manually identified and interpolated over where possible.
Smaller cosmic ray hits
were dealt with during co-addition of the extracted 1D spectra.

The 1D spectra were extracted and wavelength-calibrated
using IRAF's NOAO {\tt onedspec} package.
The extraction was done using a 3rd order Legendre polynomial function to allow
the aperture center to track any significant spectral curvature
along the dispersion axis. 
The average spectrum extraction aperture width was 9 arcseconds
(rms 2.5 arcsec), which is significantly larger than the typical
FWHM seeing ($\sim 1.5$ arcsec) for any given observation.
The 9 arcsec width corresponds, on average,
to a projected physical width of 10 kpc (rms 3.6 kpc). 
The background level was estimated in two 15-pixel wide regions
on either side of the source spectrum with a 15 pixel gap
between the center of the source spectrum and the
start of the background sampling regions.
Two iterations of $3\sigma$ rejection were done during both spectrum tracing and
background level determination to reduce susceptibility to cosmic rays.
Occasionally, spectra for other galaxies (in addition to the BCG)
fell along the slit.
We extracted these spectra as well in hopes of providing additional redshift
information for the clusters.

The extracted 1D spectra were wavelength-calibrated by extracting
identical regions of the companion arc lamp spectra obtained either
just before or just after each galaxy spectrum.
Helium-Neon-Argon arc lamps were used for these observations.
The IRAF {\tt dispcor} routine was used to perform the wavelength calibration.
We typically used a 3rd order polynomial wavelength solution.
The wavelength calibration was checked both by looking at the fit residuals
provided in the IRAF {\tt identify} and {\tt reidentify} routines,
and by confirming that the prominent night sky emission lines
appeared at their proper central wavelengths.
A final co-added 1D  spectrum for each object observed
on a given night was then produced from the individual wavelength-calibrated
1D spectra using the IRAF {\tt scombine} routine.
Any cosmic ray artifacts that may have survived the co-addition were
manually removed via interpolation using IRAF's {\tt splot} routine.

\subsubsection{Redshift Measurements}

Redshifts were measured using the IRAF-based RVSAO package {\tt xcsao}.
We used eight independent high S/N spectral templates of elliptical galaxies to
perform the cross-correlations.
These 8 templates include spectra of M32, NGC3379, NGC4648, NGC7331,
the BCG in Abell 779, and three different composite
spectra of low redshift early type galaxies.
Eight templates are chosen to allow an estimate to be made of any
systematic errors in the cross-correlation measurement. 
Regions around prominent night sky
lines (Hg, NaD, OI) and strong atmospheric OH absorption bands
were excluded from the fitting procedure.
A galaxy redshift for each object was computed by first rejecting the templates with the highest and lowest redshift
value and then averaging the results for the remaining six templates.
For nearly all our high S/N spectra, however,
all eight templates yielded consistent redshift values.
The typical dispersion between templates was 30 km s$^{-1}$
and the mean velocity error in our redshifts is 45 km s$^{-1}$.
The average \citet{td} cross-correlation R-value,
which quantifies the significance of the peak in the normalized
cross-correlation function between the galaxy and template spectra
is 8.6, with values ranging from 6 to 15 for the BCG candidates.

About 5\% of our spectra have emission lines 
(only 8 of these emission line systems are BCGs). 
We used the IRAF routine {\tt rvidlines}
to measure the redshifts of these objects,
We typically were able to identify between 8 to 10
emission features in each spectrum in which emission was present.
The velocity error in a typical emission-line based redshift is 30 km s$^{-1}$.
Table \ref{tab:ztable} lists the IDs, celestial coordinates,
heliocentric redshifts and errors for galaxies, as well as
the mean \citet{td} $R$ values.

The mean absolute-value velocity difference for $\sim200$ objects
with multiple observations
is 39 km s$^{-1}$ with a standard deviation of 41 km s$^{-1}$.
The mean absolute-value velocity difference between our
redshift measurements and that from SDSS DR7 \citep{dr7} for 82
galaxies in common between the two surveys
is 33 km s$^{-1}$ with a standard deviation of 31 km s$^{-1}$.
In both comparisons, any potential systematic shifts
are comparable to or less than the scatter in the common measurements
and are also comparable to or less than the individual measurement errors.

\begin{figure*}[!t]
\plotone{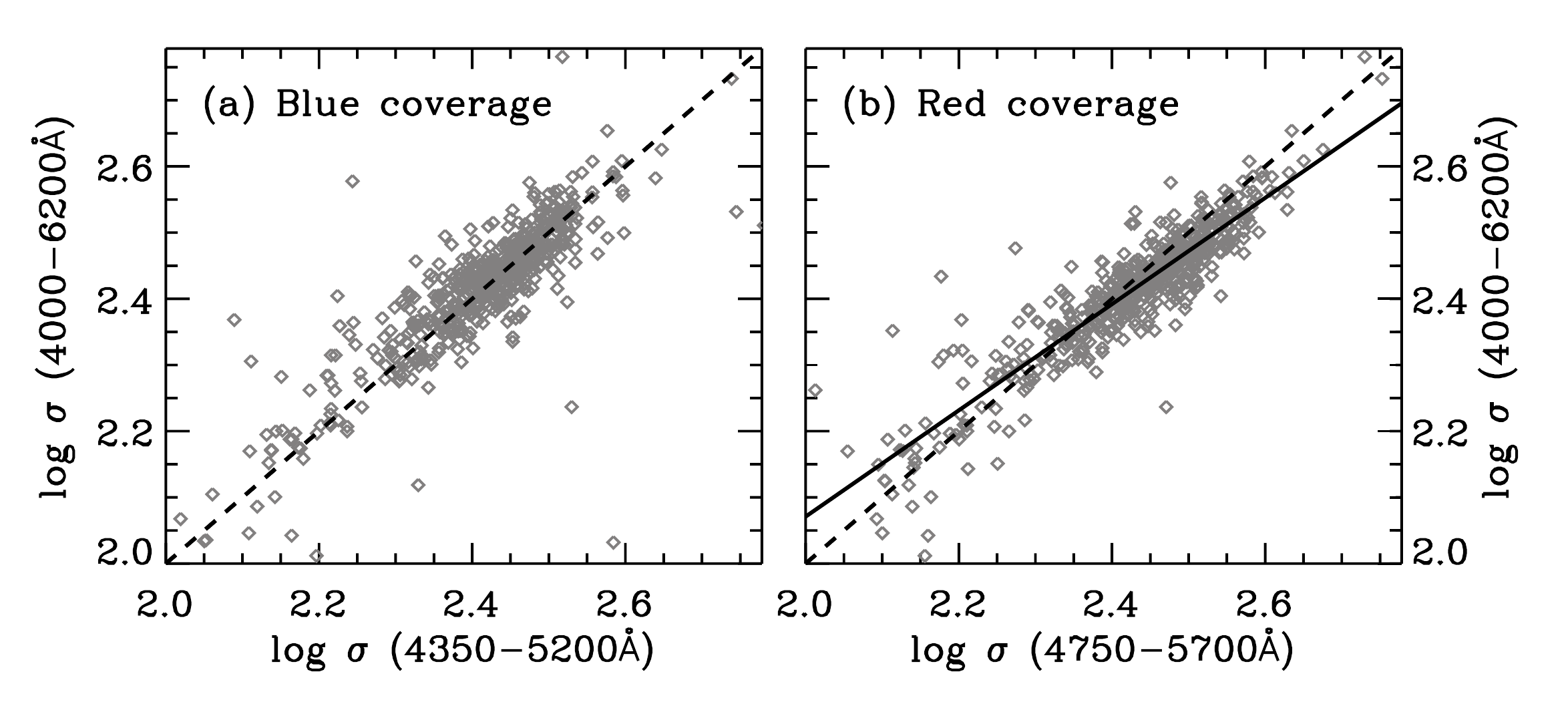}
\caption{Comparing velocity dispersions measured from the ``full''
  spectral range versus those measured from the more limited ``blue''
  or ``red'' wavelength coverage.  Open diamonds show individual
  galaxies in our sample.  Dashed lines show a one-to-one relation.
  When only the blue wavelength range is used, the recovered
  velocity dispersions agree with those derived from the full spectral
  range (panel a).  When only the red wavelength range is used, the
  measured $\sigma$ values are biased high for high-$\sigma$ galaxies
  (panel b).  We use a linear fit
to the correlation to correct $\sigma$ values measured
  in the red when the full spectral range is not available (solid
  line).
}
\label{fig:vdisp_bias}
\end{figure*}

\subsubsection{Velocity Dispersion Measurements}\label{sec:vdisp_measurements}

We measured central stellar velocity dispersions
from the extracted one-dimensional spectra
using a ``direct'' penalized pixel-fitting
method, as implemented the IDL code
pPXF\footnote{http://www-astro.physics.ox.ac.uk/$\sim$mxc/idl/}
\citep{cappellari04}.
We use the pPXF code in combination with single-burst stellar
population synthesis models \citep{vazdekis10} based on the empirical
MILES stellar library \citep{sanchez-blazquez06}, with a
Chabrier-style initial mass function (model
``Mbi1.30'')\footnote{http://miles.iac.es//pages/ssp-models.php}.  The
templates span a range of metallicities ($-1.71 < $[Z/H]$ < +0.22$)
and single-burst ages (1 Gyr $<$ age $<$ 17.8 Gyr).
As part of the fitting process, pPXF finds the linear combination
of templates that best reproduces the galaxy spectrum.
These models are
convolved with the instrumental resolution for each observing run,
which is modeled using a low-order polynomial fit to the width of the
arclines, and typically varies as a function of wavelength.  We allow
pPXF to fit for four velocity moments ($V$, $\sigma$, h3, and h4) and
use a fourth order multiplicative polynomial to account for continuum
mismatch due to imperfect spectral flux-calibration.  We mask regions
covering possible strong emission lines (the Balmer lines and
[OIII]$\lambda$4959,5007), and run the fit iteratively; on the first
run, we identify $\pm 4 \sigma$ outliers in the fit residuals as noise
spikes, mask them out, then rerun the velocity fits.

We further experimented with a number of model parameter choices that
might introduce systematic effects into the $\sigma$ measurements.
The massive BCGs in our sample have non-Solar abundance ratios (in
particular, they are enhanced in Mg, CN, and C$_2$, e.g.,
\citealt{graves07, greene13}), while the stellar population templates
have Solar-scale abundance patterns.  We experimented with masking
the absorption line regions strongly affected by these non-Solar
abundances, but this had a negligible impact on the resulting $\sigma$
measurements.  We also investigated the effects of using the pPXF
``BIAS'' keyword to push the velocity solution toward low values of h3 and h4.
This also had a negligible effect on our derived $\sigma$ values.
Fitting for only two velocity moments (i.e., setting h3 = h4 = 0)
in some cases produced a modest increase in the derived $\sigma$ values, as did
using a higher-degree multiplicative polynomial for the continuum adjustment.
There was little effect for galaxies with $\sigma \sim 150$ km s$^{-1},$ 
but increases of $\sim 20$ km s$^{-1}$ for galaxies with $\sigma \sim
350$ km s$^{-1}$ were seen.
We elected to use the fourth order multiplicative
polynomial and unbiased four moment velocity fits for our final
measurements.

By far the largest systematic effect was the choice of rest-frame
wavelength interval used in the velocity fits, with $\sigma$ values
biased by up to $\sim 40$ km s$^{-1}$ when comparing different
wavelength intervals.  Our observations were not
acquired with uniform wavelength coverage (see Table \ref{tab:specruns}).
Accordingly, we define three rest-frame wavelength ranges to use in
our analysis.  The ``full'' range of 4000--6200{\AA} was available for
nine of our 14 observing runs.  We also defined a more limited
``blue'' range of 4350--5200{\AA} for runs CT92F, CT93S, and KP92S, and
a ``red'' range of 4750--5700{\AA} for runs CT94F and CT94S.  For the
nine runs with full wavelength coverage, we measured $\sigma$ from the
``blue'', ``red'', and ``full'' wavelength separately, in order to
calibrate the effect of differing wavelength coverage as described below.

Due to imperfect data archiving,
we were only able to retrieve pixel-by-pixel error
arrays for a subset of the observing runs.
The pPXF code uses error
spectra both to penalize low-S/N or bad pixels in the fitting process
and to estimate errors in the derived parameters, such as $\sigma$.
We were able to use the iterative outlier rejection described above to
mask bad pixels, with all other pixels being assigned equal weight.
In order to estimate the uncertainties in measured values of $\sigma$,
we again resorted to iterative use of the pPXF code.  In the first
run, all pixels were simply assigned equal (and arbitrary) errors.
pPXF outputs the residuals between the best-fitting template
combination and the observed spectrum, which have Gaussian scatter
about zero.  We used the width of the scatter as an estimate of the
typical flux error per pixel, then reran pPXF using this value as the
input error for all pixels to propagate through the resulting
uncertainties in $\sigma$.  Where the true error spectra were
available, we could compare these estimated errors in $\sigma$ with
the true errors; the difference in error estimates was Gaussian with a
mean offset of 0.63 km s$^{-1}$ and width of 0.75 km s$^{-1}$.  This
means that where we could compare them, the bootstrapped error
estimates agreed with the true statistical error estimates to within 1--2
km s$^{-1}$.  This made us confident that bootstrapped errors could be
used reliably for observations whose error spectra had been lost.
Overall, the typical uncertainty in our $\sigma$ measurements is $\sim
14$ km s$^{-1}$, but varies substantially between observations,
depending on the spectral S/N and wavelength coverage.

For runs with full wavelength coverage, we compared the $\sigma$
measurements from the full 4000--6200{\AA} range to those derived from
the more limited blue (4350--5200{\AA}) or red (4750--5700{\AA})
wavelength ranges in the same spectra, as shown in Figure
\ref{fig:vdisp_bias}.  This calibration demonstrates that when only the
blue coverage is available the resulting $\sigma$ measurements are unbiased.
The mean offset between the blue and full coverage $\sigma$
values is $-3.2$ km s$^{-1}$, with rms scatter of 19.6 km s$^{-1}$ and no
clear trend with $\sigma$.  In contrast, measurements made with only
the red coverage show substantial bias; the mean offset is $+9.3$ km
s$^{-1}$ but increases to $\sim 30$ km s$^{-1}$ for the highest
$\sigma$ galaxies, with similar scatter of 20.7 km s$^{-1}$.  To put
all of our targets onto the same effective system, we fit a line that
defines the ``correction'' from the red coverage onto the full
coverage values.  This correction is applied to the $\sigma$
measurements from the CT94F and CT94S runs, which only have the red
wavelength coverage.  No correction is applied to the runs with blue coverage.

The spectroscopic observations include many repeat measurements of
individual targets, usually in different runs.  These can be used to
test the internal consistency of our $\sigma$ measurements.  Using
only $\sigma$ measurements made from the full spectral coverage, we
find that differences between repeat measurements of galaxies are
Gaussian distributed with a standard deviation of 16.5 km s$^{-1}$.
This is comparable to the expected typical statistical error of 14 km
s$^{-1}$, suggesting that the $\sigma$ measurements are stable across
the various runs.  This is not a trivial statement, given
that the observations use two different telescopes, different
instruments and instrumental configurations, and span multiple years including
instrument upgrades.  Comparing repeat observations on a run-by-run
basis, the runs showing the largest mean offsets from the rest are
KP92S ($-18.6$ km s$^{-1}$ for 16 galaxies), CT93S ($17.3$ km s$^{-1}$
for 4 galaxies), and KP94F ($-12.3$ km s$^{-1}$ for 12 galaxies).
Notice that none of these deviant runs are the ``corrected'' runs with
only red wavelength coverage.  All other runs show offsets that are $<
10$ km s$^{-1}$ from the aggregate.

Where multiple observations are available, rather than averaging the
individual measurements, we assign a ``best'' measurement for each
galaxy.  For the vast majority of the sources, the various
observations agree within the estimated 3$\sigma$ errors; for these
sources, the best measurement is the one with the smallest formal
error in $\sigma$ (i.e., that measured from the highest-S/N
spectrum).  For the six galaxies where repeat measurements show
catastrophic disagreement ($> 3\sigma$), we choose the best
observation based on the following criteria: full wavelength coverage
is preferred over limited red or blue coverage, spectra with
noticeable flux calibration or sky subtraction issues are disfavored,
higher S/N is preferred over low S/N, and better wavelength coverage
is preferred over higher S/N.
These measurements and calibrations result in a sample of 689 $\sigma$
``best'' measurements among our galaxies.

\begin{figure}[!t]
\includegraphics[width=\columnwidth]{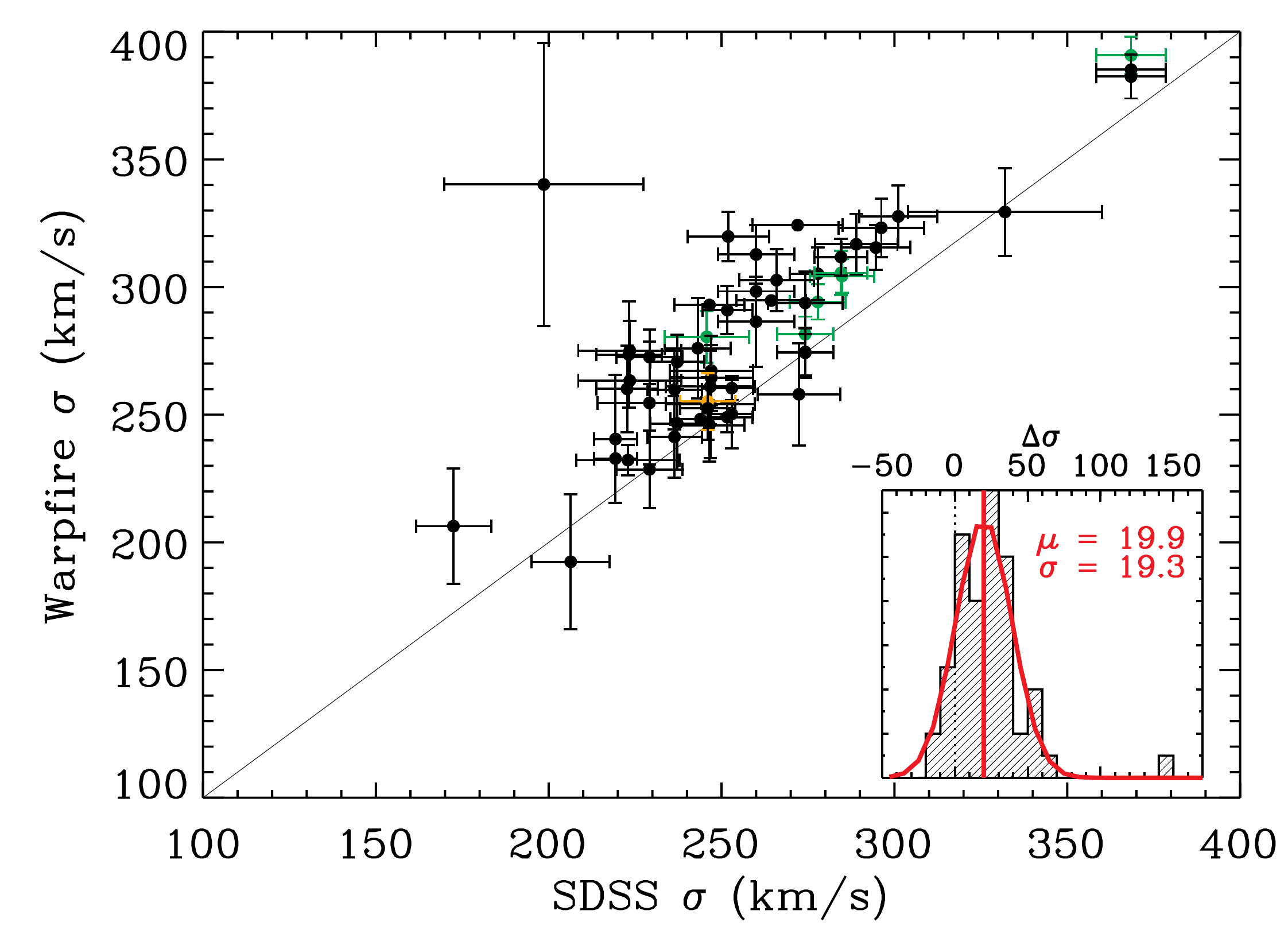}
\caption{A comparison of our $\sigma$ measurements versus those from
  the SDSS spectroscopic survey for galaxies in common.  Black, green,
  and orange data points are from the KP96S, KP93S, and KP92F runs,
  respectively.  The solid line shows the one-to-one relation.  The
  inset panel shows a histogram of the offsets between the present
  and the SDSS measurements.  These are Gaussian distributed (red
  curve) with a mean offset of 19.9 km s$^{-1}$ and scatter of 19.3 km
  s$^{-1}$.  The scatter is comparable to what is expected from the
  combined present and SDSS observational errors.  We do not correct
  our $\sigma$ measurements onto the SDSS system, but merely note that
  they are offset to higher values.  }
\label{fig:sdss_compare}
\end{figure}

Finally, 78 of our galaxies are in the SDSS spectroscopic survey,
making it possible to compare our $\sigma$ measurements to those from
the SDSS spectroscopic pipeline (\citealt{bolton}; see \citealt{am08}
for a comparison of different velocity dispersion algorithms in SDSS).
This comparison is shown in Figure \ref{fig:sdss_compare}.
The overlapping galaxy sample is mostly from the KP96S run
(black points), with a few from KP93S (green points) and one from
KP92F (orange point).  The solid line shows the one-to-one relation.
The inset panel shows a histogram of the differences between our
$\sigma$ and the SDSS $\sigma$ measurements.  These have a Gaussian
distribution with a mean offset of 19.9 km s$^{-1}$ and width of 19.3
km s$^{-1}$.
The scatter compares favorably with the estimated
statistical errors in the measurement ($\sim 17$ km s$^{-1}$ for the
typical errors from our observations
and those from the SDSS combined in quadrature) and the
internal consistency of our repeat measurements ($16.5$ km s$^{-1}$).
However, there is a significant systematic offset of $\sim20$ ${\rm km~s^{-1}}.$
The larger angular apertures ($\sim9''$) of the spectral extractions used in this work relative to
the $3''$ fiber apertures used in SDSS may explain some of the systematic shift
if the trend of increasing stellar velocity dispersion with increasing radius, like that seen in the BCG in Abell 383
\citep{newman11}, is typical.
We do not attempt to ``correct'' our values onto the SDSS
system, but note that work combining $\sigma$ measurements from
different sources and spectral reduction pipelines
must take such systematic variations into account.
Dispersion values for the BCGs and M2 galaxies are listed
in Tables \ref{tab:bcg_par} and \ref{tab:m2_par}, respectively.

\subsection{Derivation of the Cluster Redshifts}

Cluster redshifts were derived based on galaxy velocities drawn from
the database maintained by the NASA Extragalactic Database (NED),
augmented by velocities measured for the BCG candidates by us, and
SDSS Data Release 7 spectroscopy \citep{dr7} where available.
The ``biweight'' estimator of \citet{biweight} was used to calculate the
mean cluster redshifts, $V_c,$ and velocity dispersions, $\sigma_c.$
The initial calculations used galaxies within
$\pm3000~{\rm km~s^{-1}}$ of the nominal BCG
or estimated cluster redshifts, and the cluster Abell radius.
While the biweight statistic is designed
to be robust in the presence of background or foreground contamination,
we still considered it prudent to remove obvious background contamination
or other complexities, such as overlap with nearby
clusters or groups.  This was done by {\it ad hoc} inspection
of the velocity maps and histograms for each cluster.
A second robust statistic introduced by \citet{biweight}
was used to estimate the cluster velocity dispersion for clusters
with four or more velocities.

Table \ref{tab:sample} lists the final number, $N_g,$ of galaxy velocities
used to compute the mean velocity and dispersion.  This parameter
is used as a general marker for the quality of both parameters.
For some evaluations of the peculiar velocities of the BCGs within
their clusters, we will restrict the analysis to clusters with $N_g\geq50,$
to minimize the effects of the error in the mean velocity.
For analyses requiring accurate $\sigma_c,$
we require the clusters to have $N_g\geq25.$

\section{The Location of BCGs in Their Galaxy Clusters}\label{sec:loc}

\subsection{The Peculiar Velocities of BCGs}

The stereotypical image of a galaxy cluster has the BCG centrally located,
both in projected angular coordinates and radial velocity relative
to other cluster members.  Studies of clusters with rich enough velocity
sampling such that an accurate mean cluster redshift can be estimated,
however, show that the BCG may often have a significant ``peculiar velocity''
with respect to their hosting cluster \citep{zab90, mal, zab93, oh01}.
PL95 obtained the
distribution of BCG $\Delta V_1\equiv(V_1-V_c)/(1+z)$
for 42 clusters with 20 or more member
velocities, finding that the $\Delta V_1$ followed a Gaussian
distribution with $\sigma_{\Delta V_1}=264~{\rm km~s^{-1}},$ once
the error in cluster mean redshift was accounted for.
This value is $\sim0.4$ of the typical 1-D cluster velocity dispersion,
$\sigma_c=666~{\rm km~s^{-1}},$ of the same subset of clusters.
To test the hypothesis that the BCG peculiar velocities
may be related to their masses, merger histories, and ages,
we have derived the BCG peculiar velocity distribution function
and investigated the relationship of this parameter to other
BCG properties.  The distribution function is considered in this
section, while the relationship of BCG peculiar velocities
to other BCG properties will be considered later in the paper.

\begin{figure}[!t]
\includegraphics[width=\columnwidth]{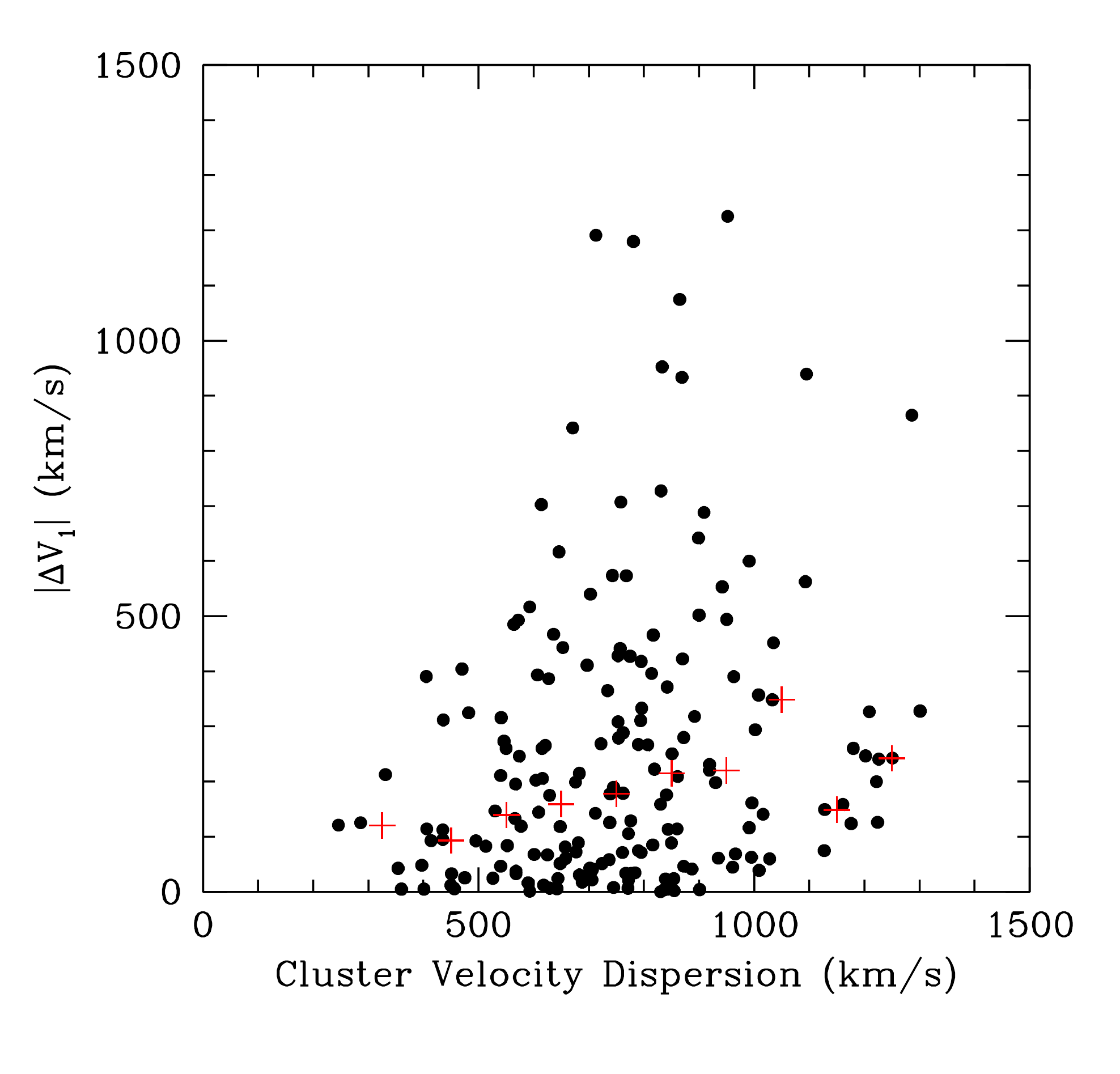}
\caption{The distribution of $|\Delta V_1|,$
the absolute value of the peculiar velocity of the BCG within the cluster
as a function of cluster
velocity dispersion, for the 178 clusters with 50 or more galaxies
with measured redshifts.  The red crosses show the {\it median} peculiar
velocity for each interval of $100~{\rm km~s^{-1}}$ in $\sigma_c$
(except for the first bin, which runs from 275 to $400~{\rm km~s^{-1}}).$}
\label{fig:dv_sig}
\end{figure}

Figure \ref{fig:dv_sig} shows the absolute values of $\Delta V_1$ as function
of cluster velocity dispersion for the 178 clusters in the present
sample that have 50 or more member velocities.
With this level of velocity information the error in the mean
cluster velocity is $\sim100~{\rm km~s^{-1}}$ or less,
allowing relatively small $\Delta V_1$ to be detected.  As can be seen,
most BCGs have $\Delta V_1$ well in excess of this error threshold.
The median peculiar velocity increases with $\sigma_c.$
A power-law fit shows
\begin{equation}
|\Delta V_1|=152\pm15\left({\sigma_c\over
600~{\rm km~s^{-1}}}\right)^{0.66\pm0.26} {\rm km~s^{-1}}.
\end{equation}
\citet{coziol} studied the
distribution of BCG peculiar velocities in a large sample of Abell clusters and
also found $|\Delta V_1|$ to increase with $\sigma_c,$ although they
did not quantify the trend.

\begin{figure}[!t]
\includegraphics[width=\columnwidth]{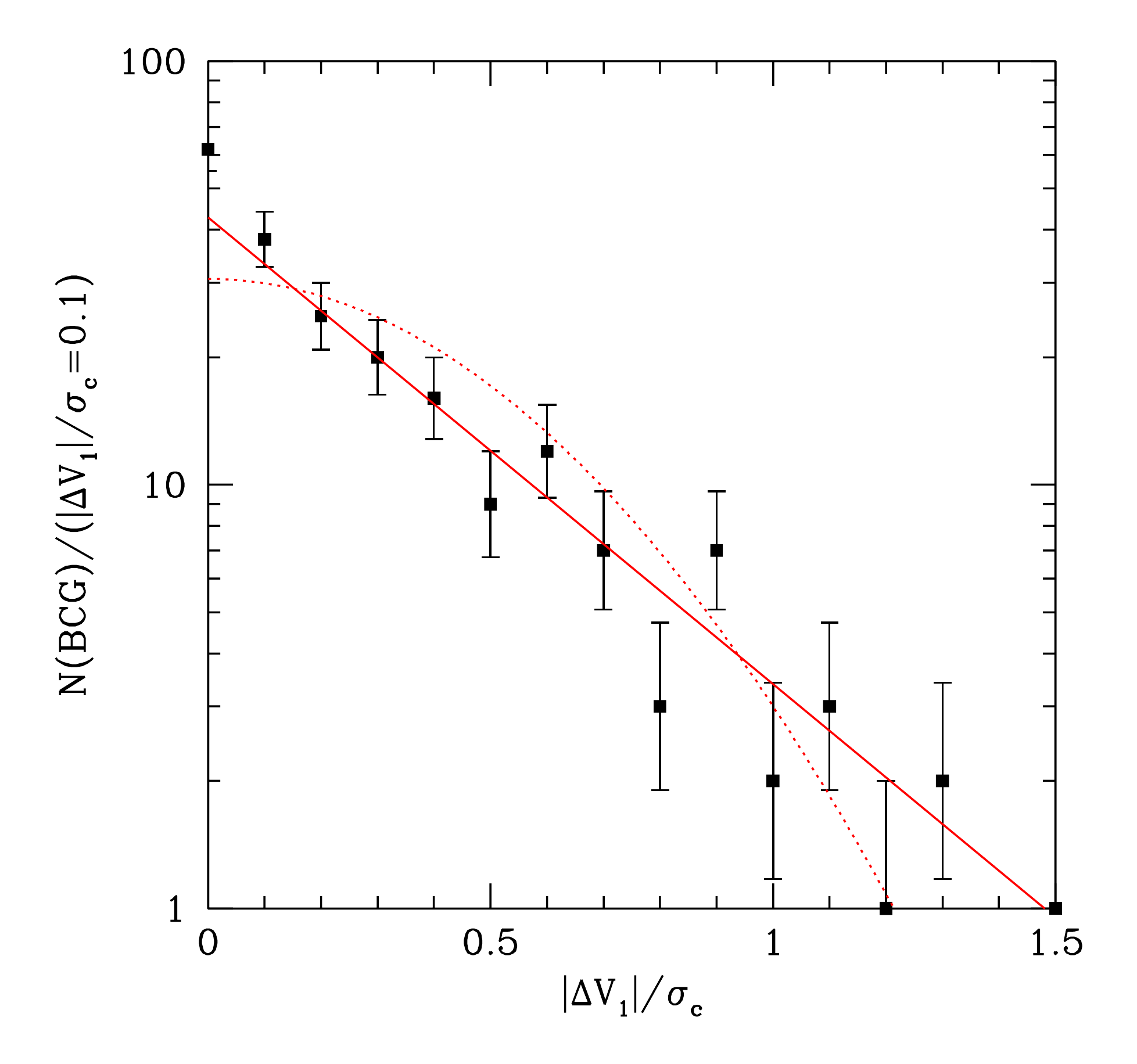}
\caption{The figure shows the binned distribution of $|\Delta V_1|/\sigma_c,$
the absolute value of the radial velocity difference between
the BCG and mean cluster velocity, normalized by the cluster
velocity dispersion, for the 178 clusters with 50 or more galaxies
with measured redshifts.  The bins are 0.1 units wide.  The solid line
is an exponential with scale-length 0.39 in $|\Delta V_1|/\sigma_c.$
The dotted line is the best-fit Gaussian distribution;
it is clearly a poorer fit.}
\label{fig:dvnorm}
\end{figure}

Figure \ref{fig:dvnorm} shows the binned distribution of $\Delta V_1$ normalized
by the cluster velocity dispersion.
Normalizing by $\sigma_c$ largely removes any dependence of
the amplitude of the peculiar velocity on the properties of the cluster itself.
Since we are restricting this analysis to clusters with 50 redshifts or
more, the errors on $\Delta V_1/\sigma_c$ will be $<1/\sqrt{50}\approx0.14.$
In normalized units, the mean $\Delta V_1/\sigma_c=0.04\pm0.04,$
with an rms dispersion of $0.49$ --- note that this number measures
a different statistical property of the distribution than does the
median peculiar velocity plotted in Figure \ref{fig:dv_sig}.
The two BCGs with the largest $\Delta V_1/\sigma_c$
values are those in A2399 and A3764, which have $|\Delta V_1|$ of 1191 and
1180 ${\rm km~s^{-1}},$ respectively, or normalized values
of 1.67 and 1.51.

The distribution of $|\Delta V_1|/\sigma_c$ is exponential in form.
Since the line-of-sight velocity distributions of galaxies in clusters
are well known to be Gaussian \citep{y77},
a random draw of any non-BCG cluster member would, of course,
echo this expectation.
An exponential velocity distribution specific to the BCGs is thus surprising.
The best-fitting exponential distribution is
\begin{equation}
\ln\left(N\right)=-2.54\pm0.18 |\Delta V_1|/\sigma_c + 3.76\pm0.20,
\end{equation}
where $N$ is the number of clusters per bin of width
0.1 in $|\Delta V_1|/\sigma_c.$
The implied exponential scale length (the reciprocal of the slope
given above) is thus
$0.39\pm0.03$ in $|\Delta V_1|/\sigma_c;$
this form and scale implies that the median $|\Delta V_1|/\sigma_c$ is 0.26.
Note that for small $|\Delta V_1|/\sigma_c,$ the observed distribution
represents the convolution of an unknown intrinsic peculiar-velocity
distribution with the error distributions
of the BCG and cluster redshifts, plus the errors in the cluster dispersions.
However, the observed distribution appears to be smooth and simple in form,
thus the intrinsic distribution is likely to
transition smoothly from BCGs with $|\Delta V_1|/\sigma_c\approx0$
to those with large $|\Delta V_1|/\sigma_c,$ where the peculiar
velocity of any given BCG is clearly significant.

The present distribution appears to be similar
to that measured by \citet{coziol}, although they did not characterize
it with any functional form.
We tested the likelihood that
the distribution was non-Gaussian using an Anderson-Darling test \citep{ad},
finding that a Gaussian distribution is strongly rejected.
For the observed rms distribution of $\Delta V_1/\sigma_c$ of 0.46,
the Gaussian is rejected at the $8\times10^{-6}$ significance level.
Deletion of the two galaxies that have the largest
relative peculiar velocities decreases the rms value of the
distribution to 0.44, and the AD-test allows a Gaussian
at $6\times10^{-5}$ significance.

It has long been known that the difference in peculiar velocities of
{\it pairs} of galaxies is exponentially distributed on small scales,
both in observations of redshift-space distortions of the two-point correlation
function and in simulations \citep{f94,marz}.
This effect has been explained in terms of the number, rather than mass,
weighting of galaxies in pair statistics \citep{dg96,j98}.
We are unaware of an equivalent study of BCG peculiar velocities in clusters.
\citet{reid}, in their analysis of redshift-space distortions
of SDSS/BOSS galaxies, identified massive halos at $z \approx 0.55$
in a $\Lambda$CDM N-body simulation containing $2048^3$ particles
in a box 677.7 $h^{-1}$ Mpc on a side.
They measured the difference in  peculiar velocity between the most dense
spherical region of radius 0.2 times the virial radius,
and the cluster overall.
Reid (private communication) finds that the distribution of this
difference is accurately exponential in a variety of halo mass bins
corresponding to rich clusters.
While it is unclear whether the Reid et al.\ identification of
the highest-density region in each halo is a good proxy for the BCG,
this result is intriguing, and it would be interesting to explore
more detailed cluster simulations in which individual subhalos
can be identified.  

The exponential distribution may reflect a dispersion in the ages
of the clusters, the timing of when the BCG was captured by the cluster,
or may simply be due to the superposition of Gaussian distributions of 
different velocity dispersions,
weighted by the BCG number distribution \citep{dg96}.
BCGs in the tail of the exponential distribution may be those in which the BCG 
arrived to the cluster in the merger of a group or subcluster relatively
recently, and not yet completely relaxed.
These ideas could also be explored in simulations,
or by looking for correlations between BCG peculiar velocity
and signatures of merging in their host clusters.  

\subsection{The Projected Spatial Location of BCGs With Respect to the X-ray Centers}\label{sec:xray}

The X-ray emission from the intracluster
medium provides insight into processes that govern the
formation and evolution of the BCGs.
Numerous investigations \citep{edge1,edge2,he97,cm98,stott}
find significant positive correlations between the total luminosity of the BCG
and the X-ray luminosity and X-ray temperature.
\citet{schom} finds that the envelope-luminosity of cD galaxies, a subset
of the BCGs, increases with total cluster X-ray luminosity.
\citet{stott} find that
the steepness of the $L_X - T_X$ relation in galaxy clusters
correlates with the stellar masses and X-ray offsets of their BCGs.
Clusters in which the offset between the BCG position and the peak
of the X-ray surface brightness distribution
is small tend to be the most regular, most massive systems
\citep{allen, smith05, hudson}.
\citet{haarsma} find that $\sim90$\% of local ($z < 0.2$)
clusters host a BCG within $\sim30$ kpc of the X-ray peak,
although their sample is small, and unlikes us,
they include a criterion of proximity to the X-ray peak
for selecting the BCG from among candidates of ``similar" brightness.
\citet{brough} find that the structure of the BCG
correlates with cluster X-ray luminosity, with the BCG envelope becoming
more extended in more luminous clusters.

Our sample is well suited to characterize the precise form
of the distribution of the spatial offset, $r_x,$
of the BCG from the peak of the intra-cluster
medium (ICM) X-ray emission, and to assess whether the spatial offset
correlates with the velocity offset of the BCG relative
to the mean cluster velocity.\footnote{Many investigators prefer to
normalize $r_x$ by a cluster over-density scale, such as $r_{200}$ or $r_{500}.$
Since these latter scales are proportional
to $\sigma_c,$ which only varies by $\sim2\times$ over the sample,
this would make little difference for the $r_x$ distribution,
which extends over three orders of magnitude.}
In addition, the availability of robust BCG profile
shape measurements allows us to determine if the BCG stellar light profile is
influenced by the spatial offset.

We cross-correlated our BCG catalog with the ROSAT-based
X-ray Brightest Abell-type Cluster Survey (XBACS; \citealt{Ebeling}).
XBACS is a flux-limited catalog derived from the ROSAT All-Sky Survey
(RASS; \citealt{Voges}).
Of the 283 Abell cluster sources listed in the VizieR version
of the XBACS sample, 111 are in common with our current survey.
An additional 70 clusters in our current sample have X-ray peak
positions from the analysis of the RASS
done by \citet{Ledlow}, which extended to an X-ray flux limit
that is $\sim$7 times lower than that used to derive the XBACS.
Chandra X-ray Observatory data were obtained as well
for 48 of the ROSAT clusters from the Archive of Chandra Cluster
Entropy Profile Tables (ACCEPT; \citealt{Cavagnolo}),
which we use in preference to the ROSAT peak positions given
the superb Chandra angular resolution.
We also searched the literature for XMM data but only 13 of the clusters
in our sample have XMM data, too few to provide independent cross-checks on
the ROSAT and Chandra samples. We thus focus our X-ray analyses on the  
above subsample of 174 clusters in our survey, using ROSAT
data for 127 clusters and Chandra data for 47 clusters.
We note that while Chandra is not a survey mission,
the Abell clusters in our survey that have Chandra X-ray temperatures
and luminosities that span the same range as the ROSAT
temperatures and luminosities.
No significant biases are introduced by including
the Chandra data in our study
of BCG dependence on the X-ray properties of their host clusters.

\begin{figure}[!t]
\includegraphics[width=\columnwidth]{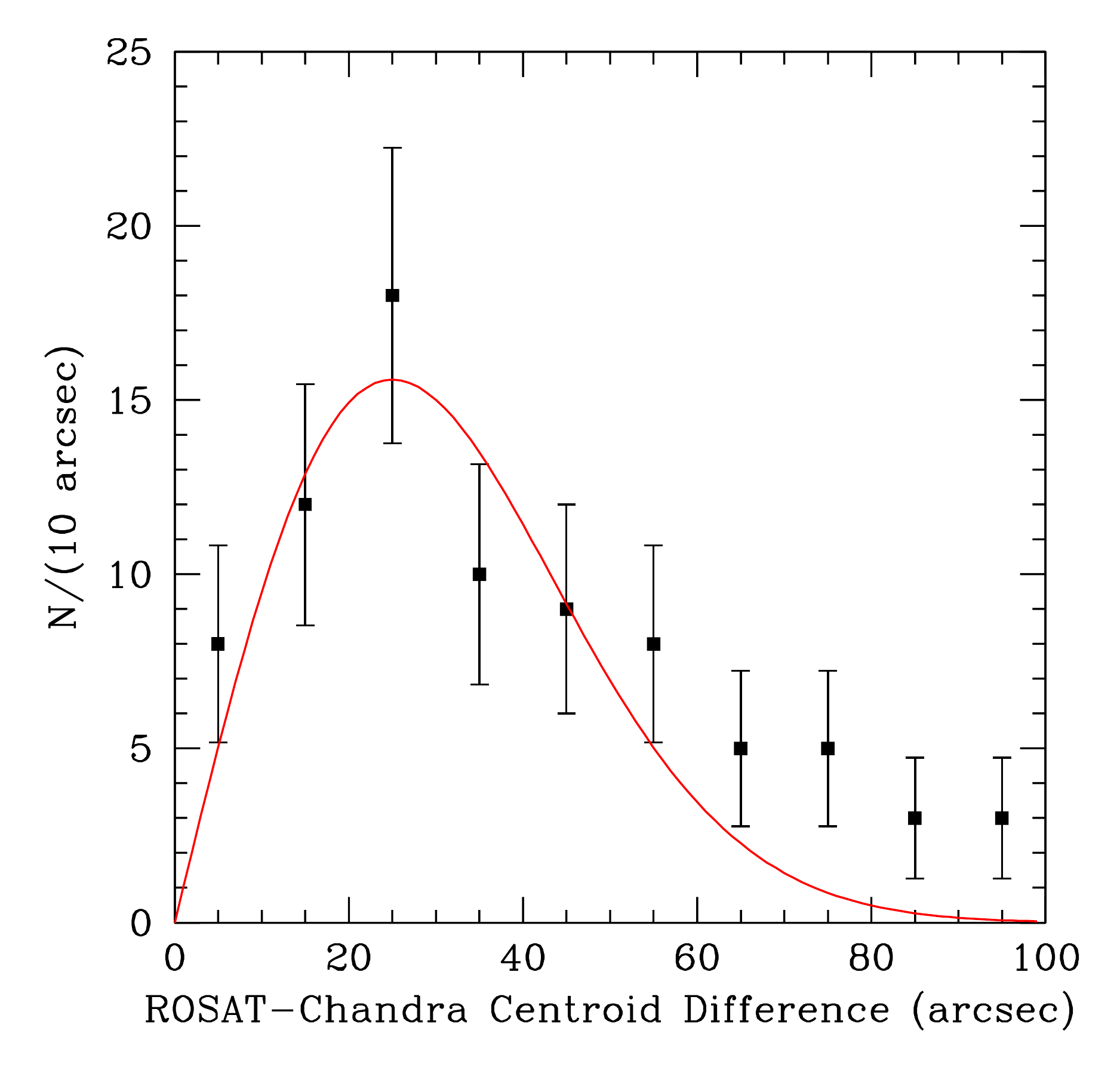}
\caption{Histograms of the angular differences between Chandra and ROSAT
centers for 101 Abell clusters with observations from both observatories.
The red line gives a best-fit Rayleigh
distribution (equation \ref{eqn:theta})
based on the assumption of a circularly symmetric Gaussian
model for the distribution of peak-location differences.}
\label{fig:rosat_vs_chandra}
\end{figure}

\begin{figure}[!t]
\includegraphics[width=\columnwidth]{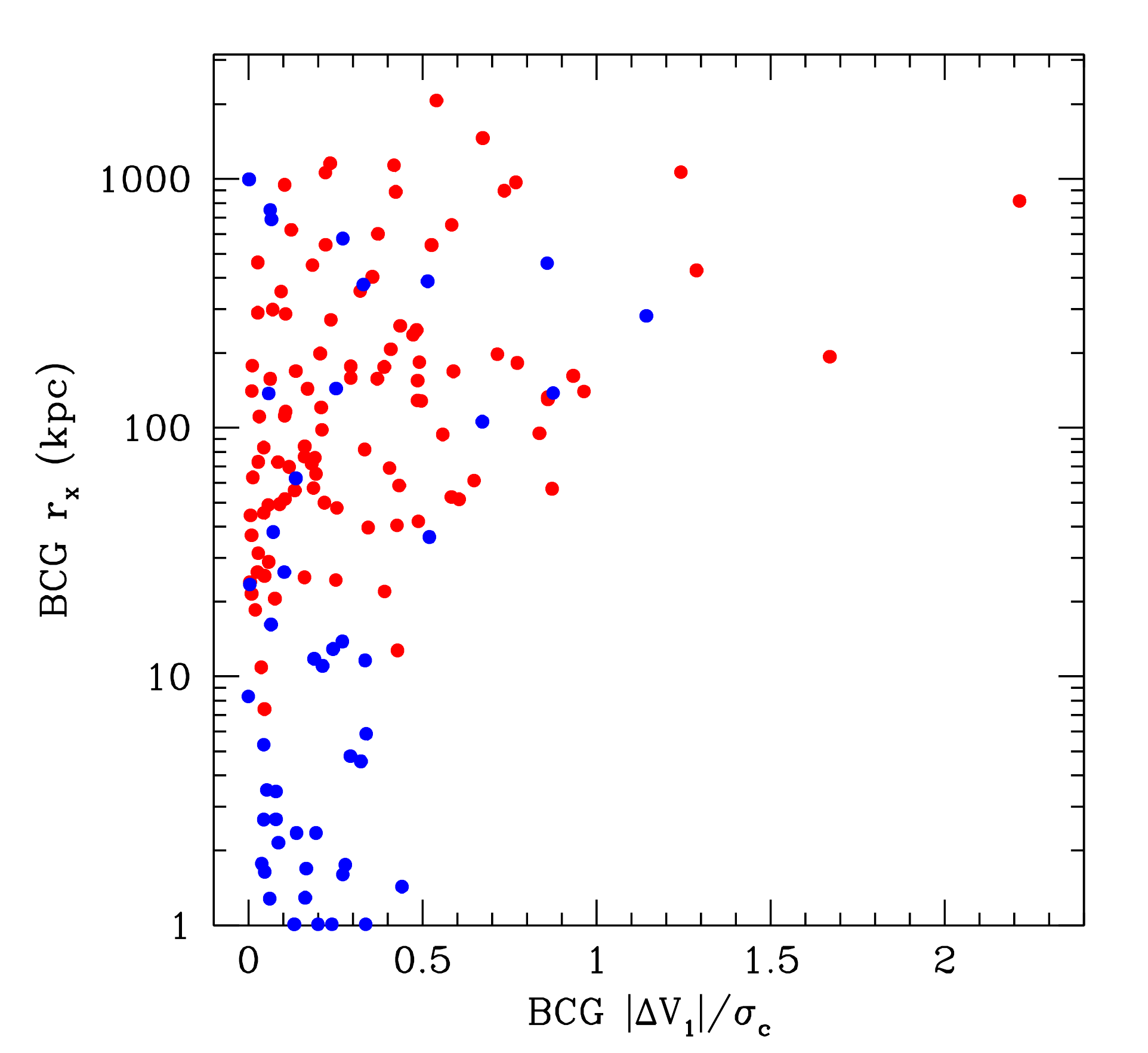}
\caption{The radial offsets of the BCGs from the cluster X-ray center
is plotted as a function of the absolute normalized peculiar
velocity of the BCGs within the cluster.
Clusters observed with Chandra are blue; those observed with ROSAT are red.
Small radial separations ($r_x<40$ kpc) always correspond to
small peculiar velocities ($|\Delta V_1|/\sigma_c<0.5$),
while large peculiar velocities ($|\Delta V_1|/\sigma_c>1$)
always correspond to large spatial offsets ($r_x>200$ kpc).}
\label{fig:drvdv}
\end{figure}

ROSAT and Chandra have very different on-axis point spread functions:
FWHM of $\sim0.5$ arcseconds for Chandra vs. $\sim1$ arcminute for ROSAT. 
We used a sample of 101 Abell clusters (not limited by the redshift limits
of our current survey) with observations from both observatories
to measure the typical difference, $\theta_x,$ between the ROSAT peak position
and the Chandra peak position.
The distribution of $\theta_x$ is shown in Figure \ref{fig:rosat_vs_chandra}.
The median and mean differences between the ROSAT X-ray position and the
Chandra X-ray position are 43 arcseconds and 69 arcseconds, respectively.
The median and mean difference between the ROSAT and the Chandra
peak X-ray positions in projected physical distance units derived
from cluster redshift information are 68 kpc and 121 kpc, respectively.

We characterize the distribution of differences between the location of the
Chandra and ROSAT X-ray peaks on the assumption that the
distribution can be modeled as a circularly symmetric Gaussian.
This will be almost entirely due to the large ROSAT PSF,
but it may be more compact than that, given that the appropriate source
of variance is the error in the ROSAT centers, rather than
the width of the ROSAT PSF itself.
The distribution of the total angular differences between the
X-ray peaks will be a Rayleigh distribution,
\begin{equation}
p(\theta_x)=\sigma_x^{-2}
\exp\left[{-\theta_x^2\over2\sigma_x^2}\right]~\theta_x~d\theta_x,
\label{eqn:theta}
\end{equation}
where $\sigma_x$ is the dispersion of the Gaussian,
as well as the peak location of the above distribution.
We fitted this form to $\theta_x$ in 10-arcsecond bins, limited
to the six bins around the peak of the distribution
shown in Figure \ref{fig:rosat_vs_chandra} to avoid the effects
of extreme outliers.
We measure $\sigma_x=25\pm3$ arcsec, or an implied FWHM of 60 arcsec
for the underlying position-error Gaussian.
The adopted distribution given by equation (\ref{eqn:theta}) for this
$\sigma_x$ is also plotted in Figure \ref{fig:rosat_vs_chandra}.

\begin{figure}[!t]
\includegraphics[width=\columnwidth]{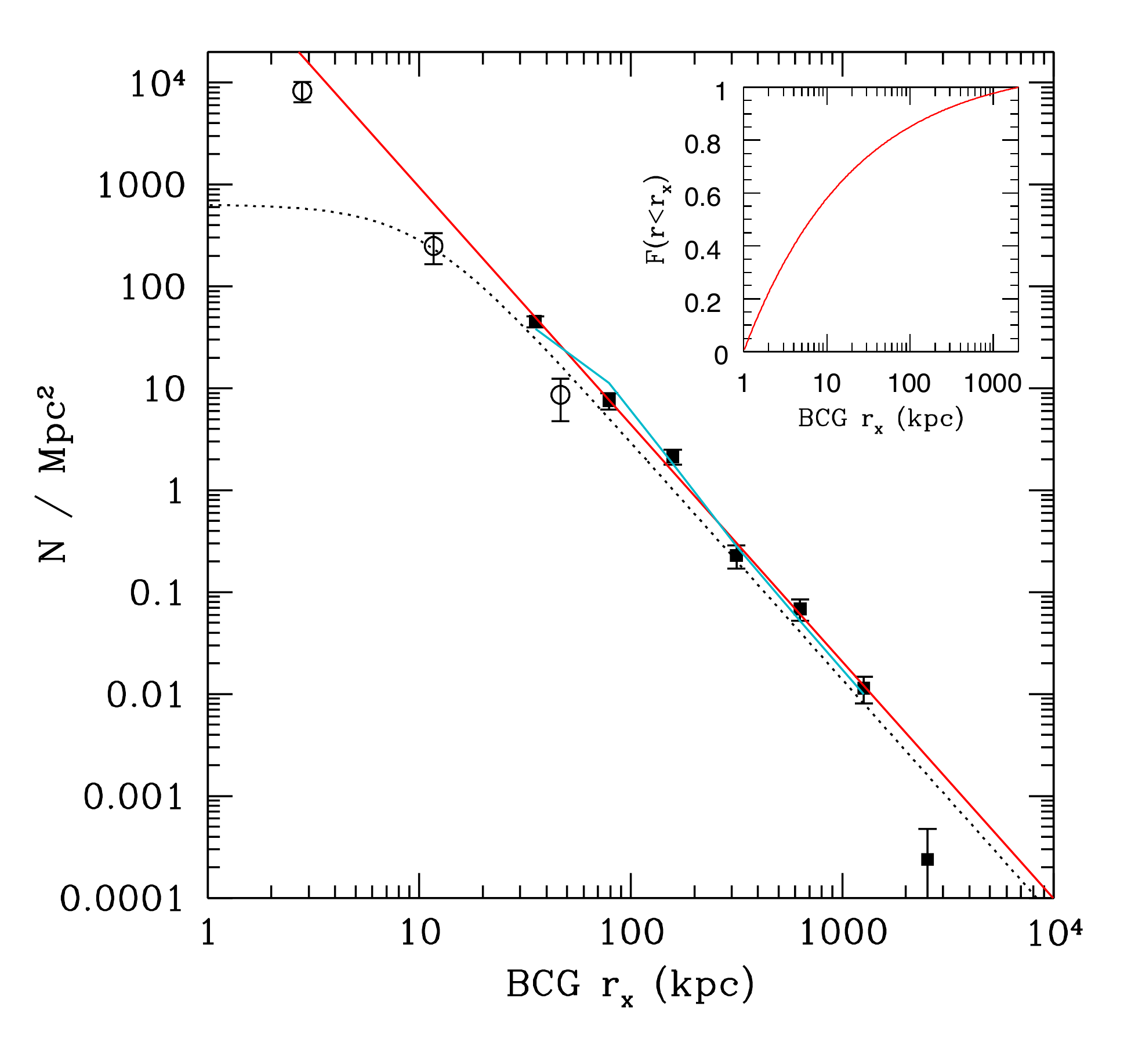}
\caption{The surface density distribution of BCGs with respect to
the cluster X-ray center.
The normalization is set to provide a unit integral over the
distribution (every cluster has one BCG).
Solid points give the density of the full set of BCGs with
measured X-ray offsets. The innermost bin extends
from the center to 50 kpc; subsequent bins have outer limits
geometrically increasing by a factor of two.
Open points are the subset of BCGs that have cluster centers
provided by Chandra, and thus are not resolution-limited on this scale.
The inner bin from this set extends from the origin to 4 kpc, with
the next two bins increasing geometrically by a factor of four.
The red line is a power-law of index $\gamma=-2.33$ fitted only to the
full-set bins.  The dotted line is a $\gamma=-2.33$ power-law with a 
10 kpc core (equation \ref{eqn:core}).
The blue line is this form as ``observed,'' blurring
it with a Gaussian with dispersion $\sigma_x$
for the fraction of the clusters observed by ROSAT.
The inset gives the cumulative integral of
the pure power-law running from 1 kpc to 2 Mpc.}
\label{fig:x_off_surf}
\end{figure}

Figure \ref{fig:drvdv} shows the offset between
the BCG position and ROSAT position as a function
of the velocity offset between the BCG and the cluster for 
the 174 clusters in our sample with ROSAT data.  The somewhat
triangular shape of the distribution of points shows that
there is an overall correlation between the BCG spatial and
velocity offsets.  In particular, $|\Delta V_1|/\sigma_c>1$ occurs
only for clusters with $r_x>200~{\rm kpc},$ while all $r_x<40~{\rm kpc}$
clusters have $|\Delta V_1|/\sigma_c<0.5.$  Stated qualitatively, BCG
that are close to the X-ray center in projection always have relatively small
velocities.  Of course, with a large enough sample, there should
be BCGs with high velocities seen in projection against the
center; however, $r_x<40~{\rm kpc}$ corresponds to
a very small portion of the projected area of the galaxy clusters.

Figure \ref{fig:x_off_surf} shows the radial distribution of
$r_x$ for the subset of BCGs with cluster X-ray centers
as a plot of BCG surface density as a function of radius.  The
normalization is set so that the cumulative integral over the
surface density for each cluster is unity (each cluster has a single BCG).
Generating this figure requires understanding the effects
of the angular resolution on the X-ray peak locations obtained with ROSAT
as well as how to incorporate the high-resolution Chandra data
into the sample.  As it happens, however, this is only a minor
issue as the ROSAT data are used mainly at larger radii and
the Chandra data are used exclusively at small radii.

The solid points in Figure \ref{fig:x_off_surf} represent
the entire X-ray sample, using both Chandra and ROSAT
together, as radial bins of $r_x.$  The innermost
bin extends from the origin to 50 kpc, thus enclosing
most of the ROSAT PSF.  Subsequent bins are rings starting at 50 kpc, with
inner and outer limits increasing geometrically by a factor of two.
The data follow a simple power-law, which
as we discuss below, appears to be the best form for the
overall BCG spatial distribution.
A log-log line fitted to the points gives the surface density as
$$
\log_{10}\rho(r_x)=(-2.33\pm0.08)\log_{10}\Bigl\lbrack{r_x\over1~{\rm kpc}}\Bigl\rbrack
$$
\begin{equation}
+5.30\pm0.19~\log_{10}\lbrack{\rm N~Mpc^{-2}}\rbrack,
\label{eqn:cusp}
\end{equation}
which is shown in the figure as the red line.

We used Monte Carlo simulations to understand the effects of the
limiting ROSAT resolution on the apparent density profile,
and to verify that the profile incorporating ROSAT
data was consistent with the
center of the distribution inferred from Chandra alone.
For an assumed surface-density profile, we drew $10^4$ ``BCGs'' for
each cluster in the X-ray subset.  For clusters observed by only
by ROSAT, we scaled the angular position-error Gaussian
to the appropriate physical resolution,
given the redshift of the cluster, and drew a point at random
from the circularly symmetric Gaussian
centered at a radius drawn from the density distribution.
For Chandra, we simply drew galaxies directly from the assumed profile.
The form we tested included an inner quadratic-core to suppress the singular
number integral as $r_x\rightarrow0,$ implied by the $\gamma<-2$ power-law,
\begin{equation}
\rho(r_x)=\rho_0\left(1+\left({r_x\over a}\right)^2\right)^{\gamma/2},
\label{eqn:core}
\end{equation}
where $a$ specifies the core scale, and $\rho_0$ the central surface density.

We did not attempt to derive $a$ formally, but simply compared the
quality of the fits obtained by varying $a$ geometrically in the
sequence of $a=5,\ 10,\ 20,$ and 40 kpc.
We fixed $\gamma=-2.33,$ given its excellent description of the
outer profile, where the ROSAT PSF would have little effect.
As it happens, even with
the large 50 kpc outer limit of the inner-most bin using the full X-ray
sample, we required $a\leq10$ kpc to obtain a satisfactory fit to the
central point.  The ``observed'' profile for $a=10$ kpc incorporating
the blurring of the ROSAT offsets
is shown in Figure \ref{fig:x_off_surf} as the solid blue line,
while the intrinsic unblurred-profile is shown as the dotted line.
The effects of the ROSAT resolution is evident on the simulated profile,
but with $a=10$ kpc, the form given by equation (\ref{eqn:core})
just matches the central point.
It is also noteworthy that the profile incorporating ROSAT data
is indeed compatible the profile inferred from Chandra data alone.

The subset of Chandra observations
underscores the conclusion that any core in the BCG $r_x$
distribution must be extremely small.  Figure \ref{fig:x_off_surf}
shows the implied surface density from the Chandra clusters alone.
Again the Chandra points were incorporated into the bins representing the
full sample, but here we can use considerably finer radial bins,
given the superb Chandra angular resolution.  The inner Chandra
bin extends from the origin to 4 kpc, with the next two bins covering
4-16 and 16-64 kpc.  The inner-most Chandra bin is fully an order
of magnitude above the density implied by the $a=10$ kpc profile.
The pure power-law fit given in equation (\ref{eqn:cusp}) did not
incorporate the central Chandra-only points, but its inward extrapolation
clearly falls only slightly above them.  In short, the BCG $r_x$
distribution shows no sign of any core or decrease in slope
as $r_x\rightarrow0.$

\subsubsection{Some BCGs Have Large X-ray Offsets}

A radial integral over equation (\ref{eqn:cusp}) gives the cumulative
distribution of the BCGs away from the cluster X-ray center, which is shown
as an insert in Figure \ref{fig:x_off_surf}.  The integral starts
at 1 kpc to avoid the central divergence, and continues out to include
a few clusters with $r_x>1$ Mpc.
The steep power-law form in equation (\ref{eqn:cusp}) unifies
two superficially different pictures of where BCGs are located
in their hosting clusters.
The median $r_x$ implied by this distribution is only $\sim10$ kpc,
and is consistent with the common impression that most BCGs
reside close to the center of the X-ray gas, and presumably to the
center of the cluster potentials.
At the same time, the distribution also includes BCGs with large
displacements from the X-ray defined center;
15\%\ of the BCGs in the present sample have $r_x>100$ kpc,
with the largest offsets reaching $\sim1$ Mpc.
Because the finding that some BCGs may be greatly displaced from the center
of the cluster potential is strongly at odds with the paradigm that
BCGs should be centrally located (at least in relaxed clusters),
we review the evidence for BCGs with large $r_x$ and
their import for understanding the formation of BCGs and clusters.

As noted in the Introduction, the first large survey of the X-ray morphology
of galaxy clusters \citep{j84} showed that the majority of systems
have well-defined X-ray cores largely coincident with the position of
a bright galaxy, typically the BCG.
However, the same study also showed that if the cluster sample was sorted
by X-ray core radius, the ensemble showed a smooth progression
to clusters with large X-ray cores not coincident with any particular galaxy.
To underscore this point we note two well-studied rich clusters
that have long been known to have BCGs markedly displaced
from the peak of the X-ray emission.
A1367, part of the present sample, is among the first examples found
of a cluster with regular X-ray morphology, but with a large
offset between the BCG (NGC 3842 in this case) and X-ray center \citep{b83};
Table \ref{tab:bcg_par} gives $r_x=354$ kpc for this cluster.
The Coma cluster (A1656) is the classic example of a rich galaxy cluster,
yet it also is a system with a large BCG/X-ray offset, having $r_x=256$ kpc.
\citet{w93} analyzed a deep ROSAT image of Coma,
and in conjunction with the positions and X-ray morphology of its two bright
central elliptical galaxies, NGC 4889 (the BCG) and NGC 4874 ({\rm M2}),
concluded that Coma was produced in a still
ongoing merger of two massive clusters.

\begin{figure*}[!t]
\plotone{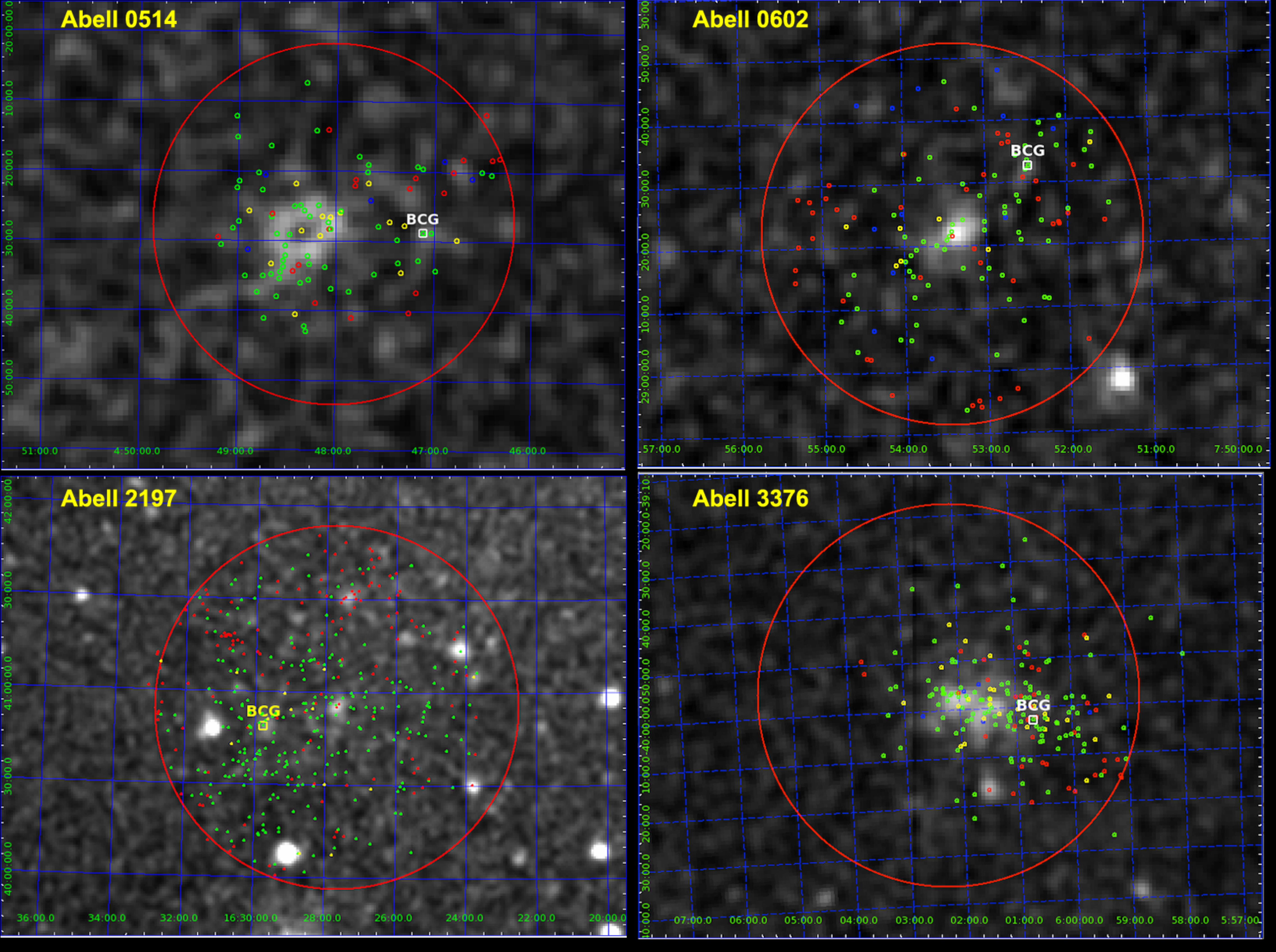}
\caption{ROSAT All-Sky Survey images (smoothed with a gaussian)
are shown for four clusters with $r_x\sim1$ Mpc.
For A0514 $r_x=1060$ kpc, A0602 $r_x=1134$ kpc, A2197 $r_x=886$ kpc,
and A3376 $r_x=995$ kpc. The large circle marks the Abell radius
centered on the peak of the X-ray emission.  The location of the BCG
is indicated.  Points mark galaxies with measured velocities, with
green for galaxies within $1000~{\rm km~s^{-1}}$ of the mean
cluster velocity, yellow for galaxies within
1000 to $1500~{\rm km~s^{-1}}$ of the mean,
red for galaxies with velocities $>1500~{\rm km~s^{-1}}$ above the mean,
and blue for galaxies with velocities $>1500~{\rm km~s^{-1}}$ below the mean.
In all four clusters the BCG clearly falls within the velocity
and spatial distributions of cluster galaxies, but are markedly
displaced from peak of the associated X-ray emission.}
\label{fig:rosat_large_off}
\end{figure*}

\citet{martel} emphasize that the BCG offset from the
cluster center of mass (which may be different from the location
of the peak X-ray emission), as well as the
velocity offset discussed in the previous section,
is a signature of the assembly of galaxy clusters by hierarchical merging.
The BCG itself may be introduced into the cluster
as part of an infalling group.
A key point is that the galaxies, X-ray gas, and the dark
matter halo of the cluster all have strongly different mechanisms
and time scales for relaxing after cluster mergers.
While we might expect the X-ray morphology of the cluster
to be disturbed by strong or recent mergers with smaller clusters or groups,
it is likely that the regularity of the X-ray gas distribution
is re-established before any new BCG introduced by the merger
is dynamically ``captured'' by the central potential.
The regularity of the X-ray morphology plus the amplitude
of $|\Delta V_1/\sigma_c|$ and $r_x$ in fact may provide means to
constrain the recent merger history of clusters.

Previous studies of the location of BCGs with respect to the peak
of the X-ray emission have produced diverse results.  \citet{patel} measured
$r_x$ for a sample of 49 clusters and found $r_x>100$ kpc in 16 systems,
or 33\%\ of the sample,
a fraction considerably larger than the 15\%\ that we found.
The positional accuracy of their centers is low,
so a large fraction of the measured offsets with $r_x\sim100$ kpc
may really be significantly smaller;
however their $r_x$ distribution has a long tail extending to
three clusters with $r_x>500$ kpc.
In contrast, \citet{haarsma} find $r_x>100$ kpc for just one cluster
out of their small sample of 33, although, as we noted above,
they also included proximity to the X-ray peak as a criterion for
selecting their BCGs in the first place.
It does appear that there is an important distinction
between searching broadly within the cluster for the BCG versus
selecting the brightest galaxy within the core of the X-ray emission.
\citet{h14} explicitly limited their search for the BCG to within
500 kpc of the X-ray peak, but found offsets out to this limit.

While we have used extensive imaging
and velocity observations to cast a wide net for the BCG in any cluster,
we have relied on the literature to provide the matching X-ray centers.
In order to understand the reliability of the largest
BCG/X-ray offsets seen in our sample,
we obtained archival Chandra or ROSAT images for the subset of
clusters with $r_x>500$ kpc.
Of the clusters with X-ray centers available,
we initially identified 29 clusters with $r_x>500$ kpc.
Of these, we accepted 22 clusters as credible systems with $r_x$
of this amplitude.
Our criterion was that the X-ray center had to fall within
the extended X-ray source closest to the BCG that was associated
with galaxies consistent with the cluster redshift.

Of the seven clusters rejected, three were cases
in which the X-ray emission was from either a foreground or
background system seen in projection close to the nominal cluster,
which itself had no detectable X-ray emission.
Since we thus had no valid X-ray center, these clusters
were dropped from the set with X-ray data.
In four clusters,
the clusters were either binary, with the BCG clearly associated
with a different X-ray component than we had assumed, or the X-ray
emission from a projected cluster at different redshift had
been selected over the X-ray emission from the nominal cluster.
In these cases we remeasured the $r_x$ with respect to the revised centers.
In one cluster, A0548, $r_x$ decreased, but still remained $>500$ kpc.
In the end we conclude that 22/174 or 12\%\ of the sample has $r_x>500$ kpc.
We show X-ray maps for four examples of clusters
with $r_x\sim1$ Mpc in Figure \ref{fig:rosat_large_off}.
The clusters have well-defined
central X-ray emission, but their BCGs are well outside of it.
 
In addition to these observational tests, we are encouraged by the
cluster formation simulations of \citet{martel}, which produce ensembles
of clusters that exhibit both the large $|\Delta V_1/\sigma_c|$ and
$r_x$ seen in the present sample.  Martel et al.\ argue that their
simulations support cluster formation by the ``merging group scenario.''
As various groups merge with the
cluster over the age of the Universe, the identity of the BCG
may change several times.  Newly arrived BCGs can be marked by high
$|\Delta V_1/\sigma_c|$ and large offsets from the
center of their clusters, which we characterize with $r_x.$

The history of many of the more massive clusters simulated by \citet{martel}
show significantly long periods during which the BCG lies at a projected
distance of more than 500 kpc or even 1 Mpc from the cluster center.
The typical value of $|\Delta V_1/\sigma_c|$ is found
to range over 0.15 to 0.31 for Abell-like clusters,
in excellent agreement with the median value
of 0.26 found for our sample in the previous section.
The maximum value of $|\Delta V_1/\sigma_c|$ seen in the simulations
may briefly exceed 1.5 in the early stages of a merger,
again in good agreement with the observational limits on the
BCG peculiar velocities.
We argue later in this paper
that additional lines of evidence support the merging-group scenario.

\begin{figure}[!t]
\includegraphics[width=\columnwidth]{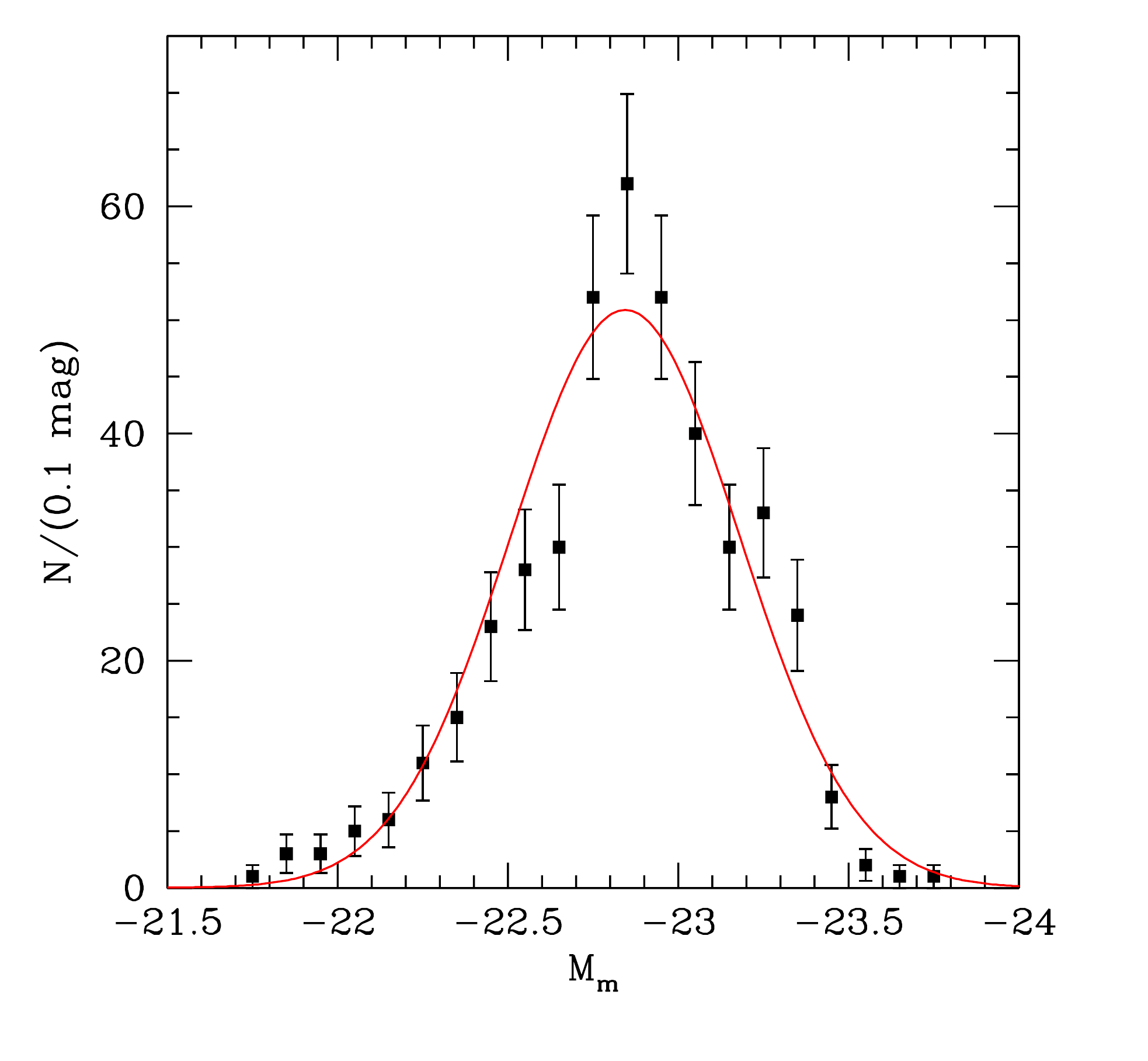}
\caption{The binned distribution of metric
luminosity, $M_m,$ for the present BCG sample.
The bins are 0.1 mags wide and the errors are from Poisson statistics.
The red line shows the best-fit Gaussian,
which has mean $M_m=-22.844\pm0.016$ and $\sigma_L=0.337$ mag ($R_C$ band).}
\label{fig:lm}
\end{figure}

\section{The Photometric and Kinematic Properties of BCGs}\label{sec:prop}

The average luminosities of BCGs have long been known to have relatively
little dispersion, allowing these galaxies to be used as 
``standard candles'' \citep{hms,s72a,s72b}.
In this section we will explore the luminosity
distribution function of BCGs and its relationship to other physical 
properties of the galaxies, such as their concentration and central
stellar velocity dispersion.  The BCG metric luminosities and
structural parameters are tabulated in Table \ref{tab:bcg_par}.  The
CMB frame has been assumed for calculation of all parameters.

\subsection{The Metric Luminosity of BCGs}

Figure \ref{fig:lm} shows the distribution of $M_m$ for the present
sample, where we have applied the extinction and k-corrections
outlined in $\S\ref{sec:imred}$ to the observed $R_C$ surface photometry.
The distribution is well fitted by a Gaussian with
mean $-22.844\pm0.016,$ and standard
deviation of $\sigma_L=0.337$ mag.
This is good agreement with $\sigma_L=0.327$ mag measured from the 15K sample
in PL95.
We have excluded three extremely faint BCGs
of the total sample of 433 BCGs from the Gaussian
fit and most of the analysis that follows.
The lowest luminosity bin plotted in Figure \ref{fig:lm} is
$-21.7>M_m>-21.8,$ which contains a single BCG.\footnote{When we
explicitly refer to $L_m$ in magnitude units we will use the
variable $M_m,$ or absolute metric magnitude.}
The three BCGs in question, those in A3188, A3599,
and A3685, all have $M_m>-21.46,$
which is yet fainter by $\sim\sigma_L;$
all are fainter than the mean $M_m$ by $>4\sigma_L.$
A3599 and A3685 have very few cluster members with velocities and
may not be real systems.

\begin{figure*}[!t]
\plotone{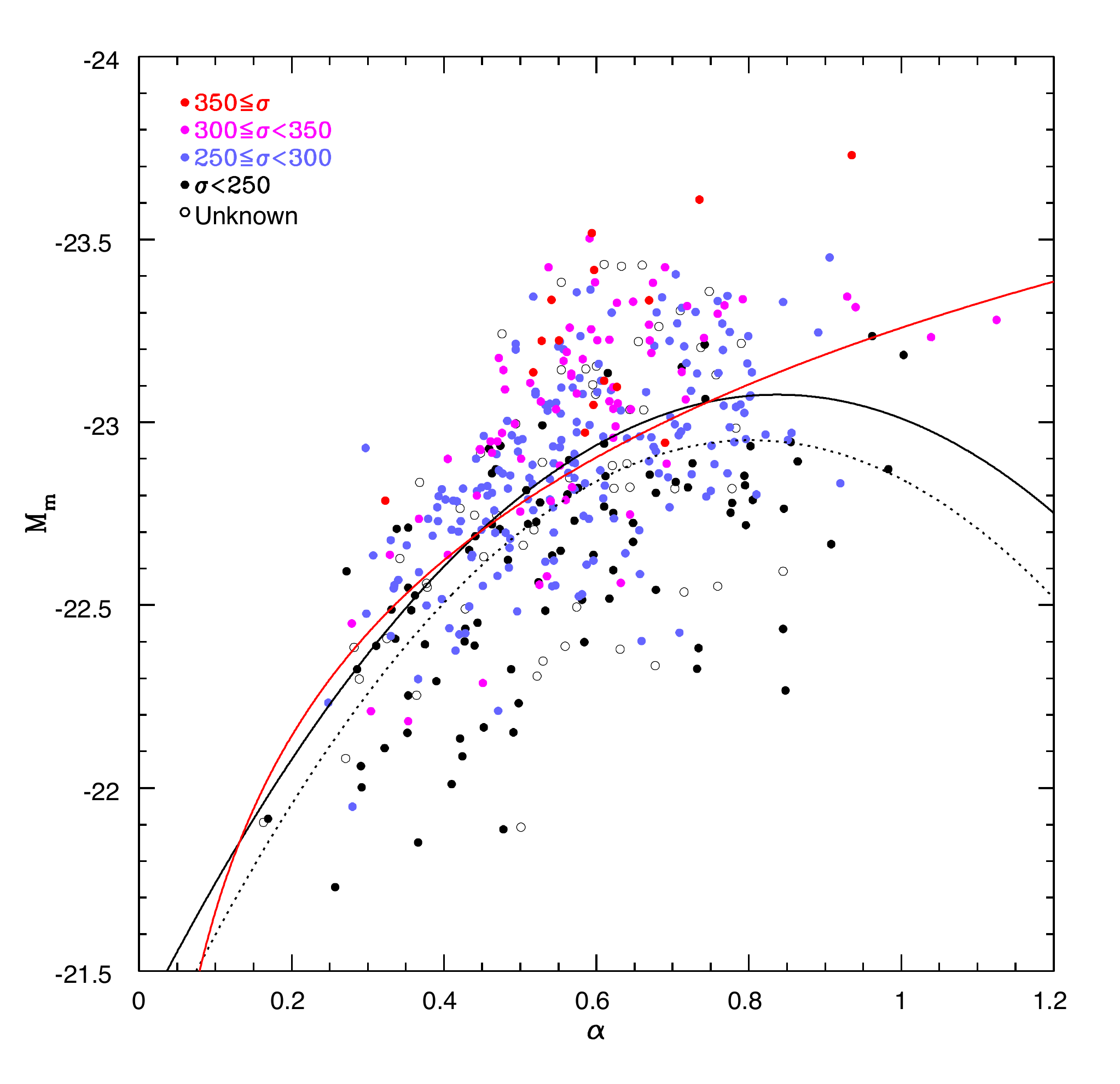}
\caption{The relationship between metric luminosity, $M_m,$ and
$\alpha$ is plotted for BCGs.
The solid black line is the mean quadratic $L_m-\alpha$
relation for the present sample, given by equation (\ref{eqn:lalp}),
while the red line fits the present sample with
a linear function of $\log\alpha$ (equation \ref{eqn:lalplog}).
The dotted line shows the quadratic $L_m-\alpha$
relation of PL95 rescaled
to $H_0=70~{\rm km~s^{-1}~Mpc^{-1}}.$  Symbols are color-coded by central
stellar velocity dispersion, $\sigma.$
Note that at any $\alpha,$ objects with higher $M_m$
typically have higher $\sigma.$}
\label{fig:lalp}
\end{figure*}

\subsection{The $L_m-\alpha$ Relationship}\label{sec:lalp}

\citet{h80} observed a sample of BCGs to refine their use as ``standard
candles'' in cosmological probes, finding that $L_m$ correlated with
the physical concentration of the galaxies.  Elliptical galaxies
have long been known to have a relationship between {\it total}
luminosity and effective radius, which is reflected in
the relationship between the metric luminosity and radial scale as well.
\citet{h80} expressed the physical concentration of the BCGs
in terms of $\alpha,$ the logarithmic slope of the variation
of $L_m$ with the physical radius of the aperture, $r,$
evaluated at the metric radius:
\begin{equation}
\alpha\equiv d\log L_m~/~d\log r~\big|_{r_m}.
\end{equation}
In addition to serving as a measure for the concentration of the BCGs,
$\alpha$ also relates the error in $L_m$ to a corresponding distance error,
$\sigma_D=\sigma_L/(2-\alpha)$ for using BCGs as standard candles.
When $\alpha=0,$ all the galaxy light is contained within the aperture,
and $L_m$ is the total luminosity, while when $\alpha=2,$
the surface brightness distribution is constant with radius, and the
metric luminosity provides no information on distance.

The $L_m-\alpha$ relationship observed by \citet{h80} showed that
$\alpha$ initially increases steeply with $L_m,$ then plateaus for
the more luminous BCGs.
LP94 and PL95 confirmed this behavior,
and quantified it by fitting a quadratic relationship
between $L_m$ and $\alpha,$
which they used as a distance indicator.
For this application, $\alpha$ is used to predict $L_m,$ which
in turn is used as a ``standard candle'' to infer the distance
to the BCGs --- in this case, $L_m$ must be the dependent,
rather than independent variable.
The $L_m-\alpha$ relationship for the present
sample is plotted in Figure \ref{fig:lalp}.  A quadratic form fitted
to all BCG but the three with $M_m>-21.5$ gives
\begin{equation}
M_m=-21.35\pm0.13-(4.12\pm0.43)\alpha+(2.46\pm0.36)\alpha^2.
\label{eqn:lalp}
\end{equation}
\vskip 8pt
\noindent We note that we weight all points equally in this fit, and
those that follow, unless we indicate otherwise.  This is because the
observed scatter around the relation of Eq.~\ref{eqn:lalp} is over an
order of magnitude larger than the formal observational errors.
The residuals about the relation are 0.267 mag rms in $M_m$ in the CMB frame,
essentially identical to the CMB-frame residuals of 0.261 mag for
the 15K sample of LP94.
In passing, we note that the quadratic form gives significantly
smaller residuals in $L_m$ than does a simple linear relation in $\alpha,$
as we first showed in PL95.
The bulk flow of the Abell cluster sample with respect to the
CMB was derived by LP94 by finding the average BCG peculiar
velocity field that minimized residuals in the $L_m-\alpha$ relationship.

The quadratic form of the $L_m-\alpha$ relationship derived
by PL95 is also shown in Figure \ref{fig:lalp}. The earlier
relation falls $\sim0.1$ mag below the present relationship.
The majority of this offset (0.06 mag) is likely to be due to
our present use of \citet{sfd} extinctions rather than the \citet{bh} values
used by PL95; the average $A_B$ values from SFD are 0.10 mag
greater than those of B\&H for the 15K sample.

The quadratic form has the unattractive feature, however, of
reaching maximum $L_m$ at $\alpha\sim0.8,$ and then predicting fainter
$L_m$ as $\alpha$ increases beyond this point where there are few BCGs
to elucidate its behavior.
This motivated us to introduce a new form,
fitting $L_m$ to a linear function of $\log\alpha.$
This new form closely parallels the quadratic form over most of the
domain, but is monotonic, and may better represent the handful
of BCGs with $\alpha>0.9.$  This form has the additional advantage
of requiring only two, rather than three parameters.
For the present sample with $M_m<-21.5,$ we measure
\begin{equation}
M_m=-23.26\pm0.03\ -\ (1.597\pm0.104)\log_{10}\alpha.
\label{eqn:lalplog}
\end{equation}
The residuals in $M_m$ are 0.271 mag rms in the CMB frame,
only 0.004 mag larger than those for the form given
in equation (\ref{eqn:lalp}).  This form is also
plotted in Figure \ref{fig:lalp} and only strongly differs from
the quadratic form for $\alpha>1,$ where we have very few galaxies.

\begin{figure}[!t]
\includegraphics[width=\columnwidth]{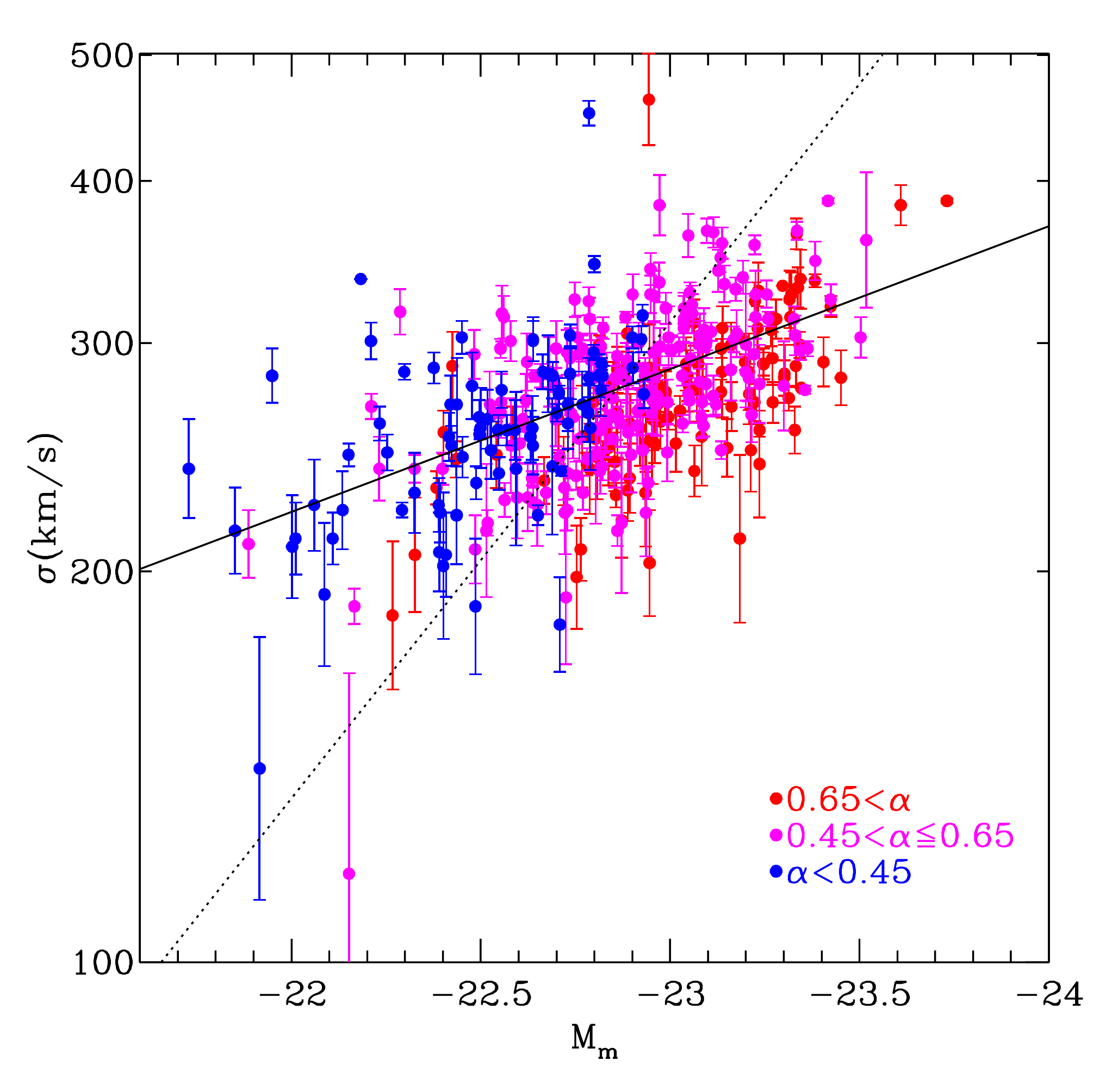}
\caption{Central stellar velocity dispersion, $\sigma,$
is plotted as a function of metric luminosity, $L_m.$
The solid line is the mean relationship between the two
parameters when $L_m$ is the independent variable (equation \ref{eqn:lsig}).
The dotted line is the relationship fitted when
$\sigma$ is the independent variable (equation \ref{eqn:sigl}).  The points
are color-coded by $\alpha.$  Note that at
any $\sigma,$ $\alpha$ tends to increase with $L_m.$}
\label{fig:l_sig}
\end{figure}

\begin{figure}[!t]
\includegraphics[width=\columnwidth]{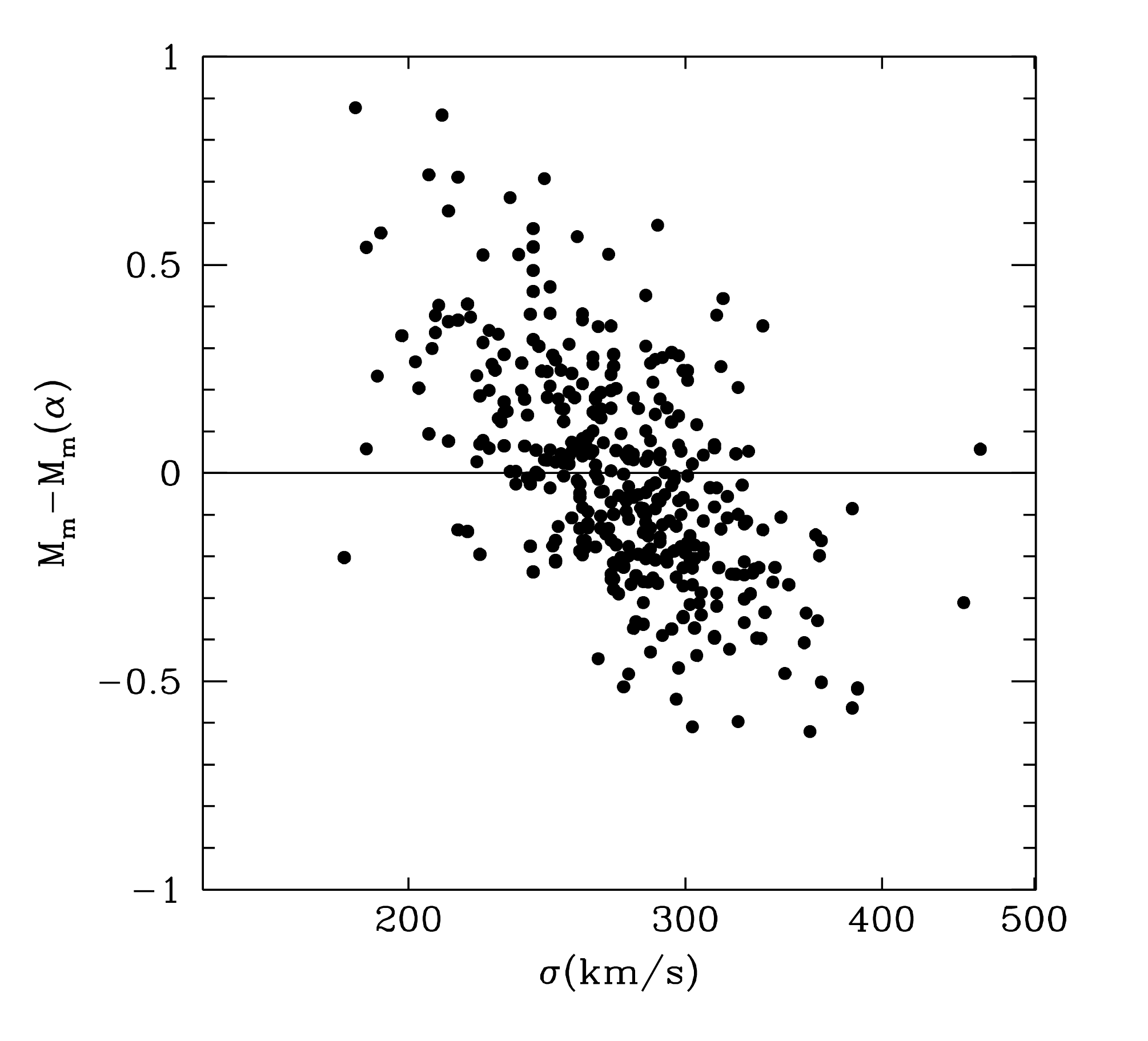}
\caption{Residuals in $M_m$ from the mean relationship
between $M_m$ and $\alpha$ given by equation (\ref{eqn:lalplog})
and shown in Figure \ref{fig:lalp}
are plotted as a function of central stellar velocity dispersion, $\sigma.$
A clear correlation is evident in the sense that positive residuals
(BCGs with fainter than the mean $M_m$ for a given $\alpha$)
correspond to low $\sigma$ and negative
residuals correspond to higher values of $\sigma.$}
\label{fig:lalp_res}
\end{figure}

\subsection{The $L_m-\sigma$ Relationship}

Figure \ref{fig:l_sig} plots the relationship between $M_m$ and $\sigma.$
While the measurement errors in $\sigma$ are subdominant to
the intrinsic scatter in this relationship,
they are not completely negligible, and we have
incorporated them into our fitting procedure, following the
methodology described in \citet{Hogg10} and \citet{Kelly11}.

If $M_m$ is treated as the independent variable, the relationship
derived from a simple least-squares fit for the 369 galaxies with $M_m<-21.5$
and measured $\sigma$ is:
\begin{align}
\log_{10}\left({\sigma\over{\rm300~km~s^{-1}}}\right) = 
\ \ \ \ \ \ \ \ \ \ \ \ \ \ \ \ \ \ \ \ \ \ \ \ \ \ \ \ \ \ \ \nonumber \\
-(0.275\pm0.023)\left({M_m \over 2.5}\right) -2.55\pm0.21,
\label{eqn:lsig}
\end{align}
\vskip 8pt
\noindent with $0.052 \pm 0.02$ scatter in $\log\sigma$, corresponding
to $\pm12\%$ in $\sigma.$
Conversely, if $\sigma$ is the independent variable, then
\begin{align}
M_m = -22.956\pm0.015-2.5(1.09\pm0.08) \times \nonumber \\
           \log_{10}\left({\sigma\over{\rm300~km~s^{-1}}}\right),
\label{eqn:sigl}
\end{align}
with an intrinsic scatter of $0.278\pm0.011$ in $M_m$,
implying that $\sigma$ is just as good as $\alpha$ is for predicting $M_m.$

The slope in equation (\ref{eqn:lsig}) is essentially the
same as the classic \citet{fj} result of $\sigma\propto L^{1/4}$ for
normal elliptical galaxies.
This suggests that the central portions of the BCGs  enclosed within
$r_m$ may have a ``normal'' relationship
between $\sigma$ and $L,$ in contrast to that between
{\it total} BCG L and $\sigma.$
As noted in the Introduction,
\citet{oh} and \citet{l07} found that $\sigma$ is only weakly
correlated with BCG {\it total} luminosity.
This behavior may reflect the putative formation of BCGs by dry mergers
of less luminous elliptical galaxies.  Simulations of this process
shows that $\sigma$ remains essentially constant over dry mergers,
with the effective radius, $R_e,$ growing rapidly with $L$ \citep{boylan}.
This associated steepening of the $R_e-L$ relation has also been seen
in BCGs \citep{l07}.

We also fitted the Faber-Jackson relationship with $M_m$
measured at $2\times$ and $4\times$ the nominal metric radius
to test the hypothesis that the relation between BCG $L$ and $\sigma$
becomes shallower as $r_m$ increases
to include a larger fraction of total galaxy luminosity,
When the metric radius is doubled, we find
\begin{align}
\log_{10}\left({\sigma\over{\rm300~km~s^{-1}}}\right) = 
\ \ \ \ \ \ \ \ \ \ \ \ \ \ \ \ \ \ \ \ \ \ \ \ \ \ \ \ \ \ \ \nonumber \\
 -(0.205\pm0.022)M_m(2r_m)/2.5 -1.94\pm0.20,
\label{eqn:sig20}
\end{align}
for the 352 BCGs that have both valid $\sigma$ and photometry at $2r_m,$
with an intrinsic scatter of $0.054 \pm 0.02$ in $\log\sigma.$
For $M_m$ measured at $4r_m$ the sample decreases to 200 BCGs,
and we measure
\begin{align}
\log_{10}\left({\sigma\over{\rm300~km~s^{-1}}}\right) =
\ \ \ \ \ \ \ \ \ \ \ \ \ \ \ \ \ \ \ \ \ \ \ \ \ \ \ \ \ \ \ \nonumber \\
-(0.147\pm0.027)M_m(4r_m)/2.5 -1.44\pm0.26,
\label{eqn:sig40}
\end{align}
with an intrinsic scatter of $0.053 \pm 0.03$ in $\log\sigma.$
In short, the slope decreases from $\sim1/4$ to $\sim1/5$
and then $\sim1/6$ as $r_m$ is doubled twice.
As even at $4r_m$ the integrated luminosity of the BCGs is still increasing,
the relationship for $\sigma$ and total luminosity would be yet shallower.

\subsection{The $L_m-\alpha-\sigma$ ``Metric Plane''}

\begin{figure}[!t]
\includegraphics[width=\columnwidth]{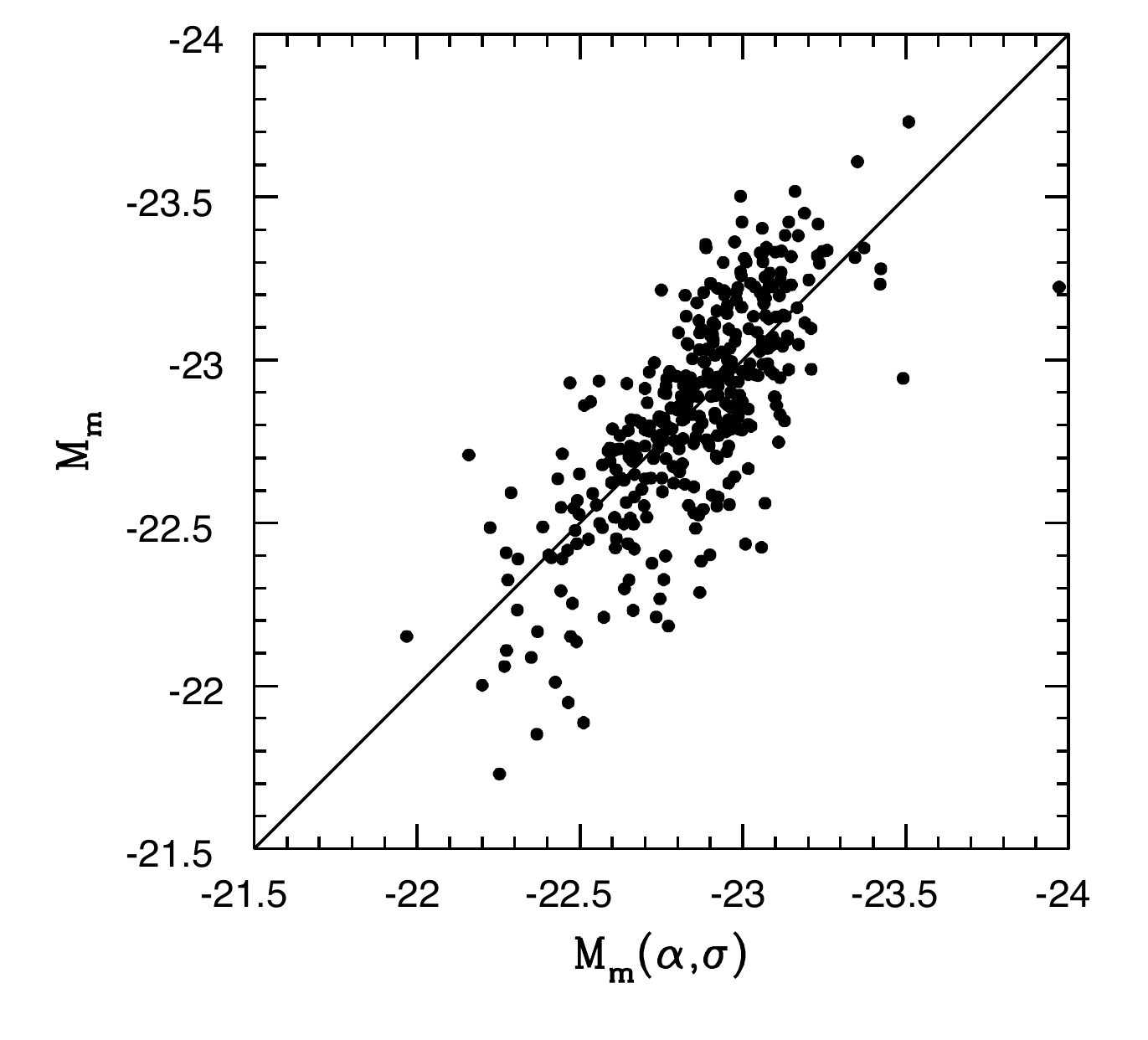}
\caption{Metric luminosity is plotted as a function
of $M_m$ estimated from $\alpha$ and $\sigma$ through
the multi-parameter relationship between the three parameters
given by equation (\ref{eqn:lalpsig}).}
\label{fig:l_al_sig}
\end{figure}

With the large 24K sample and improved knowledge of galaxy and cluster
parameters over what was available to PL95, we can now better
investigate sources of residual scatter in the $L_m-\alpha$
or $L_m-\sigma$ relationships.
In fact, given the fundamental plane relationships between
total $L,$ $\sigma,$ and $R_e$ \citep{7sfp, dnd}
for ordinary elliptical galaxies, it is not surprising to find that
a multi-parameter ``metric plane'' relationship between
$L_m,$ $\sigma,$ and $\alpha,$ has smaller scatter
than those between any two of these parameters.

The points in the $L_m-\alpha$ relationship shown
in Figure \ref{fig:lalp} are color-coded by $\sigma,$ showing
a strong gradient such that at any $\alpha,$ higher $L_m$
is correlated with higher $\sigma.$
Likewise, the color-coding of the points by $\alpha$ in the $L_m-\sigma$
plot in Figure \ref{fig:l_sig} show that at any $\sigma,$
higher $L_m$ is correlated with higher $\alpha.$
This behavior is shown more explicitly in Figure \ref{fig:lalp_res},
which plots the $M_m$ residuals from the mean $L_m-\alpha$ relationship
given by equation (\ref{eqn:lalplog}), as a function of $\sigma.$
A strong correlation is clearly evident.

Use of $\alpha$ and $\sigma$ together to predict $M_m$
for the 368 galaxies with $M_m<-21.5$
and measured $\sigma$ gives the relationship,
\begin{align}
M_m = & -23.31\pm0.03 -(1.43\pm0.09)\log_{10}\alpha \nonumber \\
             &\ \ \ \ - (2.20\pm0.17)\log_{10}\left({\sigma\over{\rm300~km~s^{-1}}}\right).
\label{eqn:lalpsig}
\end{align}
The intrinsic scatter is $0.214 \pm 0.010$ in $M_m,$
a marked improvement over the $L_m-\alpha$ and $L_m-\sigma$ relationships.
Figure \ref{fig:l_al_sig} plots observed $M_m$ as a function
of $M_m(\alpha,\sigma)$ estimated from this relationship.
This metric plane is not identical to a fundamental plane relation,
but with $L_m$ and $\alpha$ serving as proxies for $L$ and $R_e,$
it encodes similar structural information.
A detailed comparison between the metric plane
and a true fundamental plane relation for the present sample
is the subject of our second paper \citep{p2}.

\begin{figure*}[!t]
\includegraphics[width=7.5in]{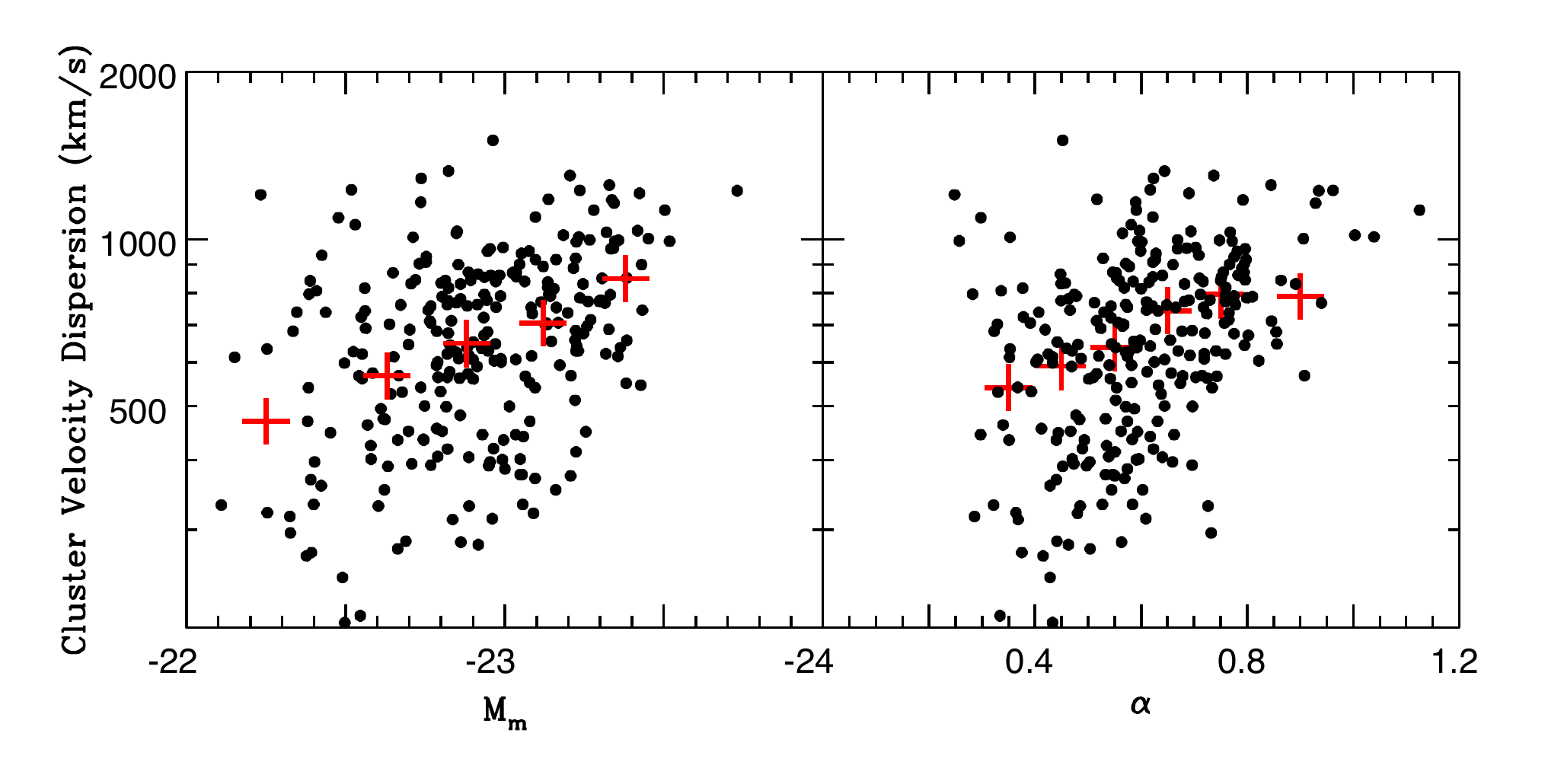}
\caption{Cluster velocity dispersion is plotted as a function
of BCG $M_m$ (left panel) and $\alpha$ (right panel)
for the 259 clusters with 25 or more galaxy redshifts.
The red crosses give the median velocity dispersion
for each 0.25 mag in $M_m$ or 0.2 bin in $\alpha.$
Both parameters increase with velocity dispersion.}
\label{fig:bcg_clus}
\end{figure*}

\subsection{The Relationship Between BCGs and Their Clusters}

\subsubsection{BCGs and the Bulk Properties of Clusters}

As noted in the Introduction, the structure and luminosity of
BCGs may be tied to the properties of the clusters as traced by
the temperature and luminosity of the associated X-ray emitting gas.
The present 24K sample suggests that BCG luminosity is also correlated with
cluster velocity dispersion, which can serve
as a proxy for cluster X-ray luminosity, $L_X,$
given the relationship between the two cluster parameters
\citep{st72, qm, wu, mgeller}.

\begin{figure}[!t]
\includegraphics[width=\columnwidth]{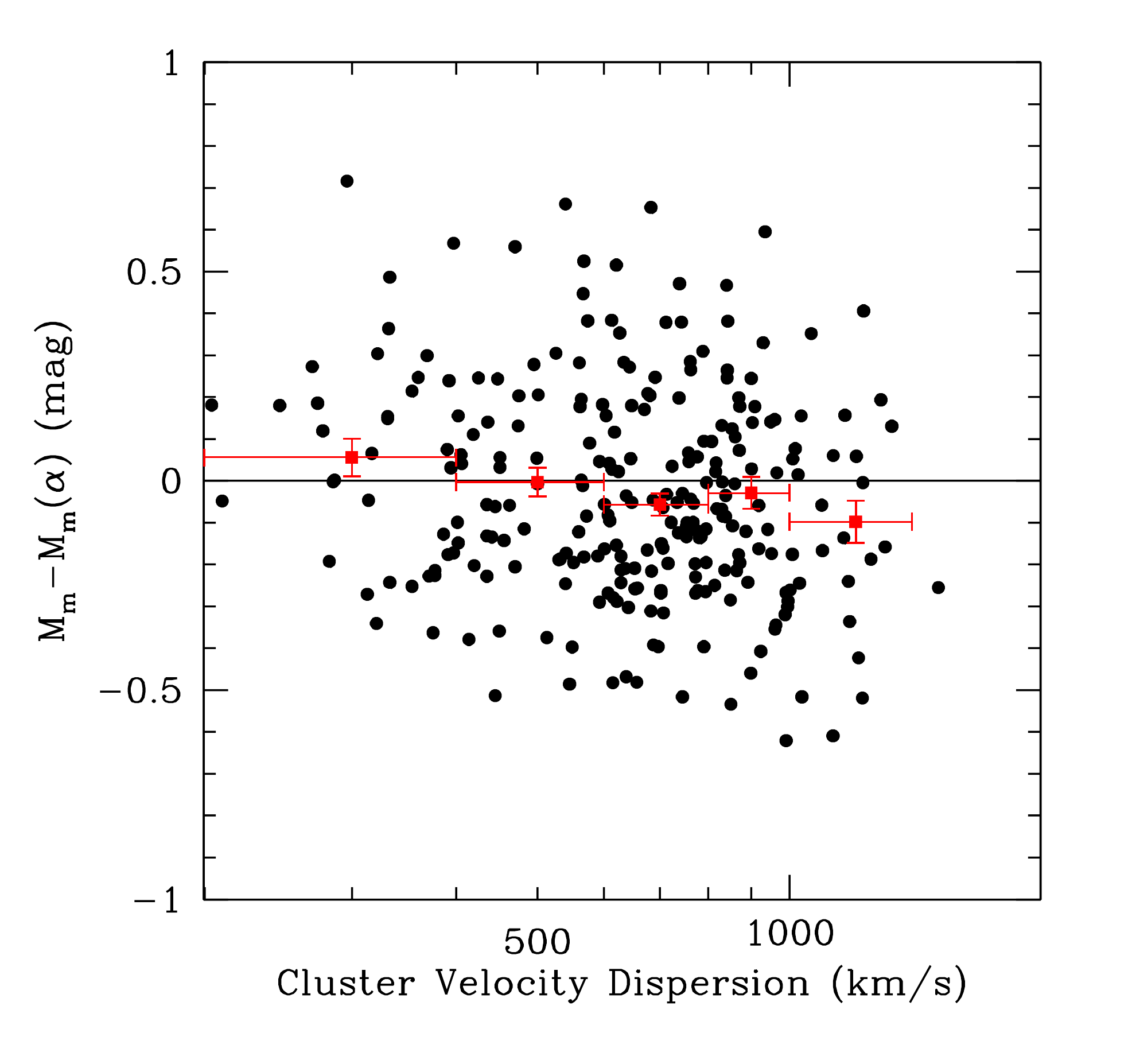}
\caption{Luminosity residuals from the $L_m-\alpha$ relationship
(equation \ref{eqn:lalplog})
are plotted as a function of cluster velocity dispersion
for the 259 clusters with 25 or more galaxy redshifts.
The red points give the mean residual in bins of width
$200~{km~s^{-1}}$ (except for the highest bin, which was
widened to include more points).  The mean of the $M_m$
residuals decreases by $\sim0.15$ mag over the range of
the cluster velocity dispersions.}
\label{fig:la_res_cd}
\end{figure}

\begin{figure*}[!b]
\includegraphics[width=7.25in]{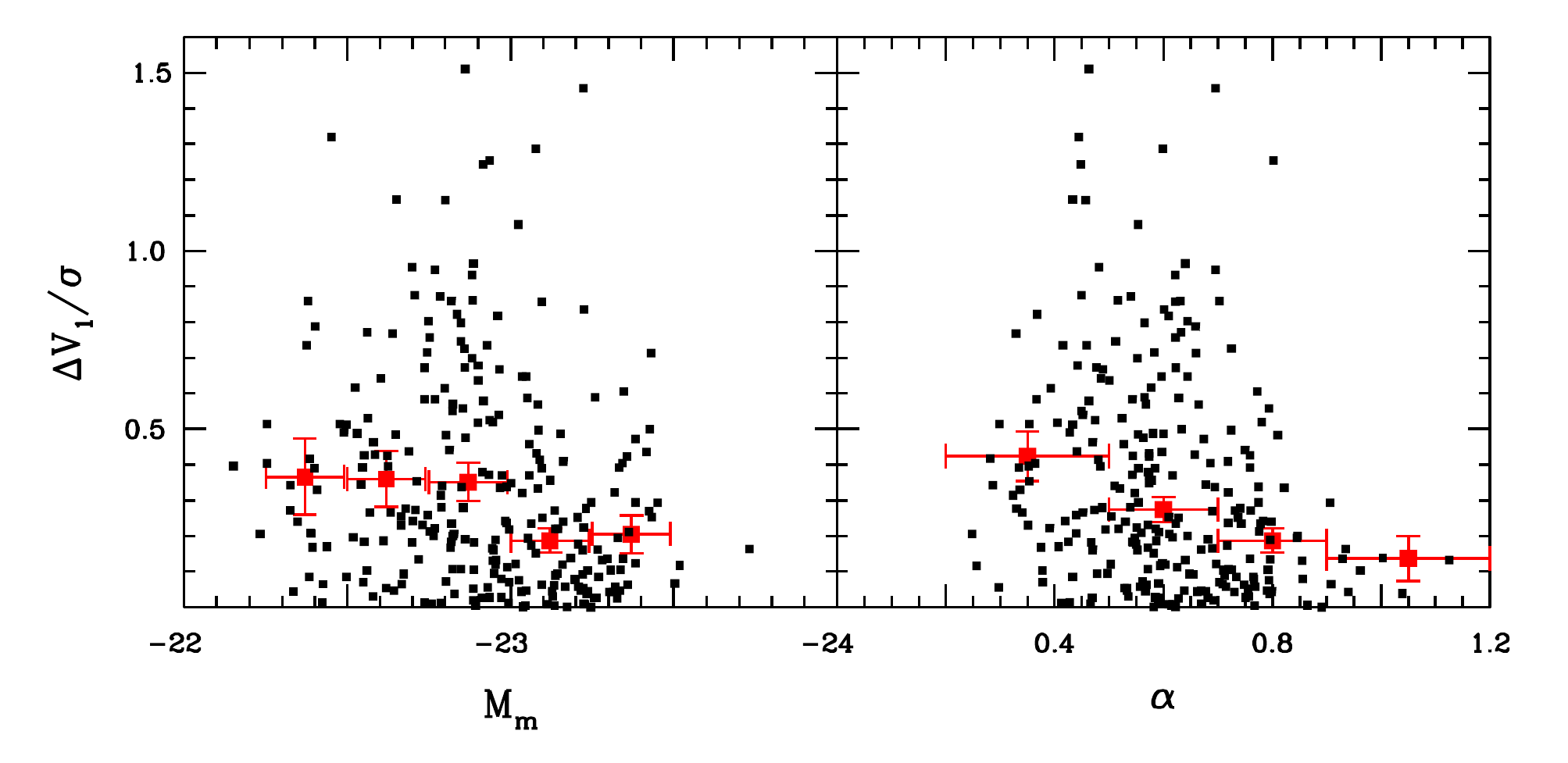}
\caption{Absolute peculiar velocity of the BCGs normalized by cluster
velocity dispersion is plotted as a function of BCG $M_m$ and $\alpha$
for the subset of clusters with 25 or more members.  The
red points gives the {\it median} $|\Delta V_1/\sigma|$ in 0.25 mag bins
in $M_m$ or 0.2 bins in $\alpha.$}
\label{fig:m1_dvn_n25}
\end{figure*}

Figure \ref{fig:bcg_clus} plots cluster velocity dispersion as
a function of both $M_m$ and $\alpha$ for the 259 clusters having 25 or more
galaxy redshifts, the minimum number needed for accurate
measurement of the velocity dispersion.  While there is considerable
scatter in the velocity dispersion at any $M_m$ or $\alpha,$
there is a clear correlation such that the median dispersion increases
by nearly a factor of two over the range of both parameters.
The relation between $\alpha$ and $\sigma_c$ in particular
echoes the relation between the shallowness
of the BCG surface brightness profile and cluster X-ray luminosity
seen by \citet{schom} and \citet{brough}.

Both correlations seen in Figure \ref{fig:bcg_clus}
raise the question of whether or not $\sigma_c$ offers
any independent information that can reduce some of the scatter
in the $L_m-\alpha$ relationship.
\citet{he97} argued that there is an $L_m-\alpha-L_X$ relationship,
which offers better prediction of $L_m$ than using $\alpha$ alone
and in doing so reduces the significance of the LP94 bulk-flow amplitude.
If we take $\sigma_c$ as a proxy for $L_X,$ we should thus
expect to see a $L_m-\alpha-\sigma_c$ relationship if
the \citet{he97} is highly significant.

Figure \ref{fig:la_res_cd} plots the residuals of the $L_m-\alpha$
relationship, given by equation (\ref{eqn:lalplog}),
as a function of cluster velocity dispersion.
The residuals of the $L_m-\alpha$ relationship
do show a barely significant correlation with cluster velocity dispersion
in the sense that brighter residuals in $L_m$ are still associated
with clusters with higher velocity dispersion.
The measured slope is $-0.23\pm0.10$ mag per dex in dispersion,
such that the mean $M_m$ residuals decrease by $\sim0.15$ mag
over the sample range of cluster velocity dispersion.

The $L_m-\alpha-\sigma$ relation (equation \ref{eqn:lalpsig}),
however, ``soaks up'' this
residual dependence on cluster velocity dispersion.  When $\alpha$ and
$\sigma$ are used to predict $L_m,$ the remaining dependence on
cluster velocity dispersion decreases to $\sim0.07$ mag/dex with no significance
over the sample range in dispersion.  Notably, the $L_m$ residuals
from this relation are already slightly better than those of the
\citet{he97} $L_m-\alpha-L_X$ relationship.
The present $L_m-\alpha-\sigma$ relation thus offers a BCG-based distance
indicator with the effects of the cluster environment removed.

Both the metric luminosity and structure of the BCGs may also be
related to the peculiar velocities of the BCGs within their hosting
clusters. Figure \ref{fig:m1_dvn_n25} plots the peculiar velocities
of the BCGs normalized by cluster velocity dispersion as a function
of BCG $L_m$ and $\alpha$ for the subset of clusters with 25 or more members.
The median peculiar velocity steadily decreases with increasing $\alpha.$
The trend of median $|\Delta V_1/\sigma|$ with $M_m$ is less clear;
however, the most luminous BCGs have relatively smaller
peculiar velocities.

\begin{figure*}[!t]
\includegraphics[width=7.25in]{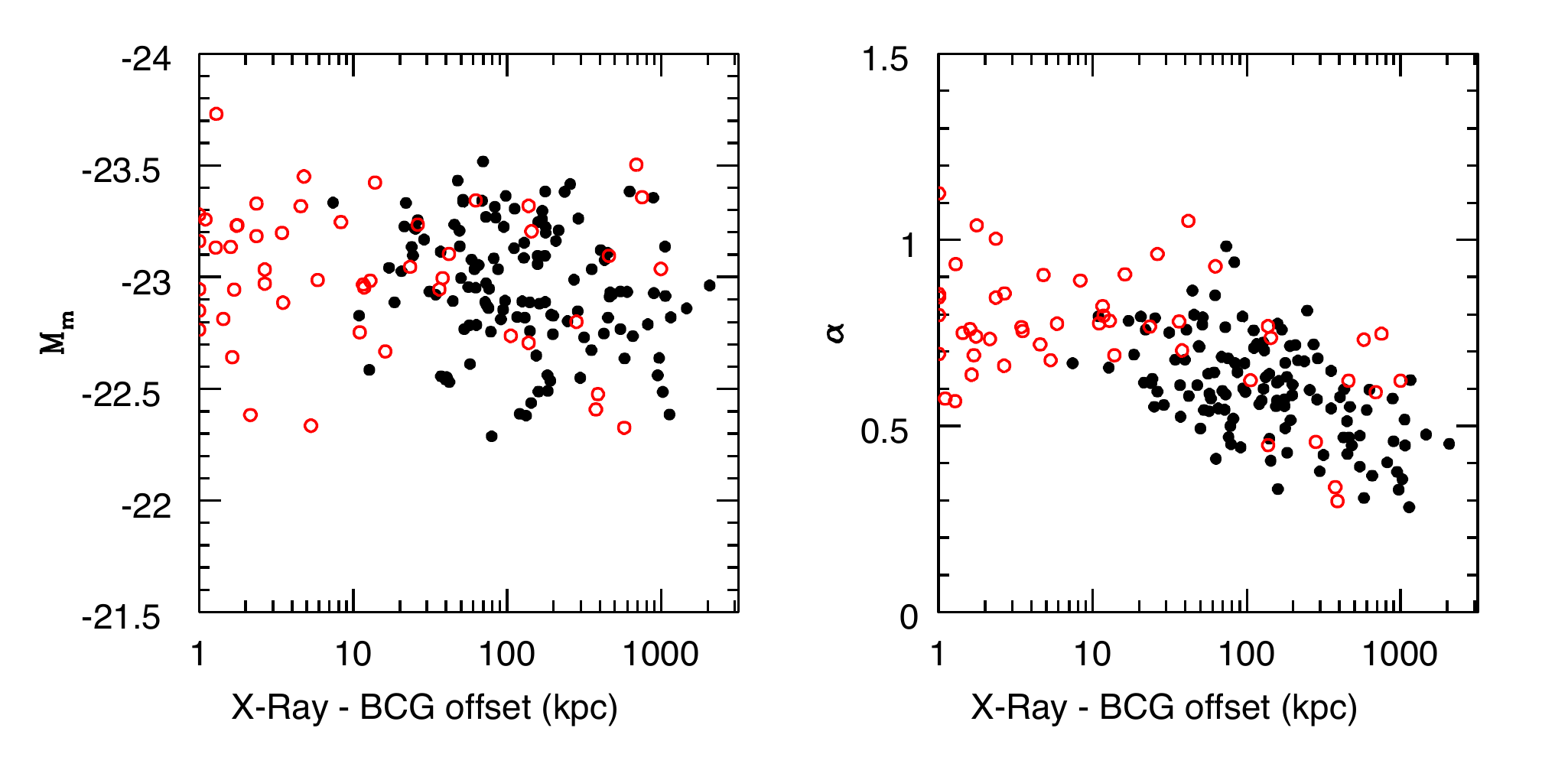}
\caption{$M_m$ and $\alpha$ are plotted as a function of distance of the
BCG from the X-ray defined center of the cluster.  Solid black symbols
indicate clusters with ROSAT measurements, while the open red symbols
indicate clusters with Chandra-based X-ray
centers.  Little dependence on $M_m$ with distance from the center
is seen, while $\alpha$ increases with decreasing distance.}
\label{fig:x_off_la}
\end{figure*}

\begin{figure}[!t]
\includegraphics[width=\columnwidth]{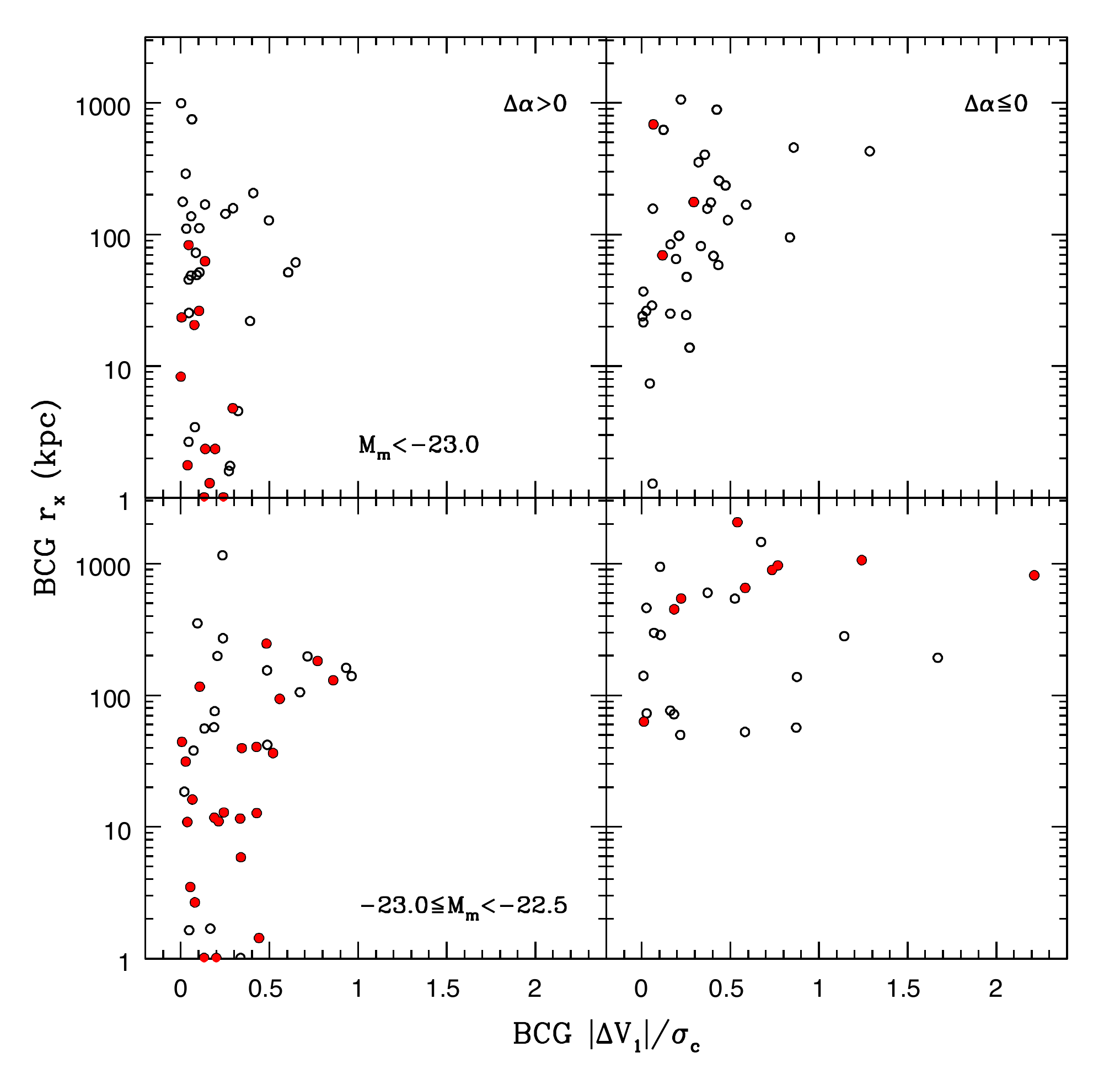}
\caption{The distribution of BCG X-ray offset, $r_x,$
and normalized velocity offset,
$|\Delta V_1|/\sigma_c,$ (for clusters with
$N\geq25$ velocities) are shown for four regions in the BCG $M_m-\alpha$ space.
The columns separate BCGs by whether or not they fall above (left)
or below (right) the mean relation of $\alpha$ on $M_m$
(equation \ref{eqn:alp_m}). 
The rows then segregate the BCGs by $M_m$ interval, with the brightest BCGs
plotted in the top row.  Red points are BCGs with $\alpha$ deviating 
from the mean $\alpha(M_m)$ relationship
by $1\sigma$ or more.}
\label{fig:xvla}
\end{figure}

\subsubsection{The Structure of BCGs and Their Positions Within the Clusters}

Figure \ref{fig:x_off_la} shows that $\alpha$ is also correlated with
the offset of the BCG within the cluster relative to
the X-ray center.  The right panel shows that $\alpha$ is clearly
larger for BCGs closer to the center of their clusters, while
left panel of the figure shows that $M_m$ is largely unrelated
to the position of the BCG within the cluster.
The former result appears to be consistent with the weak
correlation discovered by \citet{ascaso} between the effective
radii of the BCGs and their spatial offsets, with larger
BCGs being positioned closer to the centers of the clusters.

One possibility is that the increase of $\alpha$ may be dominated by
high-speed or non-merging interactions with other galaxies in the cluster.
The interactions would preferentially take place more often
at the center of the cluster and would add energy to the stellar
envelope of the BCGs, causing them to become more extended.
Without actual mergers, however, little stellar mass is added
to the BCGs, thus no overall luminosity growth occurs as the
structure of the BCGs becomes more extended.
A second hypothesis is that merging does occur as the BCG
dwells within the cluster center, but the density
of stars within the metric aperture does not increase in the process.
\citet{ho} argued that dry mergers in fact
would cause little central growth of the BCGs, a phenomenon that
is also seen in merger simulations \citep{boylan}.
Recent theoretical and observational work \citep{hop, van} indeed
argues that growth of massive galaxies since redshifts $\sim2$
is mainly in their outer envelopes.

Since both $L_m$ and $\alpha$ are related to the position
of the BCGs with respect to both the spatial and velocity centroids
of the clusters, we also examined the combined effect of the
last two parameters.   Figure \ref{fig:xvla} revisits
the plot of the BCG spatial location within the cluster, $r_x,$
versus the normalized absolute BCG peculiar velocity, $|\Delta V_1|/\sigma_c,$
which was first shown for the full sample of BCGs with
X-ray centers and accurate mean velocities in Figure \ref{fig:drvdv}.
We now split the BCGs with $M_m<-22.5$ into two luminosity bins,
\footnote{There are very few BCGs with $M_m>-22.5$ that have
X-ray cluster centers available, thus we cannot do this analysis
for the lower-luminosity BCGs.}
each of which is split further into two halves
by whether or not the galaxies are above or below
the mean relation for $\alpha,$ given $M_m,$
\begin{equation}
\alpha(M_m)=(-0.256\pm0.019)(M_m+22.5)+0.484\pm0.009.
\label{eqn:alp_m}
\end{equation}
These four subsets are shown as individual panels in
Figure \ref{fig:xvla}, with the columns separating BCGs with higher (left)
or lower (right) than average $\alpha,$ given $M_m,$ and the rows
corresponding to the two luminosity bins, with the brightest BCGs
plotted in the top row.
In any panel, we additionally note (red symbols) BCGs with
$\alpha$ residuals in excess of the rms residual,
$\sigma_\alpha$ about the mean relation;
$\sigma_\alpha=0.13$ for the full BCG sample.

Examination of the individual panels suggests that
$r_x$ and $|\Delta V_1|/\sigma_c$ indeed both work together to moderate
the structure of the BCGs.
The upper left panel in Figure \ref{fig:xvla}
contains the most luminous ($M_m<-23.0$) and extended BCGs.  There are no
BCGs with $|\Delta V_1|/\sigma_c>0.7,$ and the BCGs with $\Delta\alpha
>\sigma_\alpha,$ have $|\Delta V_1|/\sigma_c<0.3,$ or peculiar
velocities less than half of those of the BCGs in this luminosity range with
$0<\Delta\alpha<\sigma_\alpha.$
Moreover, these large $\Delta\alpha$ galaxies have $r_x<100$ kpc,
while the BCGs with $0<\Delta\alpha<\sigma_\alpha$
can have $r_x$ an order of magnitude larger.

The galaxies in the upper right panel of Figure \ref{fig:xvla}
are just as luminous as the BCGs in the panel to their left, however
this subset now has three galaxies with $|\Delta V_1|/\sigma_c>0.7,$
and a paucity of galaxies with $r_x<10$ kpc.  The three
galaxies with $\Delta\alpha<-\sigma_\alpha$ have $r_x$
an order of magnitude larger than those with $\sigma_\alpha<\Delta\alpha<0.$

BCGs in the $-23.0<M_m<-22.5$ luminosity bin largely
echo the behavior exhibited by the more luminous BCGs, although
the BCGs with $\Delta\alpha>\sigma_\alpha$ have $|\Delta V_1|/\sigma_c$
about twice as large.  The BCGs in this luminosity
range with $\Delta\alpha<\sigma_\alpha$ (right-middle panel) avoid the
centers of their clusters even more, however, with no galaxies
having $r_x<50$ kpc.  The BCGs with
$\Delta\alpha<-\sigma_\alpha$ have $r_x$ an order of magnitude yet larger.

\begin{figure}[!t]
\includegraphics[width=\columnwidth]{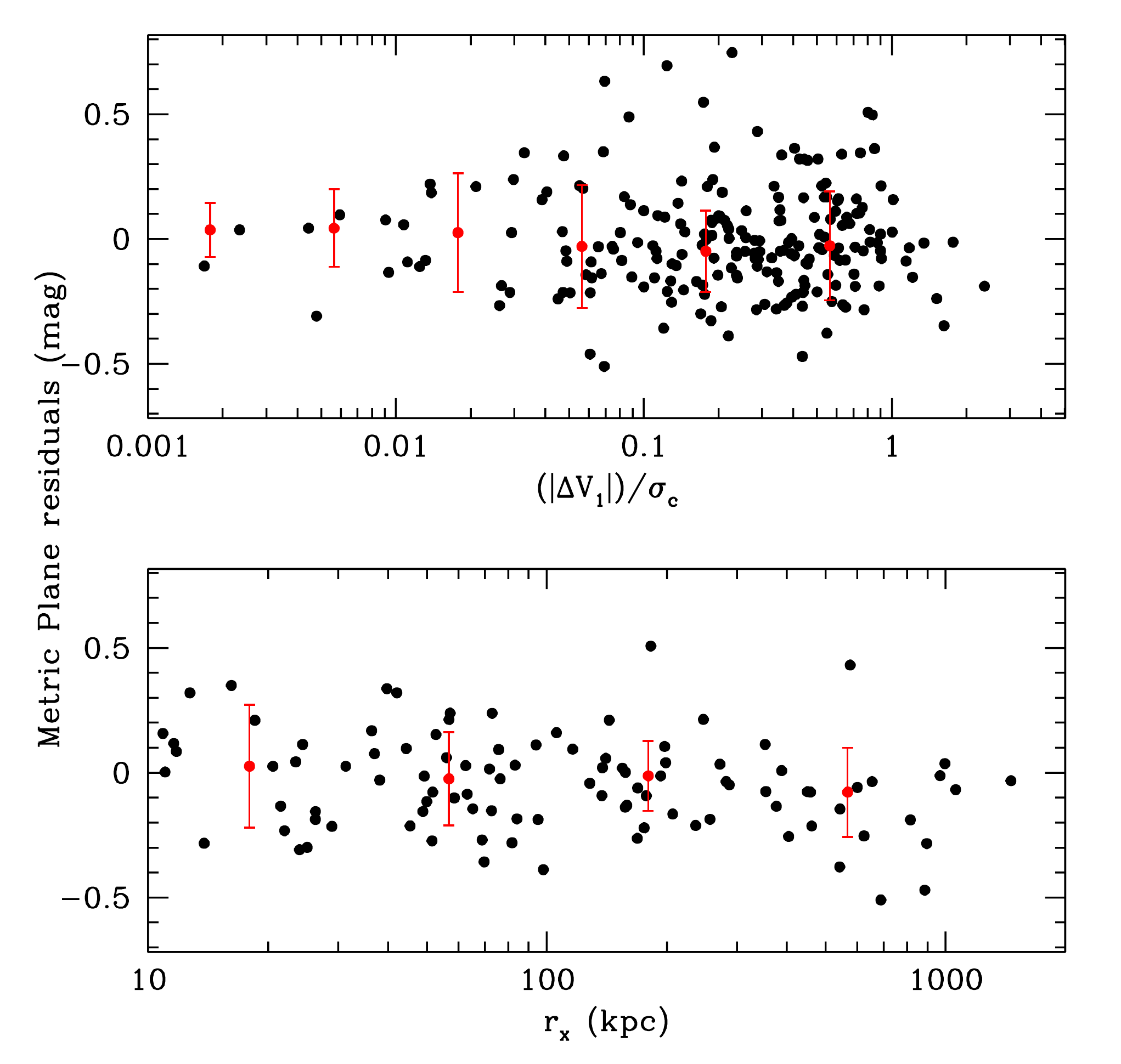}
\caption{The magnitude residuals from the metric plane relation of
Equation (\ref{eqn:lalpsig}),
shown as a function of velocity offset from the mean redshift
of the cluster (upper panel) and positional offset from the X-ray center,
where available (lower panel).
As in Figures \ref{fig:dvnorm} and \ref{fig:xvla},
the velocity offset is normalized by the velocity dispersion of the cluster.
The median values in bins, and the one-sigma widths
(as determined from the interquartile range) are also shown in red.
There is no evidence for a systematic bias, or larger scatter,
at large offset in either velocity or position.}
\label{fig:metric_res}
\end{figure}

The overall picture, again, is that $\alpha$ is larger for
the BCGs that lie closer to the spatial and velocity
center of their hosting clusters.  BCGs with $r_x<10$ kpc and
$|\Delta V_1|/\sigma_c<0.5$ are highly likely to have
markedly higher $\alpha,$ with $\Delta\alpha>\sigma_\alpha$
as compared to BCGs of the same luminosity.  Conversely,
galaxies with $|\Delta V_1|/\sigma_c>1$ always have
large $r_x$ and $\Delta\alpha<0.$  It is striking
that the BCGs with $r_x\sim50$ kpc have  larger $\alpha$
than do BCGs at larger distances from the center ---
these galaxies are still well displaced from the X-ray center,
but yet are deep enough within the potential such that
$\alpha$ has already been affected.
Conversely, position within the cluster seems to have little
effect on $L_m,$ as we already saw in
Figures \ref{fig:m1_dvn_n25} and \ref{fig:x_off_la}.

While $\alpha$ is dependent on the spatial and velocity locations
of the BCGs within the cluster,
the metric plane scatter seems to be independent of both.
Figure \ref{fig:metric_res} shows the residuals
of the metric plane (equation \ref{eqn:lalpsig}) as a function of
$|\Delta V_1|/\sigma_c$ (upper panel),
and for those objects with X-ray data, as a function of $r_x$ (lower panel).
There is no evidence for a bias or increased scatter
for objects with large offsets.
The metric plane thus again implicitly accounts for the
environmental effects of the clusters,
regardless of whether the BCGs reside in the center of the cluster
or in its outskirts.

\begin{figure}[!t]
\includegraphics[width=\columnwidth]{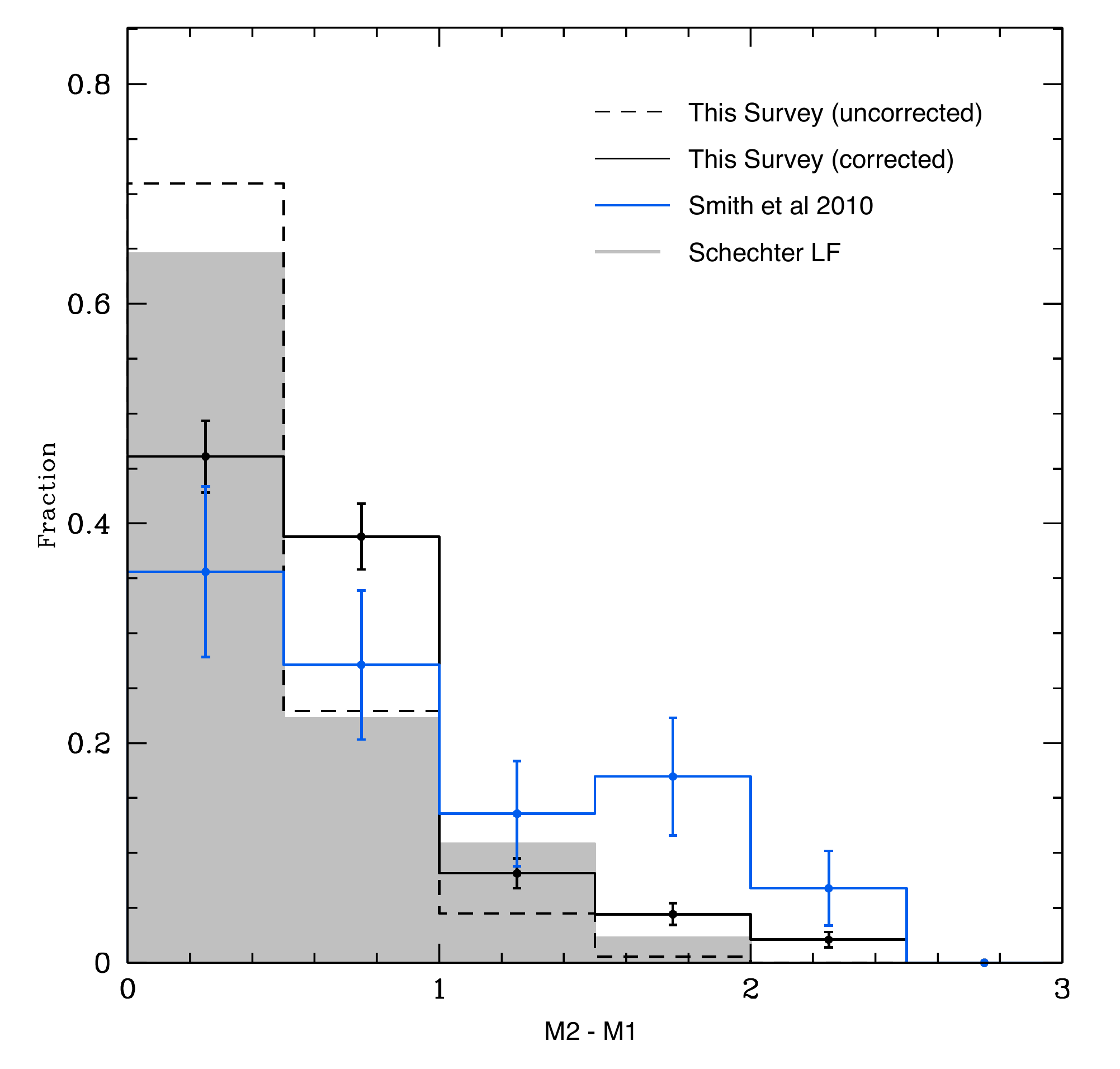}
\caption{The distribution of $M_2 - M_{1}$ for various cluster samples.
The dashed black histogram shows the $M_m({\rm M2}) - M_m({\rm BCG})$ distribution
for the 179 clusters for which we directly observed both the BCG and
the second-ranked galaxy.  $M_m({\rm M2})$ and $M_m({\rm BCG})$ are the metric
luminosities of the M2 galaxy and the BCG, respectively.
The solid black line shows this distribution corrected using a reference set
of 30 clusters with $M_m({\rm M2}) - M_m({\rm BCG})$ derived from the SDSS.
These 30 clusters were selected at random from amongst the clusters
that we did not directly observe the second-ranked galaxy.
The solid blue histogram shows the $M_2 - M_1$ distribution
from the study done by Smith et al. (2010).
The light grey shaded histogram shows the expected $M_2 - M_1$
distribution for 2000 simulated clusters with galaxy luminosities drawn from a
Schechter luminosity function.}
\label{fig:m12dist}
\end{figure}

\begin{figure}[!t]
\includegraphics[width=\columnwidth]{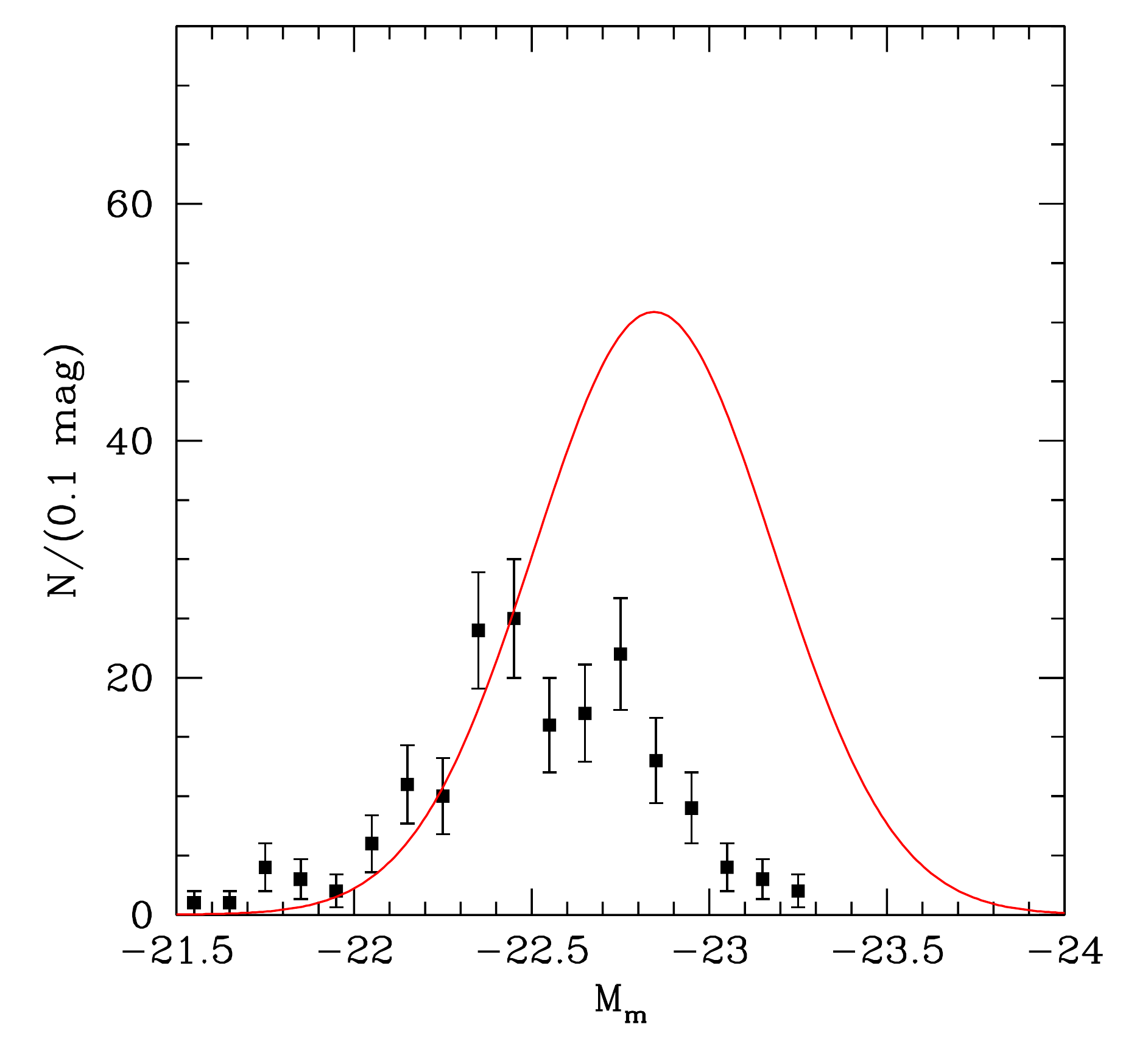}
\caption{This figure shows the distribution of the metric luminosities
of the observed M2 sample.  The red line is the Gaussian fitted
to the BCG luminosity function shown in Figure \ref{fig:lm}.}
\label{fig:m2lf}
\end{figure} 

\section{The Nature of Second-Ranked Galaxies}\label{sec:m2}

\subsection{The Photometric Properties of M2}

\begin{figure}[!t]
\includegraphics[width=\columnwidth]{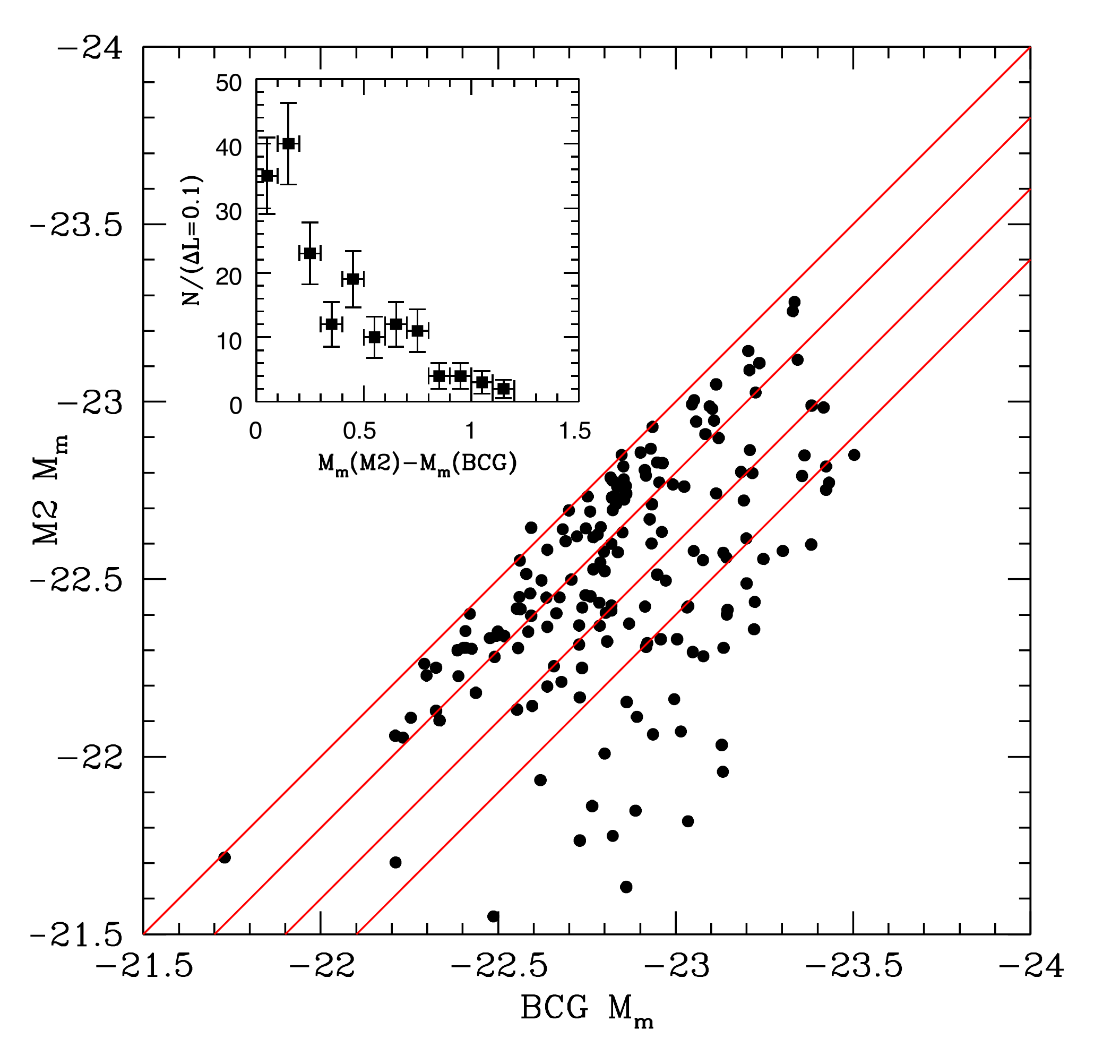}
\caption{The distribution of $M_m({\rm BCG})$ versus $M_m({\rm M2}),$
the metric luminosities of the BCG and M2 galaxies, respectively, is plotted
for the clusters with M2 observations (41\%\ of the total sample).
The upper red line line marks $M_m({\rm BCG})=M_m({\rm M2}),$ with the subsequent lines
marking offsets of $M_m({\rm M2})$ from $M_m({\rm BCG})$ in 0.2 mag steps.
The inset figure gives the histogram of the difference in metric magnitude
between the BCG and M2 in finer bins than shown in Figure \ref{fig:m12dist}.
This is equivalent to binning along the red lines in the main figure.
(The one M2 above the line is that in A3531, where the BCG becomes the
brightest galaxy in an aperture slightly larger than the nominal $r_m.$)}
\label{fig:m2m1_histograms}
\end{figure} 

\begin{figure}[!t]
\includegraphics[width=\columnwidth]{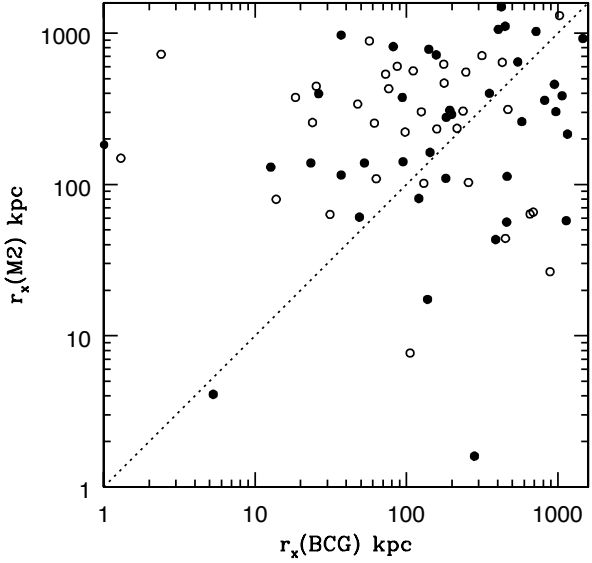}
\caption{The projected offsets of M2 from the cluster X-ray centers
are plotted against those of the BCG.  Solid symbols are clusters
for which $M_m({\rm M2})-M_m({\rm BCG})<0.3$ mag. while open symbols mark clusters
with relatively less luminous M2s.  Note that when $r_x({\rm BCG})>100$
kpc there are only three clusters with a ``rival'' M2 with
$r_x({\rm M2})<50$ kpc.  Clusters with large $r_x({\rm BCG})$
thus do not correspond to those with small $r_x({\rm M2}).$
The cluster in the figure with the smallest $r_x({\rm M2})$ is A0539.
Its BCG has a large peculiar velocity and a smaller $\alpha$ than its M2, 
but the BCG is brighter than M2 for all aperture radii.}
\label{fig:m1m2_rx}
\end{figure} 

\begin{figure}[!t]
\includegraphics[width=\columnwidth]{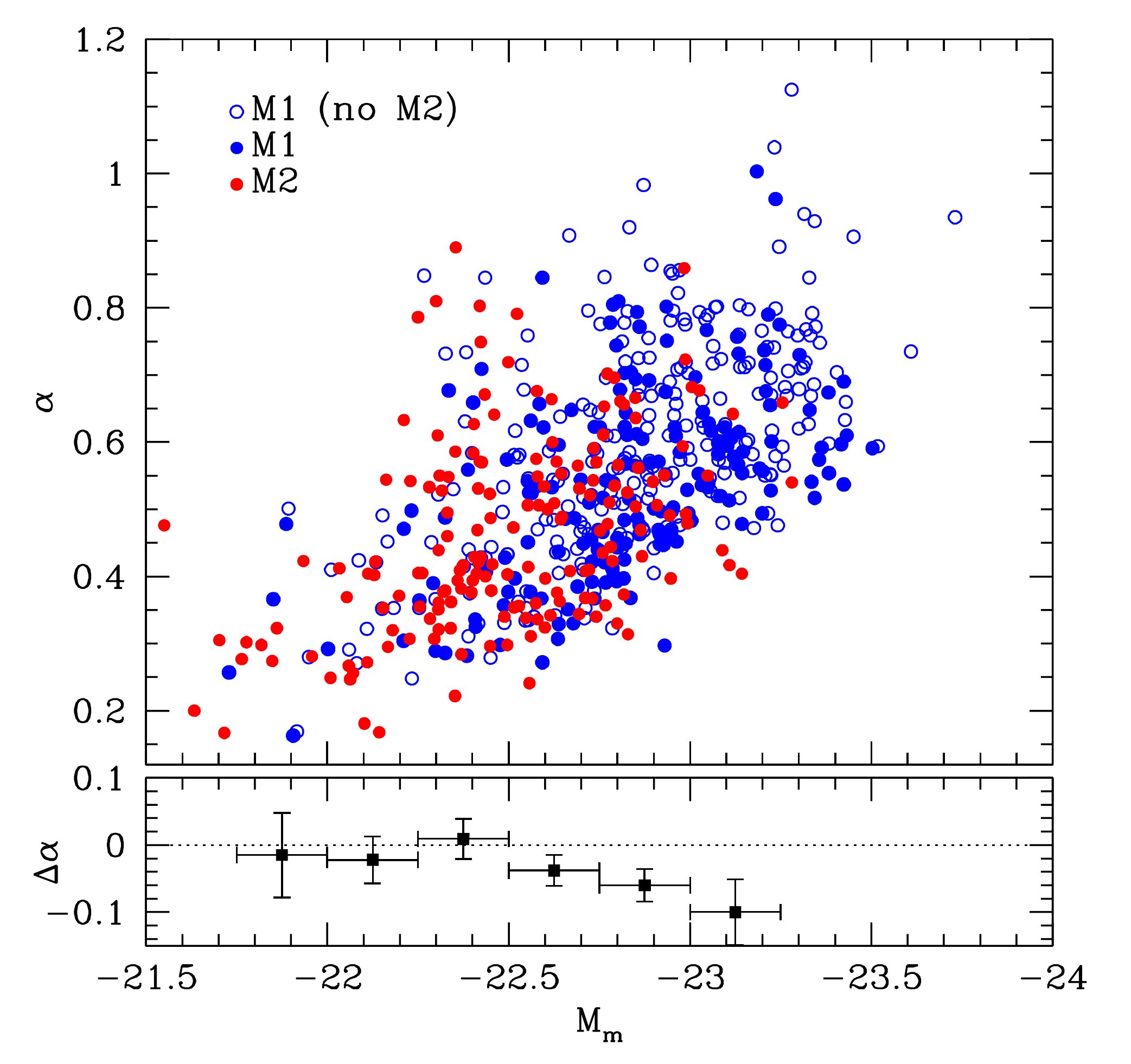}
\caption{The correlation between $\alpha$ and $M_m$
for BCGs (blue; open symbols are BCGs in clusters lacking observed M2s)
and M2s (red); $M_m$ is now shown as the independent variable.
The mean difference between the BCGs and M2 $\alpha$ in each
0.25 mag bin in $M_m$ is plotted at the bottom.
For $M_m>-22.5,$ no difference is seen between the two populations, while
the M2 have increasingly smaller average $\alpha$ than the BCGs as
$M_m$ increases in luminosity over $M_m<-22.5.$}
\label{fig:m1m2_lalp_00}
\end{figure} 

\begin{figure}[!t]
\includegraphics[width=3.5in]{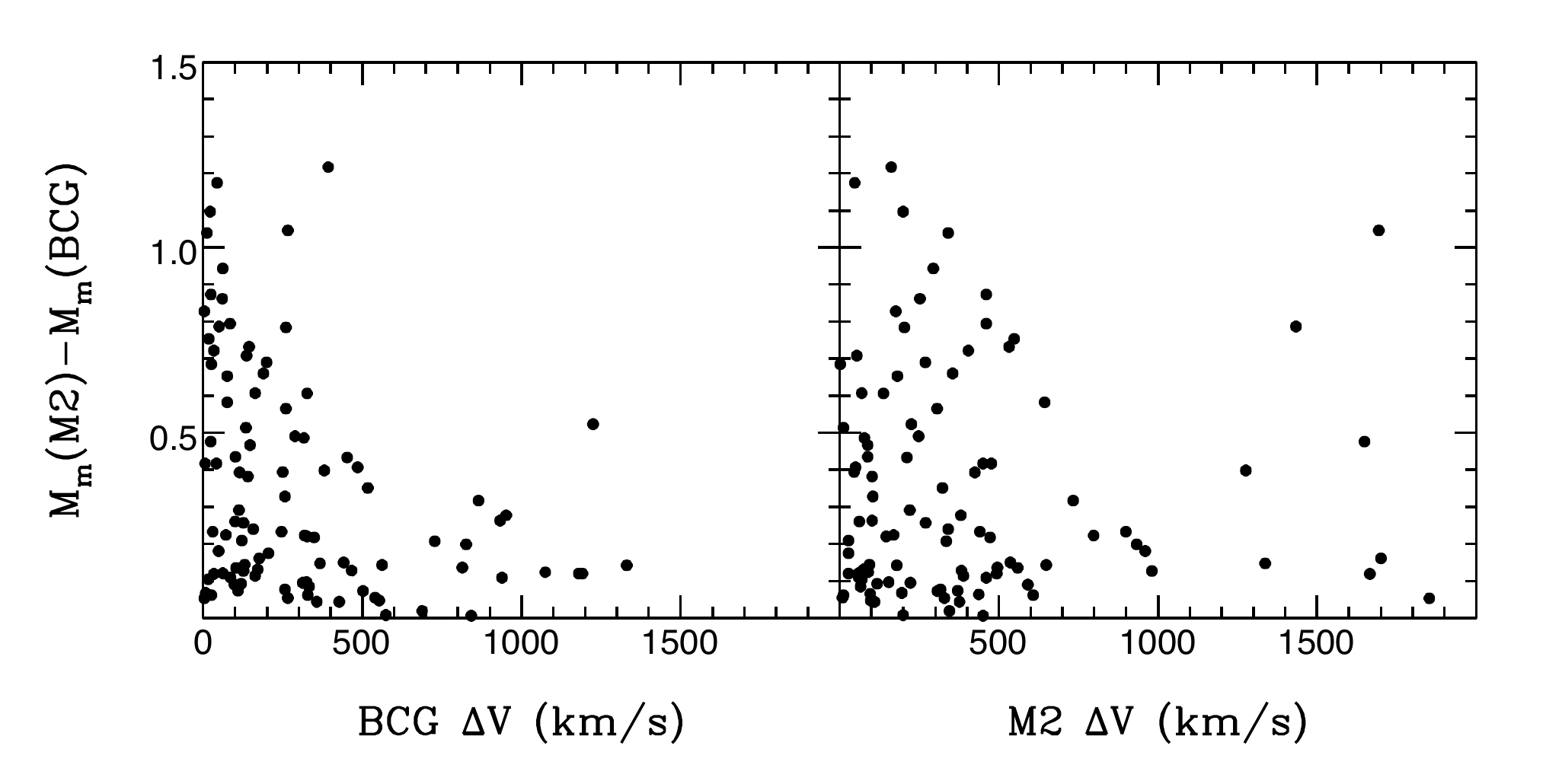}
\caption{The left panel shows the M2-BCG luminosity difference as function
of the absolute peculiar radial velocity of the BCG within the cluster.
In the right panel the luminosity difference is plotted with respect
to the peculiar velocity of M2.  Only clusters with 25 members or more
are plotted.  Note that $M_m({\rm M2})-M_m({\rm BCG})<0.5$ for nearly all clusters
in which the BCG $\Delta V_1>400~{\rm km~s^{-1}}.$}
\label{fig:m12dldv1}
\end{figure}  

\begin{figure}[!t]
\includegraphics[width=\columnwidth]{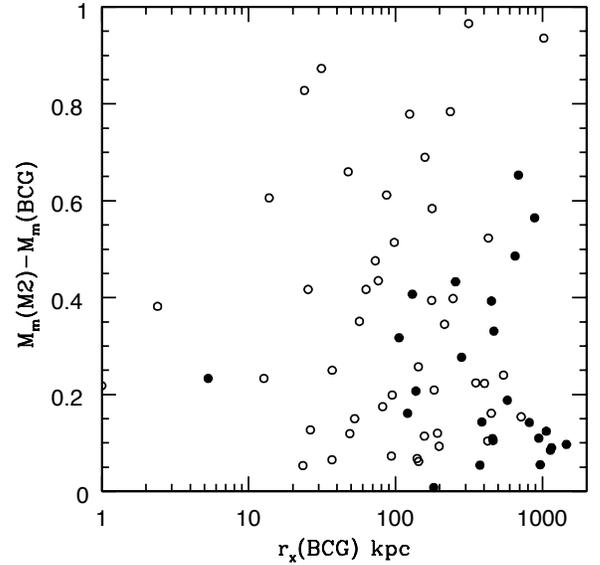}
\caption{The luminosity difference between M2 and their matching BCG,
$M_m({\rm M2})-M_m({\rm BCG})$ is plotted as a function of the offset of the BCG
from the X-ray center.  Solid symbols are clusters for which the M2
falls closer than the BCG to the X-ray center, while open symbols
mark the opposite case.
There is no relationship between the luminosity of M2 and the BCG
until $r_x<20$ kpc, where no M2s with $M_m({\rm M2})-M_m({\rm BCG})<0.2$ mag are seen.
This figure ratifies the impression from Figure \ref{fig:m1m2_rx} that
there are very few clusters with M2 closer to the X-ray center
once the BCG $r_x<100$ kpc.}
\label{fig:m1m2_dl_rx1}
\end{figure}  

\begin{figure}[!t]
\includegraphics[width=3.5in]{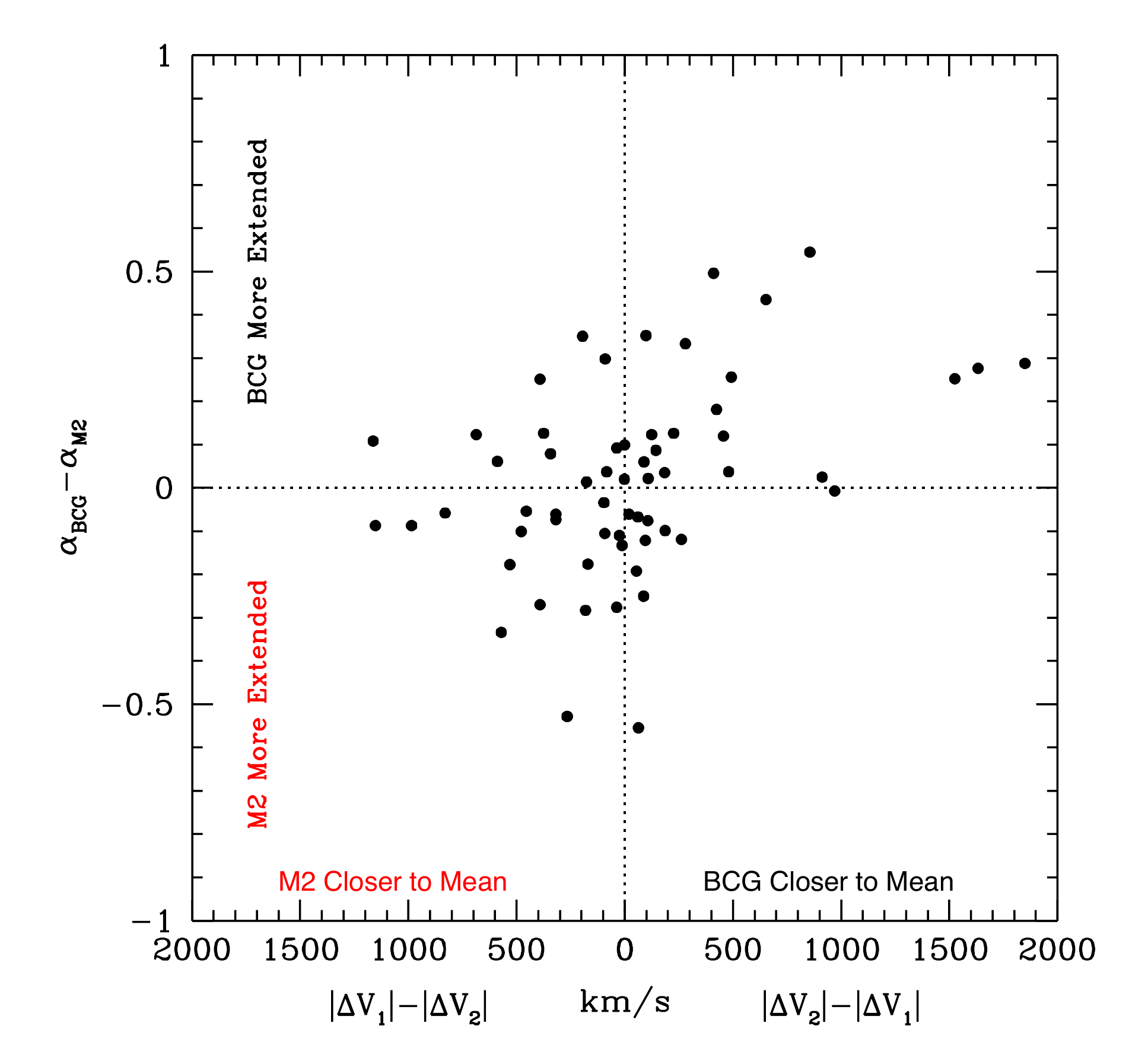}
\caption{The difference in $\alpha$ between
the BCG and M2 in a given cluster is plotted as a function
of the difference between the absolute values of their peculiar
velocities relative to the mean cluster velocity,
for M2 galaxies with $M_m({\rm M2})-M_m({\rm BCG})<0.3$ mag
and clusters having 25 or more galaxy velocities.
When the M2 falls closer to the mean velocity, typically $\alpha_2>\alpha_1,$
while the opposite is true when the BCG is the galaxy closer to the
mean velocity.}
\label{fig:m12dadv}
\end{figure} 

We imaged the second-ranked galaxies, M2, in
$\sim41\%\ $ of the clusters in the total sample.
As we described in \S\ref{sec:m2_sub}, we did this mainly when the identity
of the BCG during the initial definition of the sample was ambiguous,
thus we are less likely to have data on M2 in
cases where the BCG is {\it dominant}, i.e., considerably brighter than M2.
We assess how strongly our observational procedures bias our derived
$M_m({\rm M2}) - M_m({\rm BCG})$ distribution
by randomly selecting 30 clusters in our
15K sample for which we did not observe the second-rank galaxy.
Here, $M_m({\rm M2})$ and $M_m({\rm BCG})$ are the metric
luminosities of the M2 galaxy and the BCG, respectively.
For these 30 clusters, we used the Sloan Digital Sky Survey (SDSS)
r-band images to derive $M_m({\rm M2}) - M_m({\rm BCG})$ 
using the same photometric techniques as described in \S\ref{sec:imred}.
Under the assumption that this randomly selected sample of 30 clusters
is representative of the clusters for which we did not initially observe
the second-ranked galaxy, we derive a corrected $M_m({\rm M2}) - M_m({\rm BCG})$
distribution by drawing $M_m({\rm M2}) - M_m({\rm BCG})$ values from the SDSS
sample for the remaining $\sim59\%$ of the clusters in our sample.
The results are shown in Figure~\ref{fig:m12dist}.
The clusters for which we did observe M2 in our survey have a
$M_m({\rm M2}) - M_m({\rm BCG})$ distribution that is peaked at lower values
than the corrected $M_m({\rm M2}) - M_m({\rm BCG})$ distribution.
We show, for comparison, the $M_2 - M_1$ distribution
from \citet{smith10} and from galaxies drawn from
2000 realizations of clusters with a \citet{lf} luminosity function.
A KS test rejects consistency between our corrected
$M_m({\rm M2}) - M_m({\rm BCG})$
distribution and that from the \citet{smith10}
study at the 99\%\ confidence level.
Our selection procedure clearly introduces a significant bias that
cannot be completely compensated for.
However, the \citet{smith10} clusters are an X-ray selected sample,
and thus may favor more luminous BCGs;
their corresponding M2s may thus follow a different luminosity-offset
distribution from ours as well.
While this rules out performing analyses that require a complete
distribution of $M_m({\rm BCG})$ vs.\ $M_m({\rm M2})$, we can, however,
compare the structure of the BCG and M2 on a per-cluster basis.
It also appears that our M2 sample is close
to complete for $M_m({\rm M2}) - M_m({\rm BCG})\leq0.3$ mag.

Figure \ref{fig:m2lf} shows the distribution of the metric absolute magnitude,
$M_m$, of the present M2 sample compared to the Gaussian representation
of the BCG luminosity function shown in Figure \ref{fig:lm}.\footnote{All
the M2 parameters plotted in the figures
in this section are provided in Table \ref{tab:m2_par}.}
Interestingly, the largest differences between the M2 and BCG
distributions are seen at the bright end, rather than the
fainter end, where one might expect the selection effects
to be the most severe.   There are few highly luminous
M2s, despite a deliberate effort to include such galaxies in the sample.
For example, while there are 139 BCGs that have $M_m<-23,$
there are only nine M2s that exceed this threshold.

\subsection{The Relationship of M2 versus the BCG in the Hosting Cluster}

Figure \ref{fig:m2m1_histograms} plots the BCG versus M2 luminosities
for each cluster, as well as a histogram of the difference in the metric
magnitude between the BCG and M2 in finer bins than shown
in Figure \ref{fig:m12dist}.
In the majority of cases, the M2 systems imaged are within a few
tenths of a magnitude of the BCGs.
Indeed, the M2 in any given cluster may
be more luminous than many of the BCGs in other clusters.
We observed M2s in 54 of
the 139 clusters with BCGs with $M_m\leq-23,$ or 39\%\ of the
systems, a fraction essentially identical to that for all clusters.
At the same time, only 17/54 or $\sim30\%$ of this subset
have M2s within 0.3 mag of the BCG $M_m,$ a level at which
we consider the M2 to be a close ``rival'' of the BCG, while
60/98 or $\sim60\%$ of the BCGs with M2s observed and $-23<M_m\leq-22.5$
have an M2 that is a close rival.
The difference between the two subsets is readily evident
in Figure \ref{fig:m2m1_histograms}.
This result is consistent with the classic result from \citet{sh73} that
the most luminous BCGs are associated with relatively faint M2s.
The large luminosity differences between the BCG and M2 for
the brightest BCGs are also consistent with the
arguments of \citet{tr}, \citet{ls}, and \citet{lom} that
these galaxies are ``special'' and are not drawn from a standard
luminosity function.

Figure \ref{fig:m1m2_rx} shows that the projected offsets of the
BCGs and their corresponding M2s from the X-ray center of their clusters
follow different distributions.  There are essentially no M2s closer
to the X-ray center than the BCG once $r_x({\rm BCG})<100$ kpc.  Even
when $r_x({\rm BCG})>100$ kpc, there are only a handful of M2s with $r_x({\rm M2})<50$ kpc;
of these there are only three M2s with luminosities within 0.3 mag
of their BCGs.  In short, even though we have searched for and selected
BCGs at large distances from the X-ray centers of their clusters,
we are not overlooking a large population of bright ``central'' galaxies
that might plausibly be better choices for the dominant galaxy in the cluster.

There are other ways in which the population of M2 galaxies differs
from that of the BCGs besides just being less luminous.
Figure~\ref{fig:m1m2_lalp_00}
compares the $L_m-\alpha$ properties of the two sets.
In this case, we plot $M_m$ as the independent variable,
instead of assigning this role to $\alpha$ as we did in \S\ref{sec:lalp}.
Intriguingly, this figure echoes the conclusion suggested by
Figures \ref{fig:m2lf} and \ref{fig:m2m1_histograms} that the
differences between M2 and the BCGs are most important for the brightest BCGs.
For $M_m>-22.75,$ the M2 and BCG galaxies have essentially indistinguishable
distributions of $\alpha$ as a function of $M_m.$
It's also worth noting that given the large scatter in the
$L_m-\alpha$ relation, many of the M2s may actually have larger
$\alpha$ values than do their corresponding BCGs.
For $M_m<-23,$ however, while the mean $\alpha$ continues
to increase for BCGs, it does not for the M2s.
The few M2s with $M_m<-23$ have significantly lower $\alpha$ on average
than the corresponding set of BCGs.
Considering the results from both the previous and following sections,
this implies that the bright M2s are galaxies that are
not likely to be close to the X-ray center of their clusters.
Bright M2s with more ``normal'' $\alpha$ would be more centrally located,
and thus vulnerable to merging with the BCG.
The median $r_x$ for M2 with $M_m\leq-22.5$ is 327 kpc,
indeed showing that the brightest M2s are most likely
to be found markedly displaced from the center of the cluster.

\begin{figure}[!t]
\includegraphics[width=\columnwidth]{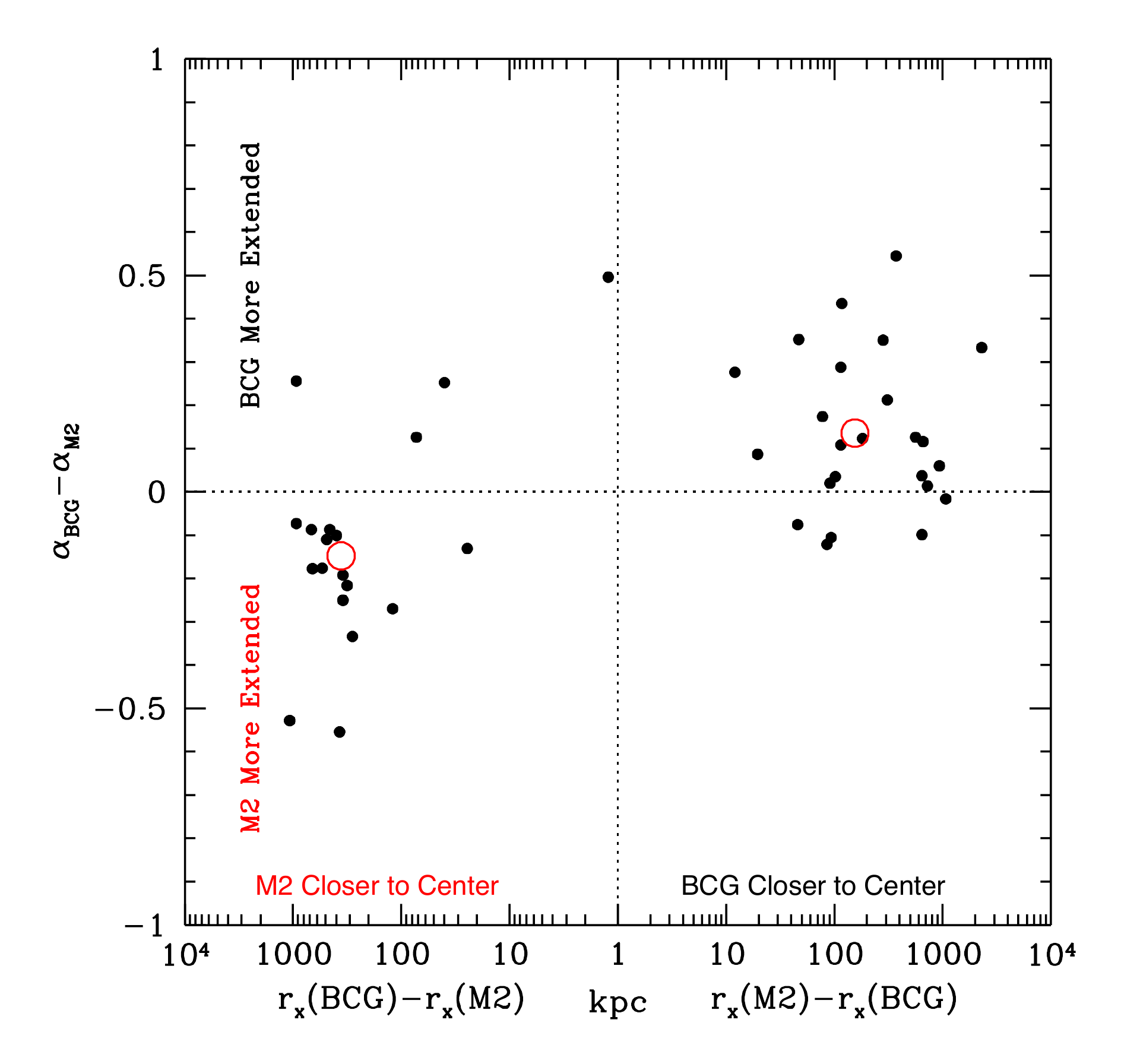}
\caption{The difference in $\alpha$ between
the BCG and M2 in a given cluster is plotted as a function
of the difference between their radial offsets from the
X-ray center in clusters where $M_m({\rm M2}) - M_m({\rm BCG}) < 0.3$ mag.
The left half of the graph plots clusters in which the BCG
is further from the X-ray center than the M2 galaxy,
while the right side plots clusters in which M2 is further away.
The red circle indicates the mean $\Delta\alpha$ for each half of the figure.
When M2 and the BCG are close in luminosity the
galaxy closest to the X-ray center is most likely to have the larger $\alpha.$}
\label{fig:m1m2_dadr}
\end{figure} 

\begin{figure}[!t]
\includegraphics[width=\columnwidth]{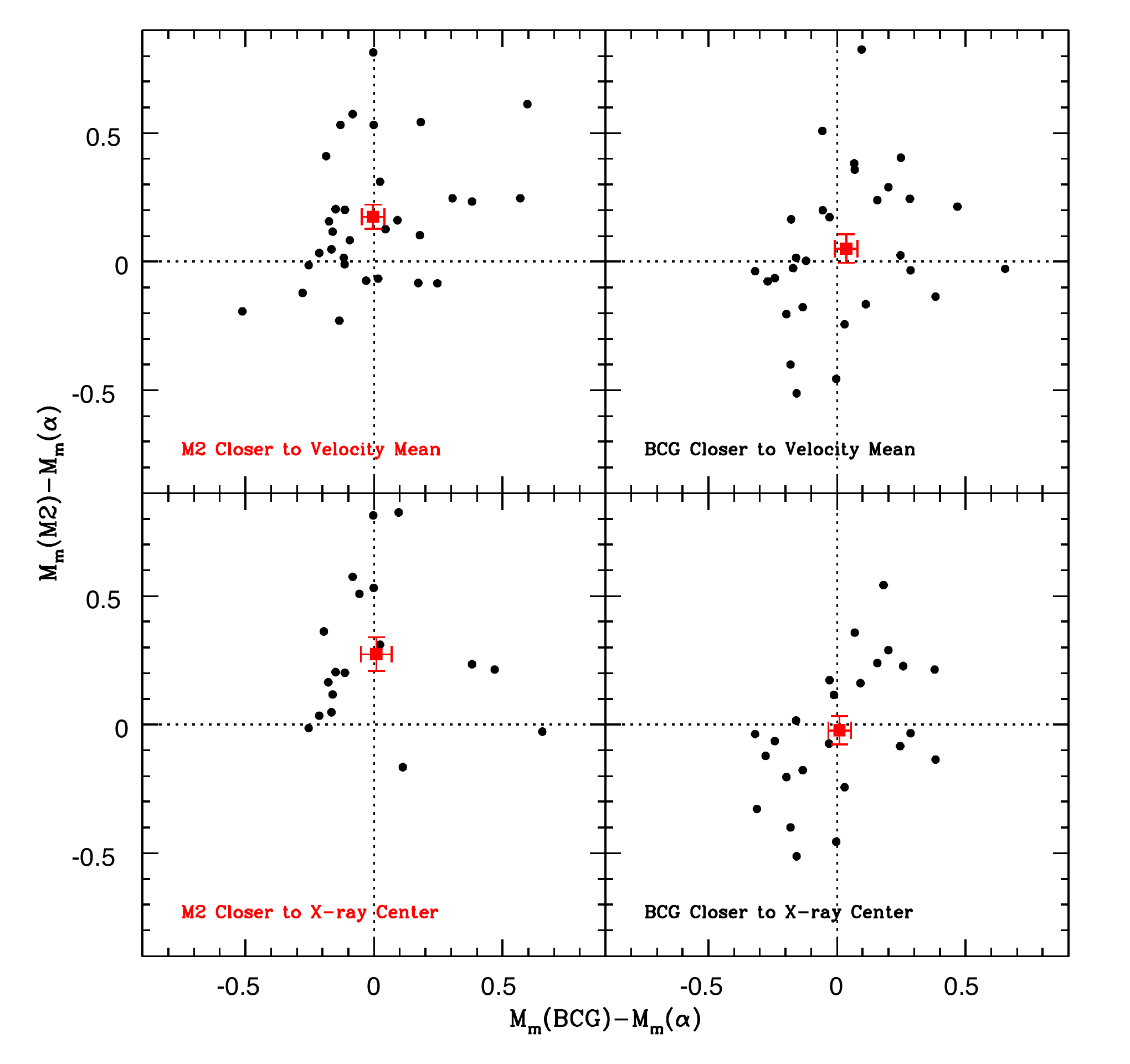}
\caption{The luminosity residuals of the BCGs from the
$L_m-\alpha$ relation (equation \ref{eqn:lalplog}) are plotted
against those of M2 for clusters in which $M_m({\rm M2})-M_m({\rm BCG})<0.3$ mag.
This set of clusters is further divided into four subsets depending
on whether the BCG or M2 is closer to the mean cluster velocity
or X-ray center.  The red symbol in each panel shows the mean
BCG and M2 luminosity residuals.  When M2 is further from the
velocity or X-ray center it appears to follow the
$L_m-\alpha$ relation.  M2s closer to the center, however,
are significantly dimmer than predicted by the $L_m-\alpha$ relation,
given their large $\alpha$ values but lower metric luminosities.}
\label{fig:m1m2_dadl}
\end{figure} 

Additional insight into the differences between M2 and the BCG
comes from considering their relative velocity and spatial displacements
within their clusters.  Figure \ref{fig:m12dldv1} plots luminosity difference
between M2 and the BCG, $M_m({\rm M2})-M_m({\rm BCG})$ as a function of either the
peculiar velocity of the BCG with the cluster (left panel) or that
of M2 (right panel) for clusters with 25 or more members (so as to
minimize the contribution of the error in the mean redshift to
the peculiar velocities).  It is striking that $M_m({\rm M2})-M_m({\rm BCG})<0.5$ mag
for nearly all clusters in which the BCG $\Delta V_1>400~{\rm km~s^{-1}}.$
This suggests a picture in which the
BCGs with high peculiar velocities are relatively recent additions
to the cluster that have not had enough time to undergo the
relaxation interactions
that would move them closer to the cluster velocity centroid.  In this
case the M2 is likely to be the ``former BCG,''
reflecting its rank prior to the infall of a more luminous galaxy.
However, a fraction of these M2s will ultimately merge with the ``new''
BCG once the latter's peculiar velocity has been reduced enough
to make merging interactions possible.
The ``new'' M2 will be less luminous than the old M2 that merged with the
BCG, thus increasing the BCG-M2 luminosity difference for that cluster.

The effect of the BCG on M2 is also visible when $M_m({\rm M2})-M_m({\rm BCG})$
is plotted as function of the BCG X-ray offset (Figure \ref{fig:m1m2_dl_rx1}).
The main signature is that there are no M2s with
$M_m({\rm M2})-M_m({\rm BCG})<0.2$
mag once the BCG gets within 20 kpc of the X-ray center.  Conversely,
there appears to be no relationship between $M_m({\rm M2})-M_m({\rm BCG})$
and $r_x$ for BCGs falling outside this radius.  The figure
also ratifies the conclusion discussed in the context
of Figure \ref{fig:m1m2_rx} that there are essentially no
M2s closer to the center than the BCG once $r_x({\rm BCG})<100$ kpc.

The structure of M2 also appears to be affected by its proximity
to the mean velocity of its hosting cluster, as is the
case for the BCGs.  Figure \ref{fig:m12dadv} compares $\alpha$ between
the BCG and M2 in clusters (with 25 or more galaxy velocities)
in which M2 is a close rival of the BCG ($M_m({\rm M2})-M_m({\rm BCG})<0.3$ mag).
On average, the M2s should have smaller $\alpha,$ given their smaller $L_m;$
however, the figure shows that this is clearly over-ridden by the effect
of the proximity of the galaxies to the velocity centroid of the cluster.
The M2s have slightly {\it larger} $\alpha$ when $|\Delta V_2|<|\Delta V_1|,$
that is when M2 has the smaller peculiar velocity, while the BCGs nearly
always have large $\alpha$ when the situation is reversed.  The
$\alpha$ differences are not symmetrical with the difference
in absolute peculiar velocities, given the initial bias for the BCGs
to have higher $\alpha,$ but the effect is clear.  The processes
that increase $\alpha$ as the BCG approaches the center
of the cluster are also in play for M2.

Likewise, when $M_m({\rm M2})-M_m({\rm BCG})<0.3$ mag,
the galaxy of the two that is closest to the X-ray center of the
cluster is more likely to have the higher $\alpha.$
Figure \ref{fig:m1m2_dadr} plots the difference
in $\alpha$ between the BCG and M2 as function of the difference between
each galaxy's offset from the X-ray center.
For the 24 clusters with $M_m({\rm M2})-M_m({\rm BCG})<0.3$ (and X-ray positions
available) and the BCG closer to the X-ray center than M2 (right half of the
figure), $\alpha({\rm M2})-\alpha({\rm BCG})=-0.14\pm0.4.$
When M2 is closer to the center (left half of the figure),
the situation is reversed, with $\alpha({\rm M2})-\alpha({\rm BCG})=+0.15\pm0.5$
for the 17 clusters in this subset.

The finding that rival M2s have larger $\alpha$ than their corresponding
BCGs despite their smaller metric luminosities when closer to the X-ray
or velocity centers can the BCGs implies that they will strongly deviate
from the $L_m-\alpha$ relation.  This is shown in Figure \ref{fig:m1m2_dadl},
which plots the BCG versus M2 luminosity residuals from the
$L_m-\alpha$ relation (equation \ref{eqn:lalplog}) for clusters
in which the BCG has $M_m({\rm M2})-M_m({\rm BCG})<0.3$ mag.
The set of clusters is divided into four subsets depending on
which of the BCG or M2 is closer to the velocity or X-ray center.
When the M2 is further from either center, both the M2 and BCG
luminosity residuals average to zero, showing that the M2s act
as essentially lower-luminosity BCGs.  Interestingly, the residuals
of both galaxies appear to be correlated for this subset.
When M2 is closer to either the velocity of X-ray center, however,
$M_m({\rm M2})$ is significantly dimmer than predicted from their large
$\alpha$ values.  M2s closer to the cluster center
than their matching BCGs cannot be simply considered to be
less luminous examples of a BCG.

\section{The Present Structure of BCGs as a Reflection of Their Origin}\label{sec:origin}

The initial motivation of this work was to define a BCG
sample to extend our earlier studies of
deviations from the smooth Hubble flow \citep{lp92, lp94}
from the 15K velocity limit to 24K.
As such, a large portion of this
paper was concerned with revisiting the complete definition of the
sample, including selection of the galaxy clusters, selection of the BCG,
measurement of the photometry, spectroscopy, and so on.
Compared to the LP94 sample, the present increase in limiting redshift
leads to a substantial increase in sample size; the 15K sample included
119 BCGs, while the 24K sample comprises 433 galaxies.
Apart from its present use to elucidate the present structure of BCGs,
the sample represents a substantial, full-sky
collection of precise BCG photometry, BCG central stellar velocity dispersions,
cluster redshifts, and cluster velocity
dispersions for nearby galaxy clusters,
all of which can be used for many other investigations.
We return to the questions posed in the Introduction, and finish with
an attempt to integrate these results into an improved picture
of the origin and evolution of BCGs.

\subsection{What Are the Properties of BCGs?}

The initial survey of BCG photometric parameters shows essentially identical
results to those presented in PL95.  The distribution of $M_m$
is Gaussian with a dispersion of 0.337 mags, and the $L_m-\alpha$
relation has scatter of 0.267 mag in the CMB frame.
We now also offer a form that predicts $M_m$ from a linear relationship
in $\log\alpha,$ rather than the quadratic form of PL95,
due to its monotonic behavior with $\alpha;$
it also provides an acceptable fit to the data.
It is notable that the random scatter in $M_m$ has not been reduced
from that presented in PL95, despite new photometry obtained for
most of the 15K BCGs, and improved cluster velocities.

The present work goes beyond PL95 by including
central stellar velocity-dispersion observations of the BCGs.  Use
of metric luminosities obtained in different aperture sizes allows
us to track the flattening of the \citet{fj} relationship between
$L$ and $\sigma$ as the aperture grows to include a larger fraction
of the total galaxy luminosity.  The relationship between $L_m$
and $\sigma$ using our standard 14.3 kpc metric radius
resembles the classic $\sigma\propto L^{1/4}$ relation, while
increasing $r_m$ $4\times$ larger to 57.1 kpc yields a much
shallower $\sigma\propto L^{1/6}$ relation.  The flatter Faber-Jackson
relation for BCGs has been discussed extensively in the literature,
but this shows that it is likely a consequence of the extremely large
envelopes of the BCGs.

The residuals of the $L_m-\alpha$ relationship correlate strongly
with $\sigma,$ thus motivating the development of a three-parameter
``metric plane'' relationship between $L_m,~\alpha,$ and $\sigma$
analogous to the ``fundamental plane'' relations.  Use of $\alpha$ and $\sigma$
together predicts $M_m$ to a precision of 0.206 mag, a substantial
improvement over the $L_m-\alpha$ relationship.
The metric plane relation also implicitly removes
correlations between the residuals of the $L_m-\alpha$ relation
and the cluster velocity dispersion.

\subsection{Where Are the BCGs Located in Their Galaxy Clusters?}

A first step in understanding the relationship of BCGs to their
galaxy clusters is to ask where they are located in the clusters.
In answering this question we have produced quantitative distribution
functions for both the projected spatial and peculiar velocities
of the BCGs with the cluster, where the spatial offset is defined with
respect to the cluster X-ray center, and the velocity with respect
to the mean cluster velocity.  The spatial offset, $r_x$ follows
a steep power-law with $\gamma=-2.33$ over three decades
in radius; there is no evidence for any core at scales $>10$ kpc.

The absolute normalized peculiar velocities
$|\Delta V_1|/\sigma_c$  of BCGs within their clusters
follows an exponential distribution, with scale length 0.39.
The spatial and velocity offsets are correlated.
Large $|\Delta V_1|/\sigma_c$ always corresponds to large $r_x$
and small $r_x$ always corresponds to small $|\Delta V_1|/\sigma_c.$

These results raise an important caveat in understanding
the relation between the BCGs and their hosting clusters.
While BCGs do prefer to reside
near the central regions of galaxy clusters,
BCGs with $r_x>100$ kpc
or $|\Delta V_1|/\sigma_c>0.5$ are common.
These outlying BCGs further follow the same metric plane as do
those closer to the center of their parent clusters.
This has important
consequences for understanding the relationship between
``intercluster light'' (ICL) and the extended envelopes of BCGs.  BCGs
are often simplistically assumed to {\it always} reside at both the
spatial and velocity center of ICL, such that it becomes
ambiguous as to where the BCG ends and the ICL picks up (see the
discussion in \citealt{l07}).  While this may be true in some cases,
it will not be true in general.

Our BCGs are defined by metric luminosity, while some authors
choose the brightest galaxy close to the X-ray center, even if it
turns out to be the M2.
The latter definition would risk the possibility of missing BCG 
that are still being accreted by rich galaxy clusters at the present epoch.
There is no question that there are first-ranked galaxies in
many clusters that are offset by large velocities and projected
separations for the X-ray defined centers.  These are most
easily understood as recently accreted additions to the cluster.
The identity of the galaxy that occupies the first-ranked position
will change as new galaxies are brought in, and may have changed
a number of times in a given cluster over cosmological history,
as is found in the simulations of \citet{deluc} and \citet{martel}.

\subsection{How Does the Cluster Environment Influence
the Properties of the BCGs?}

The relationships that we have observed
between the properties of the BCGs and their clusters support
the picture that the bulk of any given BCG is largely assembled
outside of the cluster, before the galaxy is accreted by the cluster.
The strongest effect is that $\alpha$ is clearly moderated by
both $r_x$ and $|\Delta V_1|/\sigma_c,$ such that $\alpha$
increases monotonically as the BCG lies closer and closer
to the center of the cluster.  BCGs with the largest $\alpha$
for their $L_m$ always reside close to the X-ray and velocity
center of the cluster.  Conversely, BCGs with the smallest $\alpha$
given $L_m$ are strongly displaced from the center, often
with $r_x\gg100$ kpc and $|\Delta V_1|/\sigma_c>0.5.$ 
However, $L_m$ is only weakly, if at all, related to the position of
the BCG within the cluster.

We conclude that the envelopes of the BCGs are expanded by
and perhaps even grown by interactions that become increasingly
important as the BCG spends the majority of the time in ever denser regions
of the clusters and dynamical relaxation reduces its peculiar velocity.
The fact that $L_m$ does not vary with spatial or velocity
offset from the center of the cluster argues that
the denser central body of the BCG, however, is less
affected the same processes.
Some BCGs, even at low redshifts,
have been recently accreted into the outskirts of the clusters.
Even though they may not have the classic extended envelopes
associated with, say, massive cD galaxies, their $L_m$
is already high enough that they can claim first rank over
all other galaxies within the cluster.

It is here that the properties of the M2 galaxies are important.
When M2 is close in luminosity to the BCG, we see
that which of the BCG or M2 has higher $\alpha$ depends
on which is closer to the mean cluster velocity or the center
of the cluster potential as marked by X-ray emission.
Clearly, the processes
that act on the BCG as a function of its location in the cluster
also act on M2.  In fact, the luminosity differences between the two
sets of galaxies reveal the nature of the competition for first rank.
While our sample of M2 is incomplete, we only see large
luminosity differences between them when $|\Delta V_1|$
is relatively small.  We hypothesize that when the BCG initially
enters the cluster with a high $|\Delta V_1|,$ the previous BCG,
now demoted to M2, can still survive as a close rival until the BCG
undergoes enough interactions to be captured into the central potential
of the cluster.  Conversely, when it is the M2 that has a high peculiar
velocity or has an orbit that keeps it mainly
in the outskirts of the cluster,
it can persist as close rival in luminosity to the BCG.

\subsection{A Brief History of BCGs}

Throughout the narrative our results raise the question
of the extent to which BCGs grow their luminosity inside versus outside
of the their present hosting clusters.
We see that there are some BCGs
with large $L_m,$ meaning that the assembly of their
stellar mass is essentially complete, but that also have large
$r_x$ and large $|\Delta V_1|/\sigma_c,$ which means that they
are relatively recent arrivals to their hosting clusters.
In the clusters for which this is so,
there may also be a centrally-located M2 that many would
pick as the ``true'' BCG --- certainly, that galaxy is more likely
to have the greatly extended envelope that's often associated with
the classical pictures of BCGs.
The inference is that most of the stellar mass of any given 
BCG, or at least its portion within the metric radius,
grew from the merger of progenitor galaxies outside
the rich cluster environment.
Such a scenario was suggested by \citet{m85},
who noted that the lower velocity dispersions of galaxy groups,
rather than rich clusters, made them attractive as the birthplace
of future BCGs.  The BCG may represent the merging terminus of
many of the galaxies within the group.  The merged system may
later be accreted by a rich cluster, but then may be subjected to
only minor interactions or mergers once it arrives there.
These would add energy to its envelope but little stellar mass.

On the other hand, in this study we see that
BCG $L_m$ increases markedly with $\sigma_c,$
and as we noted earlier in the Introduction, BCG luminosity
also tracks cluster X-ray luminosity and temperature.
One might presume that this only reflects the likelihood that today's rich
clusters once had the richest retinue of surrounding groups, but there
is evidence that major mergers can take place within clusters \citep{l88}.
The most luminous BCGs are unlikely to have strong M2 rivals for
the position of first rank.  Faint M2 galaxies are further uniquely
associated with BCGs with modest $|\Delta V_1|/\sigma_c.$
Whatever level of minor merging that may have taken place, this result
suggests that many of the current BCGs have cannibalized their closest rivals,
an event that would have been a major merger.
It is also a feature of dry merging that this may cause little
luminosity growth {\it within} the metric aperture \citep{ho,boylan},
while greatly extending the envelope.
The growth in $\alpha$ as the BCGs become more centrally located
argue that this is happening.

The question of whether most BCG luminosity growth occurs inside or outside of
rich clusters intersects with the issue of whether or not BCGs are ``special.''
\citet{lom} have argued that only the BCGs in the most X-ray luminous
clusters exhibit luminosities high enough such that they cannot
be explained as simply being drawn from a standard cluster
luminosity function.  This is thus a partial contradiction of the
analysis of \citet{tr} that suggested that all BCGs were created
by special processes.  In this work we see that the M2s are most structurally
similar to their corresponding BCGs when the BCGs have more modest
metric luminosities.  It is when $M_m$ starts to grow beyond $\sim-22.5$
that we begin to see their properties diverge.
We conclude that if BCGs were born in relatively small groups, it
is their accretion into rich clusters that later shapes
and grows them to their final form and special luminosities.
We will explore these themes further in follow-on papers that
will elucidate the relation of the present sample of BCGs
to the properties of giant elliptical galaxies
and measure the on-going rate of merger interactions in BCGs.

\acknowledgments

This work is based on extensive observational work conducted at Kitt Peak
National Observatory and Cerro Tololo Inter-American Observatory,
both operated by the National Optical Astronomy Observatory (NOAO).
We are indebted to the professional staff of NOAO for the
development and operation of the telescopes and instruments that
we used for this project.
NOAO is operated by AURA, Inc.,
under cooperative agreement with the National Science Foundation.
We thank Beth Reid, Megan Donahue, Kayhan G\"{u}ltekin,
Christine Jones, and William Forman for
useful conversations.
We thank Katelyn Millette for a critical reading of the manuscript.
This research has made use of the NASA/IPAC Extragalactic Database (NED),
which is operated by the Jet Propulsion Laboratory,
California Institute of Technology,
under contract with the National Aeronautics and Space Administration.
This paper has also made use of the data from the SDSS.
Funding for the SDSS and SDSS-II has been provided by the
Alfred P. Sloan Foundation, the Participating Institutions,
the National Science Foundation, the U.S. Department of Energy,
the National Aeronautics and Space Administration,
the Japanese Monbukagakusho, the Max Planck Society,
and the Higher Education Funding Council for England.

\clearpage

\LongTables

\begin{deluxetable*}{rrrlllrrl}
\tabletypesize{\scriptsize}
\tablecolumns{9}
\tablewidth{0pt}
\tablecaption{\label{tab:sample}The BCG and Abell Cluster Sample}
\tablehead{\colhead{ }&\colhead{$\alpha$}
&\colhead{$\delta$}&\colhead{$V_1$}&\colhead{$V_c$}&\colhead{$\sigma_c$}
&\colhead{}&\colhead{}&\colhead{}\\
\colhead{Abell}&\colhead{(2000)}
&\colhead{(2000)}&\colhead{${\rm km~s^{-1}}$}&\colhead{${\rm km~s^{-1}}$}&\colhead{${\rm km~s^{-1}}$}&\colhead{$N_g$}&\colhead{$A_B$}&\colhead{Notes}}
\startdata
14&  3.7957&$-23.8823$&$19280\pm 26$&$19397\pm111$&$\phantom{0}474\pm118$& 18&0.094&E473-G005\\
27&  6.1965&$-20.7324$&$15519\pm  0$&$16037\pm 90$&$\phantom{0}339\pm 79$& 14&0.094&E539-G013\\
71&  9.4947&$+29.6035$&$22044\pm 33$&$22044\pm 33$&$\phantom{0000}\ldots$&  2&0.210&\\
74&  9.5903&$-22.3449$&$19013\pm 46$&$19326\pm 56$&$\phantom{0}212\pm 59$& 14&0.091&\\
75&  9.8692&$+21.2295$&$17610\pm 42$&$17355\pm326$&$\phantom{0}731\pm511$&  5&0.146&\\
76&  9.8597&$+6.7340$&$11283\pm 17$&$11969\pm148$&$\phantom{0}491\pm120$& 11&0.179&I1565 15K\\
77& 10.1181&$+29.5558$&$21424\pm 35$&$21135\pm187$&$\phantom{0}419\pm246$&  5&0.190&U00428\\
85& 10.4602&$-9.3034$&$16581\pm 56$&$16622\pm 62$&$1009\pm 31$&263&0.162&\\
\enddata
\tablecomments{Columns: (1) Abell cluster number, (2) J2000 RA of the BCG,
(3) J2000 Dec of the BCG, (4) heliocentric velocity of the BCG,
(5) mean heliocentric cluster velocity, (6) cluster velocity dispersion (when
$N>3$),
(7) number of cluster galaxies used to derive the mean velocity and dispersion,
(8) \citet{sfd} $A_B$ extinction for the cluster, and (9) notes, which gives
the NCG, UGC, IC, and ESO catalogue designations of the BCGs where available,
and 15K denotes a BCG originally selected in the LP94 and
PL95 sample.}
\end{deluxetable*}
\clearpage

\begin{deluxetable*}{ccrl}
\tabletypesize{\scriptsize}
\tablecolumns{4}
\tablewidth{0pt}
\tablecaption{Abell Clusters Not Observed}
\tablehead{\colhead{ }&\colhead{$V_c$}&\colhead{}&\colhead{}\\
\colhead{Abell}&\colhead{${\rm km~s^{-1}}$}&\colhead{$N_g$}&\colhead{Notes}}
\startdata
A0087&$16566\pm\phantom{0}75$&166&No obs. \\
A0126&$16142\pm139$&\phantom{0}16&No obs. \\
A0160&$12854\pm\phantom{0}85$&113&No obs.	M2 misidentified in 15K sample as BCG. \\
A0396&$\phantom{0}5301\pm\phantom{0}71$&88&No obs. \\
A0419&$20644\pm127$&48&No obs.	Foreground group observed in 15K sample. \\
A0480&$\phantom{0}4363\pm200$&17&No obs. \\
A0484&$20495\pm246$&20&No obs. \\
A0524&$23330\pm156$&27&No obs.	 \\
\enddata
\label{tab:no_obs}
\tablecomments{Columns: (1) Abell cluster number, (2) mean heliocentric cluster velocity,
(3) number of cluster galaxies used to derive the mean velocity, and (4) notes.}
\end{deluxetable*}
\clearpage

\begin{deluxetable*}{rrrll}
\tabletypesize{\scriptsize}
\tablecolumns{5}
\tablewidth{0pt}
\tablecaption{M2 Galaxy Sample}
\tablehead{\colhead{ }&\colhead{$\alpha$}
&\colhead{$\delta$}&\colhead{$V_2$}&\colhead{}\\
\colhead{Abell}&\colhead{(2000)}
&\colhead{(2000)}&\colhead{${\rm km~s^{-1}}$}&\colhead{Notes}}
\startdata
0027&   6.2335&$-20.7316$&$16241\pm 47$& E539-G013; BCG by total flux\\
0071&   9.4697&$+29.6076$&$21436\pm 31$&  \\
0074&   9.9201&$-22.1967$&$19194\pm  0$&  \\
0075&   9.8897&$+21.2361$&$18494\pm120$&  \\
0076&   9.8896&$ +6.8146$&$11916\pm 31$& I1566\\
0086&  10.6667&$-21.8004$&$17639\pm 49$&  \\
0102&  12.1249&$ +1.3531$&$18613\pm 34$& U00496 \\
0116&  14.0482&$ +0.6819$&$20086\pm 60$&  \\
\enddata
\label{tab:m2_sample}
\tablecomments{Columns: (1) Abell cluster number, (2) J2000 RA of M2,
(3) J2000 Dec of M2, (4) heliocentric velocity of M2,
(5) notes, which gives
the NCG, UGC, IC, and ESO catalogue designations of the BCGs where available.
M2 galaxies that exceed the luminosity of the nominal BCG at large radii
are indicated.}
\end{deluxetable*}
\clearpage

\begin{deluxetable*}{cccrc}
\tablewidth{0pt}
\tablecaption{Summary of Imaging Runs}
\tablehead{\colhead{~}&\colhead{~}&\colhead{~}&\colhead{FOV}&\colhead{Scale}\\
\colhead{Run ID}&
\colhead{Telescope}&
\colhead{Detector}&
\colhead{($'$)}&
\colhead{($''$/pixel)}
}
\startdata
KP89F&4m&TI-2&3.99&0.299 \\
CT89F&1.5m&TI-2&3.64&0.273 \\
CT90F&1.5m&TI-2&3.64&0.273 \\
KP91S&2.1m&TEK1024&5.19&0.304 \\
CT91S&1.5m&TEK1024&7.41&0.434 \\
KP93F&2.1m&TEK1024&5.19&0.304 \\
CT93F&1.5m&TEK1024&7.41&0.434 \\
CT94S&1.5m&TEK1024&7.41&0.434 \\
KP94S&2.1m&T1KA&5.19&0.304 \\
KP94F&2.1m&T1KA&5.19&0.304 \\
CT94F&1.5m&TEK1024&7.41&0.434 \\
CT95S&1.5m&TEK2K\#4&14.80&0.434 \\
KP95S&2.1m&T1KA&5.19&0.304 \\
\enddata
\label{tab:imruns}
\tablecomments{Run ID names encode the observatory (CT=CTIO, KP=KPNO),
the year of the observations (e.g., 94 = 1994),
and the observing semester (F=Fall, S=Spring). Note that the
first five runs provided data for the 15K sample presented in PL95.}
\end{deluxetable*}
\vskip 20pt

\begin{deluxetable*}{ccllcccc}
\tablewidth{0pt}
\tablecaption{Summary of Spectroscopic Runs}
\tablehead{
\colhead{~}&
\colhead{~}&
\colhead{~}&
\colhead{~}&
\colhead{Slit}&
\colhead{Spectral Range}&
\colhead{Dispersion}&
\colhead{Spatial Scale}\\
\colhead{Run ID}&
\colhead{Telescope}&
\colhead{Detector}&
\colhead{Disperser}&
\colhead{P.A.}&
\colhead{(\AA)}&
\colhead{(\AA/pixel)}&
\colhead{($''$/pixel)}
}
\startdata
  CT92F   & 1.5-m    & GEC-10    &  KPGL-3   &  90$^{\circ}$ &  4351 $-$ 5759 & 2.449   &     0.398 \\
  CT93S   &  1.5-m   & GEC-10    & KPGL-3    &  90$^{\circ}$ & 4351 $-$ 5759  & 2.447  &      0.398 \\
  KP92S   &  2.1-m   & TI-800       &  G32         &  0$^{\circ}$ &  4275 $-$ 6247 & 2.471  &  \nodata \\
  KP92F   &  2.1-m   & Ford 3KA  &   G32        &  0$^{\circ}$ & 3344 $-$ 7957  & 2.442  &       0.76 \\
  KP93S   &  2.1-m   & Ford 3KA  &   G32        &  0$^{\circ}$ & 3309 $-$ 7949  & 2.443  &       0.76 \\
  CT94S  &   4-m      & GEC-16     &  KPGL-2  &  90$^{\circ}$ & 4629 $-$ 6280  & 2.895  &       0.73 \\
  KP94S  &   2.1-m  & Ford 3KC    &  G32         &  90$^{\circ}$ & 3706 $-$ 7850  & 2.440  &       0.76  \\
  CT94F  &   4-m     & GEC-16       &  KPGL-2  &  90$^{\circ}$ & 4723 $-$ 6367  & 2.891  &       0.73 \\
\enddata
\label{tab:specruns}
\tablecomments{Run ID names encode the observatory (CT=CTIO, KP=KPNO), the year of the observations (e.g., 94 = 1994),
and the observing semester (F=Fall, S=Spring).}
\end{deluxetable*}
\clearpage

\begin{deluxetable*}{lccrcrcc}
\tabletypesize{\scriptsize}
\tablecolumns{8}
\tablewidth{5 in}
\tablecaption{\label{ztable} Redshift Catalog}
\tablehead{
\colhead{Galaxy}&
\colhead{R.A.}&
\colhead{Dec.}&
\colhead{v$_{helio}$}&
\colhead{v$_{err}$}&
\colhead{Tonry}&
\colhead{Emission}&
\colhead{Observing}\\
\colhead{ID}&
\colhead{(J2000)}&
\colhead{(J2000)}&
\colhead{km s$^{-1}$}&
\colhead{km s$^{-1}$}&
\colhead{R value}&
\colhead{Flag}&
\colhead{Run ID}
}
\startdata
A0014-A  &   00 15 10.98 &   $-$23 52 56.2 &   19280 &   26 &  10.12 &    & CT94F\\
A0027-A  &   00 24 56.04 &   $-$20 43 53.8 &   16241 &   47 &   8.47 &    & KP95F\\
A0044-A  &   00 29 31.67 &   $+$12 02 59.0 &   21614 &   29 &  11.64 &    & KP94F\\
A0051-A  &   00 31 42.11 &   $-$23 38 45.9 &   33064 &   43 &   8.87 &    & CT95F\\
A0051-B  &   00 31 41.52 &   $-$23 38 46.2 &   33250 &   53 &   8.04 &    & CT95F\\
A0071-A  &   00 37 58.74 &   $+$29 36 12.6 &   22044 &   33 &  12.46 &    & KP94F\\
A0071-B  &   00 37 52.72 &   $+$29 36 27.5 &   21436 &   31 &  12.67 &    & KP94F\\
A0074-A  &   00 38 21.67 &   $-$22 20 41.5 &   19013 &   46 &   8.72 &    & CT95F\\
\enddata
\label{tab:ztable}
\end{deluxetable*}
\clearpage

\begin{deluxetable*}{rrrcclrc}
\tabletypesize{\scriptsize}
\tablecolumns{8}
\tablewidth{0pt}
\tablecaption{BCG Parameters}
\tablehead{\colhead{ }&\colhead{ }
&\colhead{ }&\colhead{ }&\colhead{ }&\colhead{$\sigma$}
&\colhead{$r_x$}&\colhead{}\\
\colhead{Abell}&\colhead{$\alpha$}
&\colhead{$M_m$}&\colhead{$M_m(2r_m)$}&\colhead{$M_m(4r_m)$}&\colhead{${\rm km~s^{-1}}$}&\colhead{kpc}&\colhead{Ref}}
\startdata
0014&0.983&$-22.872$&$-23.564$&$-24.069$&$219\pm15$&  73&R\\
0027&0.289&$-22.298$&$\ldots$&$\ldots$&$\ldots$&\dots&\dots\\
0071&0.397&$-22.517$&$-22.740$&$\ldots$&$262\pm14$&\dots&\dots\\
0074&0.447&$-22.927$&$-23.223$&$-23.461$&$315\pm 6$&\dots&\dots\\
0075&0.455&$-22.728$&$\ldots$&$\ldots$&$295\pm12$&\dots&\dots\\
0076&0.552&$-22.932$&$-23.329$&$\ldots$&$287\pm 2$& 467&R\\
0077&0.789&$-23.049$&$-23.589$&$-23.982$&$278\pm13$&\dots&\dots\\
0085&1.039&$-23.233$&$-23.848$&$-24.193$&$329\pm17$&   1&C\\
\enddata
\label{tab:bcg_par}
\tablecomments{Columns: (1) Abell cluster number, (2) $\alpha$ at the metric
radius,
(3) absolute metric luminosity, (4) absolute metric luminosity within $2r_m,$
(5) absolute metric luminosity within $4r_m,$
(6) galaxy stellar velocity dispersion,
(7) radial offset of the BCG from the X-ray cluster center,
(8) X-ray reference, R=ROSAT,C=Chandra.
The CMB frame has been adopted for the calculation of metric luminosities.}
\end{deluxetable*}
\clearpage

\begin{deluxetable*}{ccclr}
\tabletypesize{\scriptsize}
\tablecolumns{5}
\tablewidth{0pt}
\tablecaption{M2 Parameters}
\tablehead{\colhead{ }&\colhead{ }
&\colhead{ }&\colhead{$\sigma$}&\colhead{$r_x$}\\
\colhead{Abell}&\colhead{$\alpha$}
&\colhead{$M_m$}&\colhead{${\rm km~s^{-1}}$}&\colhead{kpc}}
\startdata
0027&0.542&$-22.229$&$224\pm 20$&\dots\\
0071&0.323&$-22.340$&$223\pm 21$&\dots\\
0074&0.408&$-22.669$&\dots&\dots\\
0075&0.528&$-22.316$&\dots&\dots\\
0076&0.397&$-22.601$&\dots&313\\
0086&0.469&$-22.864$&$260\pm 16$&235\\
0102&0.549&$-22.580$&\dots&\dots\\
0116&0.473&$-22.513$&\dots&429\\
\enddata
\label{tab:m2_par}
\tablecomments{Columns: (1) Abell cluster number, (2) $\alpha$ at the metric
radius,
(3) absolute metric luminosity, (4) galaxy stellar velocity dispersion,
(5) radial offset of the BCG from the X-ray cluster center,
The CMB frame has been adopted for the calculation of metric luminosities.}
\end{deluxetable*}
\end{document}